\documentclass[12pt]{article}
\pdfoutput=1

\usepackage{amsmath,amsfonts,amssymb,amsthm,amssymb,bbm,bm}
\usepackage{pdfsync}
\usepackage{cite}
\usepackage{graphicx,import}
\usepackage[latin1]{inputenc}
\usepackage{hyperref}
\usepackage{empheq}
\usepackage[all]{xy}
\usepackage{stmaryrd}
\usepackage{rotating}
\usepackage{color}  
\usepackage{slashed}
\usepackage{cancel}
\usepackage{hhline}
\usepackage{array, makecell}
\usepackage{comment}
\usepackage{tikz-cd}
\usepackage{hhline}
\usepackage{comment}
\usepackage{empheq}
\usepackage{booktabs} 
\usepackage{float}

\usepackage{booktabs}
\usepackage[normalem]{ulem}
\DeclareFontFamily{U}{matha}{\hyphenchar\font45}
\DeclareFontShape{U}{matha}{m}{n}{
      <5> <6> <7> <8> <9> <10> gen * matha
      <10.95> matha10 <12> <14.4> <17.28> <20.74> <24.88> matha12
      }{}
\DeclareSymbolFont{matha}{U}{matha}{m}{n}
\DeclareMathSymbol{\oright}       {2}{matha}{"69}

\numberwithin{equation}{section}

\newcommand{\p}{\partial}

\newcommand{\bit}{\begin{itemize}}
\newcommand{\eit}{\end{itemize}}
\newcommand{\bd}{\begin{description}}
\newcommand{\ed}{\end{description}}

\newcommand{\va}{\scriptscriptstyle}

\newcommand{\bc}{\begin{center}}
\newcommand{\ec}{\end{center}}
\newcommand{\lbr}{\llbracket}
\newcommand{\rbr}{\rrbracket}
\newcommand{\C}{{\mathbb C}}

\newcommand{\R}{{\mathbb R}}
\newcommand{\Z}{{\mathbb Z}}

\def\W{W}
\def\heq{\,\hat{=}\,}

\newcommand{\cL}{{\mathcal L}}

\newcommand{\cF}{{\mathcal F}}

\newcommand{\cP}{{\mathcal P}}
\newcommand{\cM}{{\mathcal M}}

\newcommand{\SU}{\mathrm{SU}}

\newcommand{\SL}{\mathrm{SL}}

\newcommand{\be}{\begin{equation}}
\newcommand{\ee}{\end{equation}}
\newcommand{\bea}{\begin{eqnarray}}
\newcommand{\eea}{\end{eqnarray}}
\newcommand{\bs}{\begin{subequations}}
\newcommand{\es}{\end{subequations}}
\newcommand{\nn}{\nonumber}
\newcommand{\la}{\label}

\newcommand{\f}{\frac}

\def\p{\partial}
\def\bC{C}
\def\bD{\bar{D}}
\def\bP{\bar{P}}
\def\bq{\bar{q}}
\def\bU{\bar{U}}

\def\bE{\bar{E}}
\def\bF{\bar{F}}
\def\bR{\bar{R}}

\def\div{\mathrm{div}}
\def\fin{\mathrm{fin}}

\newcommand{\dd}{\mathrm{d}}
\newcommand{\na}{\nabla}

\newcommand{\mean}[1]{\langle{#1}\rangle}

\newcommand{\sd}{\slashed{\delta}}

\def\a{\alpha}
\def\b{\beta}
\def\g{\gamma}
\def\d{\delta}
\def\eps{\epsilon}

\def\th{\theta}

\def\l{\lambda}
\def\m{\mu}
\def\n{\nu}
\def\x{\xi}

\def\s{\sigma}
\def\t{\tau}

\def\om{\omega}

\def\Si{\Sigma}

\def\rd{\mathrm{d}}
\def\pa{\partial }

\newcommand{\elld}{\ell_{\mathrm{div}}}
\newcommand{\varthetad}{\vartheta_{\mathrm{div}}}

\newcommand{\scri}{\cal I}

\newcommand{\ov}[2]{\overset{\scriptscriptstyle {#2}}{#1}}
\newcommand{\bmw}{BMSW }
\newcommand{\bmsw}{\textsf{bmsw }}
\newcommand{\bms}{\textsf{bms }}
\newcommand{\ebms}{\textsf{ebms }}
\newcommand{\gbms}{\textsf{gbms }}

\begin{document}

\title{\Large{\bf 
The  Weyl  BMS group and Einstein's equations  }}

\author{ Laurent Freidel$^1$\thanks{lfreidel@perimeterinstitute.ca} , 
Roberto Oliveri$^2$\thanks{roliveri@fzu.cz} ,
Daniele Pranzetti$^{1,3}$\thanks{dpranzetti@perimeterinstitute.ca} , 
Simone Speziale$^4$\thanks{simone.speziale@cpt.univ-mrs.fr}
}
\date{\today}

\date{
\small{\today\\
\bigskip
\textit{
$^1$Perimeter Institute for Theoretical Physics,\\ 31 Caroline Street North, Waterloo, Ontario, Canada N2L 2Y5\\ \smallskip
$^2$CEICO, Institute of Physics of the Czech Academy of Sciences,\\
Na Slovance 2, 182 21 Praha 8, Czech Republic\\ \smallskip
$^3$ Universit\`a degli Studi di Udine,
via Palladio 8,  I-33100 Udine, Italy
\\ \smallskip
$^4$CPT--UMR 7332, CNRS, 13288 Marseille, France}}}

\maketitle

\begin{abstract}
We propose an extension of the BMS group, which we refer to as  Weyl BMS or \bmw for short, that includes 
super-translations, local Weyl rescalings and arbitrary diffeomorphisms   of the 2d  sphere metric. 
After generalizing the  Barnich--Troessaert bracket, we 
show that the Noether charges of the \bmw group provide
a centerless representation of the \bmw Lie algebra
at every cross section of null infinity. 
This result is tantamount to proving that the
 flux-balance laws for the Noether charges
 {imply} the validity of  the asymptotic Einstein's equations at null infinity.
 The extension requires a holographic renormalization procedure, 
 which we construct without any dependence on background fields. The renormalized phase space of null infinity reveals new pairs of conjugate variables.
 Finally, we show that  \bmw group elements label the gravitational vacua.

\end{abstract}

\vspace{.5cm}

\begin{flushright}
\emph{In memory of Ezra Ted Newman, \\ and the exploration of the heavens that he so passionately pursued}
\end{flushright}
 
\newpage
\tableofcontents

\section{Introduction}\la{sec:Intro}
Since the seminal work of Emmy Noether \cite{Noether:1918zz}, revealing the deep connection between symmetries and charge conservation, the investigation of the symmetries in Nature has driven the most successful and dramatic breakthroughs in all fields of physics. Despite this long history, the study of symmetries does not cease to reveal surprising features of physical systems. One of the most notable examples is surely the central role of symmetries in gravitational systems.

The study of the gravitational group of symmetry at null infinity, the BMS group, dates back to the sixties   \cite{Bondi:1960jsa, BMS, Sachs62,Newman:1962cia}   (see \cite{Winicour16} for a review). 
However, about fifty years later, a surprising
relationship between super-translation symmetry, the soft graviton theorem of Weinberg \cite{Weinberg:1965nx}, and  the displacement
memory effect \cite{Christodoulou:1991cr, ThorneB} was discovered \cite{Strominger:2013jfa} and elaborated in terms of a
so-called infrared triangle \cite{Strominger:2014pwa} 
(see \cite{Strominger:2017zoo, Compere:2018aar} for reviews).
{{Shortly after, the existence of a  subleading soft graviton theorem was shown in
\cite{Cachazo:2014fwa} and it was related to the spin memory effect by \cite{Pasterski:2015tva}. This  subleading soft graviton theorem  was interpreted as a Ward identity for  an extension of the  original BMS group  \cite{Campiglia:2014yka, Flanagan:2015pxa}, called the generalized BMS group, featuring super-Lorentz transformations. This group was  studied canonically in \cite{Compere:2018ylh}, where it was shown that the vacua transitions generated by the generalized BMS group are related to the refraction memory effect.
A proposal for a possible relation between vacua transition and  spin memory effect was given in \cite{Himwich:2019qmj}. 
Overall, this seems to suggest a picture of an infrared square rather than a triangle  \cite{Compere:2018ylh}.
}}
 More recently, a connection between this generalized symmetry group and another extension of BMS called extended BMS
\cite{Barnich:2009se, Barnich:2011mi } was revealed through the shadow transform \cite{Donnay:2020guq}.
The new framework of ``celestial holography'' has originated from these  correspondences. It represents an approach aimed at defining the quantum gravity $S$-matrix by combining CFT techniques and the BMS symmetries of null infinity  \cite{Kapec:2014opa, Kapec:2016jld, Pasterski:2016qvg, Strominger:2017zoo, Donnay:2020guq}.

Simultaneously, BMS-like symmetries have been investigated at finite distances for non-extreme black hole horizons \cite{Donnay:2015abr, Donnay:2016ejv,Donnay:2019jiz}, as well as more general null boundaries \cite{Hopfmuller:2016scf, Wieland:2017zkf, Chandrasekaran:2018aop,Wieland:2020gno}.
This renewed interest in understanding the symmetries of gravity requires developing a general framework that provides clear guidelines to relate the transformation properties of (quasi-local) physical observables to the dynamical content of the gravitational field equations. 

With this in mind, 
the notion of corner symmetry algebra for local gravitational sub-systems \cite{DonnellyFreidel, Freidel:2015gpa, Freidel:2016bxd} has been proposed and provides an algebraic definition of  gravitational subsystems.  
In fact, a new approach to quantum gravity has been advocated in \cite{Freidel:2018pvm, Freidel:2019ees, Freidel:2020xyx, Freidel:2020svx, Freidel:2020ayo,  Donnelly:2020xgu, Livine:2021qwx, Chen:2021vrc}, where states of quantum geometry  are  built
as representation states of the corner symmetry algebra.

Since corner symmetries label states of the system  without reference to time, one has to wonder: How is the dynamics encoded? The proposal first sketched in \cite{Freidel:2019ees} is to recast the quantum dynamics of the gravitational field in terms of flux-balance laws for the corresponding quantum charges.  It then becomes crucial to understand how the corner symmetry algebra extended by super-translations  relates to the flux-balance laws associated with the presence of (null) boundaries. 
The key idea developed here and formulated more generally in \cite{Freidel:2021cbc} is to identify the flux-balance laws with a canonical representation of the symmetry algebra.
This relates the local-holographic description of the bulk gravitational dynamics and the boundary symmetry algebra. 
In the present paper we focus on developing these connections for physical observables on the  celestial sphere, since asymptotic null infinity provides a perfect arena to test the viability and the potential of this new approach.

At the core of this program is the idea that each local gravitational sub-system can be understood as an open Hamiltonian system \cite{Troessaert:2015nia}, where the passage of gravitational (and matter) radiation across the interface between subregions determines the evolution in time of the charges associated to the corner symmetry algebra.
In such a case, there are two fundamental problems to address: first, one needs to choose a  split of the Hamiltonian action of diffeomorphisms between charge and flux;  second, we have to find a notion of bracket that defines the action of symmetry transformations and the time evolution dynamics of the phase space variables, like the Poisson bracket for a closed Hamiltonian system.

Our construction's capstone, developed in \cite{Freidel:2021cbc}, is the introduction of the \emph{Noetherian split}, which allows us to define charges and fluxes in an unambiguous way from a given  Lagrangian.
This split can be viewed as a formalization of the Wald--Zoupas prescription \cite{Wald:1999wa}. This construction builds upon the correspondence between boundary Lagrangian and corner symplectic potential \cite{Freidel:2020xyx}, the   concept of anomaly introduced in \cite{Hopfmuller:2016scf} and the important work \cite{Chandrasekaran:2020wwn}, 
who derived the relationship between anomaly and cocycles. 
The Noetherian split is also instrumental for the derivation, based on first principles \cite{Freidel:2021cbc}, of a new \emph{canonical bracket} of the symmetry charges that generalizes the one introduced by Barnich and Troessaert in \cite{Barnich:2009se,  Barnich:2010eb, Barnich:2011mi}.
The new bracket resolves the ambiguity related to the handling of the non-integrable contributions and the presence of 2-cocycles \cite{Barnich:2007bf,Barnich:2011mi, Barnich:2013axa, Compere:2020lrt,  Adami:2020ugu, Ruzziconi:2020wrb, Alessio:2020ioh, Chandrasekaran:2020wwn, Fiorucci:2020xto}.
In particular, we show following \cite{Freidel:2021cbc, Chandrasekaran:2020wwn} that the bracket provides a faithful and centerless representation of the algebra of vector fields generating the symmetry transformations. 
This  is an essential and necessary step towards quantization:
to have a well-defined quantization prescription, one needs to have a direct equivalence between path integral and canonical quantization, which means that the choice of Lagrangian should determine the charge algebra. And we also need to have a symmetry algebra with no field-dependent cocycles, i.e., we need a symmetry algebra not a symmetry algebroid.
These semiclassical properties are achieved by our construction.

The flux-balance laws for mass and (angular) momentum aspects play a crucial role in general relativity's physical interpretation. These are usually derived using Einstein's equations (EEs) \cite{Compere:2019gft}. Covariant phase space methods \cite{Kijowski1976ACS, Gawdzki1991ClassicalOO, Crnkovic:1986ex,Ashtekar:1990gc,Lee:1990nz,Wald:1999wa} then elevate these laws to relations among canonical generators of the asymptotic symmetries.
Here we show that covariant phase space methods are much more powerful if adequately used: the flux-balance laws can be \emph{derived} 
from the symmetries of the phase space structure and properties of the symplectic 2-form. In other words, working off-shell of the EEs, we show how the charge bracket defined on a general corner is related to Einstein's tensor's projection on the given 2-sphere.

The connection between symmetries and EEs means that the bigger the symmetry group the more equations of motion can be accessed. If one uses at null infinity the original BMS group, then only the evolution equations for the energy can be derived.
To obtain the  momentum evolution equations, it is required instead to use  the generalized BMS group \cite{Campiglia:2014yka,Compere:2018ylh, Campiglia:2020qvc}. To have access to the rest of the asymptotic equations, one  needs to further extend the asymptotic symmetry group  to include conformal transformations of the metric on the 2-sphere. 
This leads us to the introduction of the {Weyl BMS group or, shortly, the} \bmw group, whose infinitesimal generators are shown to satisfy a Lie algebra, which corresponds to the maximally extended sub-algebra of the full diffeomorphism algebra \cite{Ciambelli:2021vnn}  of null infinity $\scri$ associated to a sphere embedded in $\scri$. 
{The \bmw group merges together the two different BMS extensions  \cite{Barnich:2010eb, Compere:2018ylh}  proposed in the  literature, and we show explicitly how these previous extensions can be recovered from it}.
{The larger \bmw group gives us access to more equations of motions. In particular, }
the renormalized charge bracket for  every generator in the \bmw group allows us to recover up to eight  of the ten asymptotic EE, and 
not  just the three flux-balance equations for energy and momentum at $\scri$.

Moreover, the explicit inclusion of conformal transformations in the symmetry group of null infinity allows us to introduce a new Weyl charge, disentangling the super-rotation and super-boost components of the  super-Lorentz transformations canonically studied in \cite{Compere:2018ylh}. In this way, the momentum charge becomes a Hamiltonian  charge with vanishing symplectic flux, providing a preferred notion of angular momentum (see e.g. \cite{Bonga:2018gzr, Compere:2019gft, Ashtekar:2019rpv, Elhashash:2021iev, Chen:2021szm, Compere:2021inq} for a recent debate on this ambiguity).
An important reflection of this feature of the \bmw group is the fact that the orbits of Minkowski spacetime, under the finite action of the group generators, 
yield
non-equivalent vacua
of asymptotically flat spacetimes whose degeneracy is labelled also by the sphere diffeomorphisms, in addition to  the super-translation and the conformal transformation fields as shown in \cite{Compere:2016jwb}.

In the language of \cite{Ashtekar:1981sf, Ashtekar:1981bq}, this conformal extension means that the universal structure on $\scri$ is reduced from a null vector and degenerate metric (original BMS) to  a (thermal) Carroll structure (BMSW)
\cite{Chandrasekaran:2018aop, Ciambelli:2018wre, Ciambelli:2019lap,Donnay:2019jiz}; see also \cite{Herfray:2020rvq, Herfray:2021xyp} for an intrinsic and conformally invariant  geometrical description of null infinity.

As for previously considered extensions of the BMS group \cite{Barnich:2009se, Campiglia:2014yka, Compere:2018ylh}, the \bmw  extension is also confronted with radially divergent charges at face value. While the derivation of the Einstein's equations is  insensitive to these divergences, it is however an important consistency requirement that all charges associated with asymptotic symmetries
should be finite. By exploiting the results of \cite{Freidel:2020xyx}, which relate the introduction of a  boundary Lagrangian  to a shift of the symplectic potential by a corner term in an unambiguous manner, 
we perform a phase space renormalization  for boundary conditions admitting the \bmw symmetry group.  We explicitly show that all Noether charges and fluxes become finite.
This way, the renormalization procedure extends the results of Compere et al. \cite{Compere:2018ylh, Compere:2020lrt} from the generalized BMS to the \bmw group and it reveals an extra pair of conjugate variables parametrizing the phase space of null infinity represented by the 2-sphere scale factor and the Noetherian energy charge aspect. At the same time, 
quasi-local and asymptotic charges are now on same footing thanks to renormalization, and insights from the study of the phase space of null hypersurfaces \cite{Reisenberger:2007ku, Reisenberger:2012zq, Parattu:2015gga, Hopfmuller:2016scf, Reisenberger:2018xkn, Hopfmuller:2018fni,Oliveri:2019gvm} can be imported to study extensions of the symplectic structure of $\scri$
\cite{Campiglia:2015yka, Campiglia:2020qvc}.

The outline of the paper is as follows.
We start with a comprehensive summary of this paper and its companion paper \cite{Freidel:2021cbc} in Section \ref{sec:summary}. This section is self-consistent, and gives a detailed account of the results obtained.
Section \ref{SecGeo} provides a geometric interpretation of the asymptotic expansion of the components of the Bondi metric. Section \ref{sec:WBMS} contains the derivation \emph{ab initio} of the \bmw group and its relation with the original, extended and generalized BMS groups.
In Section \ref{sec:tetrad}, we introduce the Einstein--Cartan Lagrangian, which we use to compute the Noether charges and fluxes associated to the \bmsw generators in section \ref{sec:charges} and \ref{sec:fluxes}, respectively. Section \ref{sec:FB2} provides the derivation of the asymptotic EEs from the flux-balance relations, and Section \ref{sec:div-pot} shows the renormalization procedure for charges and fluxes. We also provide the construction of the boundary Lagrangian that yields the Noetherian construction of the Barnich--Troessaert charges. We conclude with a discussion and future endeavours in Section \ref{sec:Conc}. Four appendices collect technical results about asymptotic expansions in Appendix \ref{AppExp}, the derivation of the flux formulae in Appendix \ref{AppC}, variation of the fields in Appendix \ref{Appvariation}, and about Weyl scalars in Appendix \ref{AppF}.

\vspace{0.8cm}

\paragraph{Notation \& nomenclature:}
We use units $8\pi G=c=1$.
We use Greek letters for spacetime indices, uppercase Latin letters $\{A, B, C, \dots\}$ to label coordinates over the two-dimensional sphere, uppercase Latin letters $\{I, J, \dots\}$ for internal Lorentz indices, and lowercase $\{i,j, \dots\}$ for the dyad on the sphere. The notation $T_{\mean{AB}}$ means the symmetric trace-free components of a given tensor $T$. Given a vector field $V=V^\m\p_\m$, we use the notation $V[\cdot]=V^\m\p_\m (\cdot)$.  
We denote   $\dot{F}=\pa_uF$ and $F'=\pa_r F$. Equations evaluated on-shell of the equations of motion are characterized by the symbol  $\hat{=}$. 
The symbol $\stackrel{\scri}=$ means that the right-hand side is evaluated at future null infinity $\scri$. A bar over a quantity denotes its leading asymptotic value. We denote by  $o(r^x)$ all the terms of order $n<x$ and by $\mathcal{O}(r^x)$ a term of order $n\leq x$, with $x,n\in \Z$. 

We call the  charge associated to the $\mathrm{Diff}(S)$ component of the boundary symmetry group ``momentum charge'' rather than ``angular momentum charge'', as customary in the previous literature.
This different nomenclature reflects the (local) holographic perspective advocated in \cite{Freidel:2019ees, Freidel:2019ofr, Freidel:2020xyx, Freidel:2020svx}, where the corner symmetry charge associated to diffeomorphisms tangent to the corner 2-sphere, being that at finite or infinite distance, has a clear interpretation as a momentum charge. Furthermore, from the point of view of fluid/gravity correspondence \cite{Penna:2015gza, Penna:2017bdn},  this charge has the interpretation of the fluid momenta.

\section{Summary of the results}\la{sec:summary}

In this section, we describe some results of \cite{Freidel:2021cbc} which are needed in the following, and summarize the main results of this paper, namely
\begin{enumerate}
\item the extension of the BMS group  including Weyl conformal rescaling;
\item the derivation of the asymptotic Einstein's equations at null infinity from flux-balance laws;
\item the definition of a procedure for phase space renormalization, yielding finite expressions for the canonical charges and fluxes;
\item the Noetherian derivation of Barnich--Troessaert BMS charges by a shifted Lagrangian;
\item the labelling of the vacua degeneracy spanned by the \bmw group by four arbitrary functions on the sphere, corresponding to super-translations, sphere diffeomorphisms and  conformal transformations. 
\end{enumerate}

Let us start by introducing some main elements of the covariant phase space formalism. We  denote by $\rd$ the spacetime differential and by $\delta$ the field space differential.
As shown  in \cite{Freidel:2020xyx}, using the Anderson homotopy operators \cite{anderson1992introduction}, it is possible to associate a unique symplectic potential $\theta =\theta_{L}$ to a given Lagrangian $L$.
This potential is such that $\delta L = \rd \th - E$, 
where $E$ labels the field equations,
and it is used to define 
the pre-symplectic 2-form $\om=\d\th$. To be more precise, $\th$ and $\om$ are the integrands of the symplectic potentials and 2-form, but with slight abuse of language we will refer to them in the same way. The actual pre-symplectic  2-form is the integral on a 3d hypersurface $\Si$,
\be
\Omega :=\int_\Si \delta \theta. 
\ee
In this paper, we assume that $\Sigma$ is an  hypersurface with  boundary $S :=\pa \Sigma$, which we call the corner.
We do not have to restrict $\Sigma$ to be space-like or null.

Given a vector field $\xi$ in spacetime, 
 we denote $\iota_\xi$ the vector field contraction on spacetime forms 
and $I_\xi$ the field space contraction on field space forms.
 We also denote ${\pounds}_\xi$ the Lie derivative and 
 $\delta_\xi $ the corresponding field variation.
Both Lie derivatives and field variations  are related to the interior products by the Cartan's magic formula
\be
{\cL_\xi}= \rd \iota_\xi +\iota_\xi \rd, \qquad \delta_\xi = \delta I_\xi +I_\xi \delta.
\ee
For the gravitational field space, we have $\delta_\xi g_{\m\n} = \cL_\xi g_{\m\n}$.
In the following, we will be interested in field-dependent diffeomorphisms\footnote{See, e.g., \cite{Adami:2020ugu, Ruzziconi:2020wrb}  for an application in lower dimensional gravity of a field-dependent redefinition
of the generator parameters in order to make the charges integrable.}
, hence $\d\xi\neq 0$.
In this case, it is necessary to consider also the operator $I_{\delta \xi}$ which denotes 
 the field contraction along a  form-valued vector. 
Field-dependent diffeomorphisms appear inevitably when working in a fixed gauge, as is customary done in the study of asymptotic symmetries. 
Indeed, suppose we impose a set of gauge-fixing conditions $ F^i(g_{\mu\nu})=0$.
Obviously, a field-independent vector field will not preserve the gauge, but a subset of field-dependent 
diffeomorphisms  such that $\delta_\xi F^i = {D_\xi}^{i}(F)$, where $D_\xi$ is a linear differential operator, will. This is what happens, as we will discuss below,  when we chose the (partial) Bondi gauge $g_{rr}= g_{rA}=0$.

A key ingredient in the construction of \cite{Freidel:2021cbc} is the {\it anomaly operator} $\Delta_\xi $, first introduced in \cite{Hopfmuller:2016scf} and also studied in \cite{Chandrasekaran:2020wwn}. As shown in more details in \cite{Freidel:2021cbc}, anomalies necessarily appear when one introduces boundary Lagrangians that depend on the extrinsic geometry, such as the normal or its derivative.
Given a form $\om$ both in spacetime and field space, its  anomaly is given by the difference between the field space action and the spacetime Lie derivative
\be
\Delta_\xi \omega, \quad\mathrm{with}\quad
\Delta_\xi:=(\delta_\xi -{\pounds}_\xi - I_{\delta \xi}).\label{anodef}
\ee  
In this paper, we concentrate on the Einstein--Cartan formulation,
which is  a gravity formulation defined by a {\it covariant} Lagrangian, and thus we have 
\be\la{DL}
\Delta_\xi L=0\,.
\ee
This covariance property is in general broken by the phase space renormalization procedure, where in general a non-covariant boundary Lagrangian needs to be introduced. We will come back to this important aspect shortly.

\subsection{Flux-balance relation}\la{sec:FR-rel}
On-shell of the  field equations of motion we have $E \heq 0$, which implies the vanishing of the  constraints  $C_\xi$ defined by 
\be
\rd C_\xi  =I_\xi E\,,
\ee
explicitly
 \be\la{Constraints}
 C_\xi= \xi^\mu G_{\mu}{}^\nu \epsilon_\nu, \qquad \epsilon_\mu = \iota_{\pa_\mu}\epsilon,
 \ee
 where $G_{\mu\nu}$ is the Einstein tensor and $\epsilon$ is the volume form.
In situations where \eqref{DL} is satisfied, we can define the {\it Noether charge aspect}
\be
 \rd q_\xi  \hat{=} I_\xi\theta -\iota_\xi L\,, 
\ee
whose integral is the {\it Noether charge},
\be\la{NQ}
Q_\xi :=\int_S q_\xi\,. 
\ee
We also introduce  the  \emph{Noetherian flux}
\be\la{NF}
\cF_\xi :=\int_S \left(\iota_\xi \theta + q_{\delta \xi}  \right)\,,
\ee
so that the fundamental canonical relation\footnote{ The contraction $I_\xi \Omega= {\delta_{\xi}}\, \lrcorner\, \Omega $ is also denoted 
$\Omega( \delta_{\xi},\delta)$ in the literature, and sometimes  $-\sd H_\xi$.}
\be\label{Flux0}
- I_\xi \Omega \heq  \delta Q_\xi  - \cF_\xi
\ee
is satisfied.

 The main result of  \cite{Freidel:2021cbc} (which build upon the results of   \cite{Chandrasekaran:2020wwn})  is to use the covariant phase space
formalism to show how the choice of a Lagrangian defines uniquely a Noetherian split for charges\eqref{NQ} and fluxes \eqref{NF} which allows one to define flux-balance laws in terms of a bracket available for open Hamiltonian systems, even in the presence of anomalies.  
 While the general derivation in  \cite{Freidel:2021cbc,Chandrasekaran:2020wwn} accounts for the presence of anomalies in both the Lagrangian and symplectic potential, we are initially interested in an application of this bracket in the case where these vanish, like in the  Einstein--Hilbert or Einstein--Cartan formulations. It is important to realize though that, even in the simpler case where a covariant Lagrangian is available, the anomaly of the Noether charge {\it does not vanish}, and one has in general 
  \be\label{QAnomaly}
\Delta_\xi Q_\chi=
Q_{\delta_\chi\xi}
-Q_{\lbr\xi,\chi\rbr},
\,
\ee
 where
   the modified bracket $\lbr \cdot ,\cdot \rbr $ is defined by
  \be\la{Lie-mod}
  \lbr\xi,\chi\rbr:= [\xi, \chi]_{\mathrm{Lie}} + \delta_\chi \xi - \delta_\xi \chi\,.
 \ee
This modification of the Lie bracket is necessary in order to take into account the field dependence of the transformation generators, such that the commutator of two field space variations is again given by the a symmetry transformation
 \be\label{maincom}
 [ \delta_\xi,  \delta_\chi]= -\delta_{ \lbr \xi,\chi\rbr}\,.
 \ee
The key relation \eqref{QAnomaly}
can be used to show that the bracket 
  \be\label{BTb}
  \{Q_\xi,Q_\chi\}_{L}:= \delta_\xi Q_\chi - I_\chi \cF_\xi
  + \int_S \iota_\xi \iota_\chi L
 \ee
satisfies the  off-shell {\it  flux-balance relation}
 \be\la{TheBra}
 \boxed{\,\,  \{ Q_\xi, Q_\chi\}_L+Q_{\lbr\xi,\chi\rbr}=-\int_S\iota_\xi  C_\chi\,.\,\,}
 \ee
 {
The bracket \eqref{BTb} is a generalization of the bracket first considered by Barnich and Troessaert \cite{Barnich:2009se,  Barnich:2010eb, Barnich:2011mi}, as it contains an extra term depending on the Lagrangian. The expression \eqref{BTb} is valid for non-anomalous Lagrangians and can be generalised to include Lagrangian anomalies. The generalized Barnich--Troessaert bracket \eqref{BTb} has been explicitly shown in  \cite{Freidel:2021cbc} to satisfy the  Jacobi identity in full generality, including the case when anomalies are present.
Moreover, it satisfies the following {two} essential properties.
}

 First, it is \emph{independent} of the choice of  boundary Lagrangian
 and corner symplectic potential $(\ell,\vartheta)$. {Namely,} if we consider 
  a change of Lagrangian $L'=L+\rd\ell$ and 
  symplectic potential $ \theta'-\theta =  \delta \ell-\rd \vartheta $,  the new Noether charge and flux  are related to the old ones by the following shifts
\be \label{trans1}
Q'_\xi-Q_\xi =  \int_S (\iota_\xi \ell-I_\xi \vartheta ),
\qquad 
\cF'_\xi-\cF_\xi 
=  \int_S \left(\delta \iota_\xi \ell-\delta_\xi \vartheta \right) .
\ee
What is remarkable in these formulae is the fact that, even if the boundary Lagrangian and corner symplectic potential break the covariance property of the original pair $(L, \theta)$---like, in general, it is the case for holographic phase space renormalization (see  Section \ref{sec:Ren})---  all the anomaly contributions finally drop out of the shift, as shown in \cite{Freidel:2021cbc}.  
The shifts  of charge and flux  \eqref{trans1} preserves the fundamental  relation \eqref{Flux0} and they leave the LHS of  \eqref{TheBra}  invariant, namely
 \be\label{brack3}
  \{Q'_\xi,Q'_\chi\}_{L'} + Q'_{\lbr \xi,\chi\rbr} = 
   \{Q_\xi,Q_\chi\} + Q_{\lbr \xi,\chi\rbr} \,.
 \ee
 It is important to keep in mind though that the shifted bracket is now modified by the boundary Lagrangian anomaly; explicitly, the shifted bracket now reads
 \be\la{bran}
  \{Q'_\xi,Q'_\chi\}_{L'}=\delta_\xi Q'_\chi - I_\chi \cF'_\xi
   +K^{L'}_{(\xi,\chi)}\,,
 \ee
 where
 \be\la{K}
 K^{L'}_{(\xi,\chi)}= \int_S\left(   \iota_\chi  \Delta_\xi\ell -\iota_{\xi}  \Delta_\chi\ell\right)+ \int_S \iota_\xi \iota_\chi L'\,.
 \ee
This expression of the cocycle component of the bracket in terms of the boundary anomaly  was first derived by Speranza and Chandrasekaran in   \cite{Chandrasekaran:2020wwn} (see also \cite{Francois:2020tom} for the inclusion of anomalies in the study of covariant phase space with boundaries in gauge theories).
This means that the flux-balance relation \eqref{TheBra} is  also preserved under change of boundary Lagrangian. 
This invariance property guarantees that it can be used also in the case where the expressions for charges and fluxes are formally divergent. In fact, as we will anticipate in the next section, the phase space renormalization procedure amounts to a shift by a  boundary Lagrangian so that the shifted symplectic potential $\theta'$, and the corresponding shifted charges and fluxes \eqref{trans1}, become finite. In this way, 
one can use the bracket  \eqref{BTb}  at finite distance and see that all the potentially divergent terms in the flux-balance law cancel each other so that one can then take the limit to null infinity. 
 This strategy is indeed available in the Einstein--Cartan formulation with zero cosmological constant and it will be worked out in Section \ref{sec:FB2}.

The second main property of the bracket \eqref{BTb} is the fact that it provides a \emph{ representation} of the  commutator \eqref{maincom} when on-shell of the constraints. 
 In particular, the flux-balance relation \eqref{TheBra}, when applied to null infinity, explicitly connects the symmetry algebra on the celestial sphere to the asymptotic Einstein's equations at $\scri$.  In other words, Einstein's equations at null infinity can be expressed as the demand that the Noether charges corresponding to the \bmw transformations, defined in Section \ref{sec:WBMS}, form a representation of the \bmw  algebra under the bracket \eqref{BTb}  on {\it every} cross section of $\scri$.
 One of the results of this paper is indeed to explicitly show that demanding 
\be\la{Flux1}
 \{ Q_\xi, Q_\chi\} = -Q_{\lbr\xi,\chi\rbr}
\ee 
{\it implies} 
the Einstein equation projected along the pair $(\xi,\chi)$.
Finally, the relation \eqref{TheBra} can equivalently be written as 
\be
\d_\xi Q_\chi + Q_{\lbr\xi,\chi\rbr} \heq I_\chi \cF_\xi - \int_S \iota_\xi \iota_\chi L,
\ee
which expresses how the charges evolve under the infinitesimal translation by $\xi$, making it manifest why it is a flux balance relation.

\subsection{Future null infinity}\la{sec:null-infinity}

In order to present our main results concerning the  renormalized phase space at  $\scri$, it is convenient to introduce first the Bondi formalism for the parametrization of future null infinity  and a comparison between the boundary conditions we will impose  and the original BMS boundary conditions.
We use Bondi-Sachs coordinates $x^\mu=(u,r,\sigma^A)$, where $u$ labels outgoing null 
and twist-free geodesic congruences, $r$ is a parameter along these geodesics, which measures the sphere's radius, 
and $\s^A$ are coordinates on the 2-sphere.
The metric in these coordinates can be conveniently parametrized as follows \cite{Bondi:1960jsa, BMS, Sachs:1962wk}\footnote{With $F=V/(2r)$ in terms of Sachs' original parametrization.}
\be \label{eq:BondiMetric}
\rd s^2 = -2e^{2\beta} \rd u \left( \rd r  + F \rd u\right) + r^2 q_{AB} \left(\rd \sigma^A -U^A\rd u\right)\left(\rd \sigma^B -U^B\rd u\right).
\ee
The metric $q_{AB}$ and its inverse $q^{AB}$ can be used to raise and lower the indices of tensors on the 2-sphere. We denote $D_A$ its Levi-Civita covariant derivative, and $R_{ABCD}=q_{A[C}q_{D]B}R$ its Riemann tensor and Ricci scalar.

The Bondi gauge conditions are   
\be\label{Bondig}
g_{rr}=0,\qquad g_{rA}=0,\qquad \pa_r \sqrt{q}=0, 
\ee
where $q=\det q_{AB}$. This gauge-fixing allows one to study the limit to future null infinity taking $r\rightarrow\infty$. The original BMS boundary conditions, or fall-off conditions, are \cite{Bondi62,Sachs:1962wk}
\be\label{BondiAC}
g_{ur}= -1+\mathcal{O}(r^{-2}),\quad 
g_{uA} = \mathcal{O}(1), \quad  g_{uu}=-1+{\cal O}(r^{-1}),  \quad q_{AB}=\ov{q}{\circ}_{AB}+{\cal O}(r^{-1}),
\ee
where $\ov{q}{\circ}_{AB}$ is the metric of the round 2-sphere. 
We wish to consider more relaxed boundary conditions, with the goal of finding a larger extension  of the BMS group. 
To that end, we keep the first two conditions in \eqref{BondiAC}, but allow for an arbitrary leading order metric $\bar q_{AB}$ in the last. The asymptotic form of the Einstein's equations then requires to relax also the condition on $g_{uu}$ for consistency. Accordingly, our proposal is 
\be\label{ExtAC}
g_{ur}= -1+\mathcal{O}(r^{-2}),\qquad 
g_{uA} = \mathcal{O}(1), \qquad  g_{uu}=\mathcal{O}(1),  \qquad q_{AB}=\mathcal{O}(1).
\ee
From these asymptotic conditions, the metric coefficients have the following fall-off behavior,
\begin{subequations}\la{eq:FallOff}
\begin{align}
F&= \bar{F}- \frac{  M}{r}+o(r^{-1})\,,\\
\beta&=\frac{\bar\beta}{r^2}+o(r^{-2})\,,\\
U^A&=\frac{\bar{U}^A}{r^2}- 
\frac{2}{ 3 r^3} \bar{q}^{AB}\left(\bP_B+ \bC_{BC}\bU^C +\pa_B\bar\beta\right) +o(r^{-3})\,,\la{UA}
\\ 
q_{AB} &= \bar{q}_{AB} + \frac{1}{r}C_{AB} + \frac{1}{r^2}\left(D_{AB}+\f14 \bar{q}_{AB}C_{CD}C^{CD} \right)
+ \f1{r^3} E_{AB} +  o(r^{-3}).
\end{align}
\end{subequations}
All functions here depend a priori on the 2-sphere coordinates and on retarded time $u$.
$M$ is the Bondi mass aspect, and 
the asymptotic expansion of $U^A$ is chosen to insure that the momentum aspect $\bP_A$ is the BMS Noether charge associated with super-Lorentz transformations; more on this below.
The 2d tensors 
$(C_{AB}, D_{AB}, E_{AB})$ are symmetric and traceless as a consequence of the determinant gauge condition. The tensor $C_{AB}$ is (twice) the asymptotic shear of the null geodesic congruence. 
The tensor $D_{AB}$ is known to vanish if we demand smooth fall off, namely absence of logarithm terms (see, e.g., \cite{Winicour16}). We will follow this assumption in this paper, and thus set from now on $D_{{AB}}=0$.
The tensor $E_{AB}$ appears as the dominant term in the expansion of one of the components of the Weyl tensor, see App. \ref{AppF}.

Notice that the asymptotic expansion of the metric coefficients listed above is consistent to inverse cubic order. To see this one assigns a notion of conformal weight  which is such that $[r]=-1$ and $[\rd s^2]=-2$. This implies that $[q_{AB}]=0$, $[U^A]=0$ and $[C_{AB}]=1$ while  $[F]=2$, $[\beta]=1$.
This means that to achieve a cubic order expansion we only need to expand $F$ to order $r^{-1}$ since $[M]=3$ and similarly we only need to expand $\beta$ to order $r^{-2}$.

Indices in the asymptotic expansion are raised and lowered using the leading order metric $\bq_{AB}$. We denote $\bD_A$ and $\bar R_{ABCD}$ the associated covariant derivative and Riemann tensor.
With this expansion, the asymptotic Einstein's equations give the following relations for the leading order terms (see section~\eqref{sec:FB2} below  for details)
\begin{subequations}\la{asym-EE}
\begin{align}
\pa_u \bar{q}_{AB}&=0\,, 
 &\textsf E_{\bF}&:= \bar R -4 \bar{F} = 0\,,\la{asym-EE1} \\
 \textsf E_{\bar \beta}&:=  \bar\beta + \frac{1}{32 }  \bC_{AB}\bC^{AB}=0, 
& \textsf E_{\bU^A}&:=  \bar{U}^A+ \frac12\bar{D}_B \bC^{AB}=0. \la{asym-EE2}
\end{align}
\end{subequations}
The first equation can be understood as a boundary condition, rather than a dynamical equation. 
The remaining three equations restrict the number of constraint-free data that one has to assign, and have therefore a completely dynamical meaning.\footnote{We also notice that the first two imply that the leading order of $g_{uu}$ and $q_{AB}$ is $u$-independent.} 
 As shown in Section \ref{sec:FB2}, the time-derivative of  \eqref{asym-EE2} can be understood as flux-balance laws, associated to conformal rescaling transformations. 
 The rest of the asymptotic equations are the
conservation  equations $\textsf E_{ M}=0=\textsf E_{{\bP}_A}$
for the mass and momenta aspects, with
\begin{align}
\textsf E_{ M}  &:=\dot{M} - \f14 \bD_A\bD_{B} N^{AB} -\f12 \bar{\Delta}\bar{F} + \f18 N_{AB}N^{AB}  ,\label{EEM}\\
\textsf E_{{\bP}_A} &:= \dot{\bP}_A -\bD_A M - \f18 \bD_A \left( C^{BC}N_{CB}\right) - C_{AB}\pa^B \bar{F} \cr
&\quad -\f14 \bD_C \left( \bD_A \bD_B C^{BC} - \bD^C\bD^{B}C_{AB}\right)\cr
&\quad -\f14 \bD_B \left( N^{BC}C_{AC} - C^{BC}N_{AC}\right) + \f14 N^{BC}\bD_A C_{BC}\,.\la{EEP}
\end{align}
These can also  be understood as flux-balance laws
associated to super-translation and {sphere diffeomorphism, respectively}.

There are two other declinations of the BMS boundary conditions in the literature.
In \cite{Barnich:2009se,Barnich:2010eb,Barnich:2016lyg}, the authors allow for a divergent term in $g_{uu}$ of  order $r$, while restricting the leading order metric to be conformally related to the round metric $\mathring{q}_{AB}$. In this set-up, the asymptotic Einstein's equations allow for a $u$-dependence in the conformal factor. We will refer to these boundary conditions as the \emph{extended} BMS boundary conditions and the corresponding symmetry group as the extended BMS group.
Another set of boundary conditions proposed in 
\cite{Campiglia:2014yka, Compere:2018ylh} is similar to the one described here. Except that it fixes the determinant of the boundary metric $\sqrt{\bq}$ while allowing for an arbitrary variation of its conformal class {$\sqrt{\bq}\bq^{AB}$}.
We will refer to these boundary conditions as the \emph{generalized} BMS boundary condition and the symmetry group as the generalized BMS group.

Our proposal is more general than \cite{Campiglia:2014yka, Compere:2018ylh}, in that we allow also the determinant of the leading metric to vary. This introduces additional divergences in the symplectic potential, but we will show that it is possible to renormalize them as well, thus extending the results of \cite{Compere:2018ylh} {in the case of the generalized BMS group}.
However, it does not go as far as introducing a linear divergence in $g_{uu}$ and thus a time dependence of the leading order metric. One reason for this is that $g_{uu}=\mathcal{O}(1)$
appears to be sufficient to describe the physics of compact binaries \cite{Blanchet87,Blanchet:2020ngx}.\footnote{The Einstein's equations can be perturbatively solved in harmonic gauge under the assumption of no incoming radiation \cite{Blanchet87}. Transforming to Bondi or Newman--Unti coordinates one obtains $g_{uu} = \mathcal{O}(1)$  \cite{Blanchet:2020ngx}.
} 
The consequences of these different boundary conditions on the asymptotic symmetry group are discussed in Section~\ref{sec:WBMS}.

 \subsection{Phase space renormalization}\la{sec:Ren}

If a  Lagrangian 
has a  symplectic potential 
that diverges in an asymptotic limit, it can be tempting to strengthen the fall-off conditions in order to remove such divergences. However, doing so 
arbitrarily restricts the phase space definition and can set to zero quantities of physical interest.
One may ask whether a more general procedure of renormalization exists, that allows one to identify from first principles what are the weakest fall-off conditions acceptable. This is the idea behind the holographic renormalization procedure.

Holographic renormalization techniques for conserved charges  have been developed in the context of AdS/CFT in  \cite{deHaro:2000vlm, Papadimitriou:2005ii, Compere:2008us,Compere:2020lrt}. 
These techniques have been revisited in the context of flat space asymptotic by \cite{Freidel:2019ohg} for QED in all dimensions and most notably by \cite{Compere:2018ylh} for the case of 4d gravity. These methods rely on adding 
 to the gravitational action a set of  covariant boundary counter-terms, and have been applied to the case of $  \Lambda$-${\rm BMS}$ in 4 dimensions in \cite{Compere:2020lrt}. The main new ingredient is the focus on the renormalization of the symplectic potential and not simply on the renormalization of the Lagrangian. 
 Here we follow an approach closer to
\cite{Compere:2018ylh}, where the phase space renormalization is defined by the addition of a boundary Lagrangian and related corner symplectic potential \cite{Freidel:2020xyx} such that the shifted pair $(L', \theta')$ is finite. Our approach relies on the choice of a particular foliation of the boundary induced by the choice of Bondi coordinates, but it  does not use background fields.

Let us summarize the results of the procedure worked out in detail in Section \ref{sec:div-pot}.
Suppose that one starts from a covariant Lagrangian $L$ with covariant symplectic potential $\theta$
  that possesses some divergences.
  The goal is to 
  define a finite renormalized Lagrangian $L_R$
  with symplectic potential $\theta_R$ by the subtraction of a boundary divergent Lagrangian $\ell_{\mathrm{div}}$ with  corner symplectic potential term $ \varthetad$. 
The subtraction terms may have non-vanishing  
  anomalies $\Delta_\xi \elld$ and  $\Delta_\xi \varthetad$ 
  -- let us stress again that the source of these possible anomalies is the field dependence of the transformation parameters and eventually the coordinates dependence, and not due to the introduction of background structures; moreover, the anomalies do not enter the expressions  \eqref{trans1} for the shifted charges and fluxes.
 The renormalized quantities are then 
 \be\la {ren1}
 L_R := L+\rd \ell_{\mathrm{div}}, \qquad \theta_R :=\theta - \rd \vartheta_{\mathrm{div}}
 + \delta \ell_{\mathrm{div}}\,.
 \ee
 We can thus define, from the general expressions in \eqref{trans1},  the renormalized  Noether charge 
\be
 Q_\xi^R 
 =Q_\xi +\int_S  (\iota_\xi \ell_{\mathrm{div}} -I_\xi \vartheta_{\mathrm{div}} )
 \, ,\la{Qren}
\ee
and the renormalized Noetherian flux
\be
\cF^R_\xi
=\cF_\xi+ \int_S (\d\iota_\xi \elld-\d_\xi \varthetad )\,.\la{Fren}
\ee
 In Section \ref{sec:div-pot}, we  show that the divergent term of the symplectic potential at $\scri$ can be written as
 \be\label{thdvarth}
 {\theta}_{\mathrm{div}} =  
 {\rd} \varthetad\, ,
 \ee
with $\elld=0$  on-shell of $\pa_u \bar{q}_{AB}=0$ and
\be
\varthetad =  \left(  \frac{r^2}{2}\delta\sqrt{\bq}  -\f{r}4  \sqrt{\bq} C^{AB} \d \bar q_{AB}    \right)  \rd^2\sigma
+r \bar\vartheta \wedge \rd u\,,
\ee
 where $\bar\vartheta$ is a one form on $S$ such that 
 \be 
\rd \bar\vartheta = \frac12 \delta \left(\sqrt{\bq}\bR 
\right).
\ee
Its explicit expression is given in \eqref{vartt}.

Furthermore, our phase space renormalization procedure yields a finite expression for the 
symplectic 2-form  at null infinity\footnote{In the Bondi coordinates we will use below, we pick an orientation with $\epsilon = e^{2\beta} r^2\sqrt{q}\, \rd u\,\rd r \, \rd^2\sigma$,  which means that 
 $\epsilon_u = e^{2\beta}r^2 \sqrt{q} \,\rd r\,  \rd^2 \sigma$ while
 $\epsilon_r= -e^{2\beta}r^2 \sqrt{q} \,\rd u\, \rd^2\sigma$ contains a minus sign.} 
\begin{empheq}[box=\fbox]{align} \la{Omega}
\Omega^R_{\scri} = -\int_{\scri} \delta \theta_R^r \,\epsilon_r 
=\int_{\scri} \bigg[&\frac14 \d N_{AB}\!\curlywedge\! \delta (\sqrt{\bq}\, C^{AB})
 - \frac12 \d\left( \bF  C_{AB}   +D_{\langle A} \bU_{B\rangle}  \right)\!\curlywedge\!  \delta \left( \sqrt{\bar{q}}\,  \bq^{AB} \right)\cr
  &+\d \left(M -{ \frac12}  D_A\bU^A\right) \!\curlywedge\! \delta\sqrt{\bq}
 \bigg]\dd u ~\dd^2 \sigma\,,
 \end{empheq}
 where we use the curly wedge symbol to denote a wedge in field space.
If we restrict to the original BMS boundary conditions, only the first pair of conjugate variables in \eqref{Omega} remains and we recover the original radiative phase space of \cite{Ashtekar:1981bq, AshtekarReula}. The \bmw extension proposed here allows us to reveal two extra pairs  so that the phase space at $\scri$ on which the \bmw group acts is parametrized by the set of  conjugate variables
\be
 N_{AB}\leftrightarrow\sqrt{\bq}\,C^{AB}\,,\quad 
 \left(\bF  C_{AB}   +D_{\langle A} \bU_{B\rangle}\right)   \leftrightarrow\sqrt{\bq}\,\bq^{AB} \,,\quad 
\left( M -{ \frac12}  D_A\bU^A\right) \leftrightarrow \sqrt{\bq}\,.\la{pairs}
\ee
Note that, in each pair, one of the variable is always a density. The second pair, which contains the complex structure on the sphere, as $\sqrt{\bq}\,\bq^{AB} $ is conformally invariant,\footnote{ If one denote $\epsilon_{AB}$ the Levi-Civita symbol, the sphere complex structure is $ J_A{}^B: = 
\sqrt{\bq}\,\epsilon_{AC}\bq^{CB}$, and we can check that it satisfies 
$J^2=- 1$ and $D_A J_B{}^C=0$. } was already revealed in \cite{Compere:2018ylh}. The symplectic potential of \cite{Compere:2018ylh} can be obtained up to a corner term\footnote{ The relationship between the generalized \bms symplectic structure of Compere et al. \cite{Compere:2018ylh} and ours is $ \Omega_{\bmsw}= \Omega_{\bms} - \f14 \int_{\pa \scri}\delta C^{AB} \curlywedge \delta C_{AB}$ after imposing $\delta\sqrt{\bq}=0$.} 
from ours by imposing $\delta\sqrt{\bq}=0$.
The third pair contains the  sphere conformal scale $\sqrt{\bq}$, which is conjugate 
to the renormalized  Noetherian energy charge aspect (see \eqref{QTR} below).

\subsection {BMSW group and  Einstein's equations}

The \bmw group, which is studied in more details in Section \ref{sec:WBMS},  is given by the semi-direct product $\left(\mathrm{Diff}(S) \ltimes \R_W^S\right) \ltimes \R_T^S$
and it is generated by the vector fields $\left(\xi_{T} ,\xi_{W},\xi_{Y}\right)$  given by (see \eqref{xi} for the complete expression)
\begin{subequations} \label{BMSW}
\begin{align}
\xi_T&:=
T\pa_u + o(1)
 \la{xiT-S} \\
\xi_\W
&:=  \W( u  \pa_u - r \pa_r)  + o(1), \la{xiW-S}\\
  \xi_{Y}
  &:= Y^A \pa_A 
\,. \la{xiY-S}
\end{align}
\end{subequations}
where the $o(1)$ refers to terms that vanish when $r\to\infty$. 
$T$ and $W$ are time independent functions on $S$, while $Y=Y^A\pa_A$ is a time independent vector field on $S$.
We will refer to $\xi_T$ as the super-translation, 
$\xi_W$ as  the Weyl super-boost, and $\xi_Y$ as the $\mathrm{Diff}(S)$ vector fields.
A general \bmw vector field is then given by  $\xi_{(T, W,Y)}:= \xi_T +\xi_W+\xi_Y$.

The \bmsw Lie algebra structure is defined by
\be\la{xi-Lie-2}
\lbr \xi_{(T_1, W_1,Y_1)}, \xi_{(T_2, W_2, Y_2)}\rbr= \xi_{(T_{12}, W_{12},Y_{12})}\,,
\ee
where 
\begin{subequations}
\begin{align}
Y_{12}&=[Y_1,Y_2]_{\mathrm{Lie}},\\
\W_{12} &= 
 Y_1[\W_2] -Y_2[\W_1],\\
T_{12} &=  Y_1[T_2] - \W_1 T_2  -(Y_2[T_1]-\W_2 T_1),
\end{align}
\end{subequations}
and $[\cdot,\cdot]_{\mathrm{Lie}} $ denotes the sphere Lie bracket.

The renormalization procedure, complemented by the use of the asymptotic Einstein's equations for $\bF$ in \eqref{asym-EE1} and  for $\bU^A$ in \eqref{asym-EE2} (necessary to renormalize $Q_Y$),  yields the following finite expressions for the  energy (or super-translations), Weyl and momentum (or diff($S$)) Noether charges:
\begin{align}
Q_T^R &=\int_{S}  \sqrt{\bar{q}}\,  T  \left( M - \f12 \bD_A \bU^A \right)\,,\la{QTR}\\
Q_W^R &=\int_S \sqrt{\bq} \,\W\, \left[4 \bar \beta
+ u   \left(M
-\frac 1{2} \bD_A \bar{U}^A\right)
\right]
\,,\la{QWR}\\
Q_Y^R &= \int_S \sqrt{\bq}~ Y^A\left(
 \bar{P}_A  + 2 \bD_A \bar\beta   
 \right)\,,\la{QYR}
\end{align}
where  $(T, W,Y^A)$ are  arbitrary functions on the 2-sphere. 
The full renormalized Noether charge then reads
\begin{align}
Q^{R}_{(T, W, Y)}=\int_S\sqrt{\bq} 
\left[
(T+uW) \left( M-\f12 \bD_A \bU^A\right)
+(2\W-\bD_A Y^A)\,2\bar \beta
+Y^A
 \bar{P}_A    
\right]\,.\la{Qfull}
\end{align}

The charges \eqref{QTR}, \eqref{QWR}, \eqref{QYR}, together with the corresponding finite expressions for the Noether fluxes $\cF^R_T, \cF^R_W, \cF^R_Y$, given in Section \ref{sec:div-pot}, can be used to have a well defined notion of  charge bracket  at null infinity so that the validity of  the asymptotic Einstein's equations is \emph{equivalent} to having a {\it centerless} representation of the residual diffeomorphism algebra at {\it each} cross-section of $\scri$, namely
\be\la{Lie-R}
 \{ Q^R_\xi, Q^R_\chi\} {=} -Q^R_{\lbr\xi,\chi\rbr}\,.
\ee
This result extends the one of \cite{Compere:2020lrt},  where the  generalized BMS surface charge algebra of \cite{Campiglia:2014yka, Compere:2018ylh} was shown to close only at the corners $u\rightarrow \pm\infty$ of $\scri$. In particular, it shows that the \bmw charge algebra is realized as the  symmetry algebra of asymptotically flat spacetimes at {\it all times}.

One important subtlety is that the \bmsw charge associated to the sphere diffeomorphisms $\mathrm{diff}(S)$  is \emph{not} the same as the \bms super-Lorentz charge appearing in  \cite{Compere:2018ylh}.
For instance, the Noetherian flux $\cF^R_Y$ for \bmsw $\mathrm{diff}(S)$ vanishes, while the \bms flux $\cF^{R-\bms}_Y$ does not.
This can be seen explicitly from the fact that  a \bms vector field can be written as 
\be
 \xi^\bms_T= \xi_{T}\,,\quad \xi^\bms_Y=\xi_{Y} +  \xi_{\W=\f12 \bD_A Y^A}\,.\la{xixi0}
 \ee
 By linearity this means that  a \bms super-Lorentz transformation is the sum of a \bmsw sphere diffeomorphism transformation plus a Weyl super-boost.
 In particular, it implies that the  \bms super-translation and super-Lorentz charges are  
 \be 
 Q^{R-\bms}_{T} = {Q}^R_{T}, \qquad 
 Q^{R-\bms}_{Y} = Q^R_Y+ {Q}^R_{\W=\frac{1}2 \bD_AY^A} .
 \ee
 Given this relation, it follows that the BMS charges also provide
 a centerless representation of the canonical algebra as in \eqref{Lie-R}.
 
 This conclusion seems to be in  contrast with the statements of \cite{Barnich:2011mi,Flanagan:2015pxa, Compere:2018ylh} that the \bms  charges constructed by Barnich--Troessaert 
 possess a non-trivial 
 field-dependent 2-cocycle.
 It is also  in tension with  the statement of \cite{Compere:2020lrt}  that  a centerless representation exists only at the corners of  $\scri$.
 The Barnich--Troessaert charges are given by 
 \be\la{Qbms}
{Q}^{\textsf{BT}}_{(T, Y)}=
\int_S\sqrt{\bq} \left(
2\tau M
+Y^A  \bar{P}_A     
 \right)\,,
\ee
where $\tau=T+\f u 2 \bD_A Y^A$.
They were first derived in \cite{Barnich:2011mi} using the Barnich--Brandt formalism \cite{Barnich:2001jy} and later in \cite{Flanagan:2015pxa} from a split following  the prescription laid out by Wald--Zoupas \cite{Wald:1999wa}.

To resolve the tension, we show that these charges are in fact Noether charges, obtained through the prescription \eqref{trans1} with boundary Lagrangian 
\be\la{ellbms}
\ell = \sqrt{\bq}
\left(M-\f 1 8 C^{AB} N_{AB}\right)
\rd u\wedge \rd^2\sigma\,.
\ee
Furthermore, we prove that the non-central 2-cocycle appearing in \cite{Compere:2020lrt} when computing the Barnich--Troessaert bracket of the charge \eqref{Qbms} is nothing but the extra contribution \eqref{K} in the bracket \eqref{bran} due to the non-covariance of the boundary term \eqref{ellbms}.

Therefore, we can understand the Barnich--Troessaert split between integrable and non-integrable terms \cite{Barnich:2011mi} as a {\it Noetherian split} for the boundary Lagrangian  \eqref{ellbms}, and
the appearance of a 2-cocycle  in the charge bracket at a general cross section of $\scri$ is simply a reflection of the fact that the original Barnich--Troessaert bracket does not take into account the contribution \eqref{K}, {which originates from the fact that} the boundary Lagrangian \eqref{ellbms} has an anomaly. In other words, if we wish to work with 
the Barnich--Troessaert charge bracket, the only way to avoid the appearance of a 2-cocycle is to apply the Noetherian split with a covariant boundary Lagrangian. 

In the case of the Einstein--Cartan (or Einstein--Hilbert) Lagrangian, the only charge yielding a centerless representation of the \bmsw Lie algebra  
is the  Noether charge \eqref{Qfull}. The analog charge for  the generalized BMS case is the Noether charge  given by  \eqref{Q-0}.

The  option we propose to resolve the appearance of cocycles when working with a non-covariant Lagrangian is instead to  include the anomalies in the definition of the charge bracket as in \eqref{bran}, in order to insure that the bracket does not involve field-dependent cocycles. This proposal allows us to introduce  anomalies in a consistent manner {\it both} at the level of  the charges and the bracket.

Finally, the vacuum sector can be constructed from the orbits of the \bmw group by acting with the group elements
 \be
g_{(T,\W,Y)} = e^{\delta_T}e^{\delta_\W} e^{\delta_Y} 
\ee
on the  phase space at null infinity. The infinitesimal transformations $\delta_T, \delta_\W, \delta_Y$ are the ones generated respectively by the vector fields \eqref{BMSW} and given explicitly in Appendix \ref{Appvariation}. The vacua $|T,\W,Y\rangle = \hat{g}_{(T,\W,Y)} |0\rangle$ are thus labelled by the two sphere diffeomorphisms arbitrary functions $Y^A$, in addition to the  super-translations $T$,  and  conformal transformations $W$. 
We expect each factor entering the vacua definition to be related to a memory effect: The super-translation $T$ accounts for the displacement memory effect \cite{Christodoulou:1991cr,ThorneB}; the diffeomorphism accounts for a change of asymptotic conformal structure 
which could lead to a memory effect generalizing the spin memory effect \cite{Pasterski:2015tva}. 
{We expect the Weyl rescaling factor $W$ to be related to the refraction memory effect \cite{Compere:2018ylh}, 
 as this determines the vacuum value of the News tensor (see Section \ref{sec:vac}). These open questions will be investigated elsewhere.
 }

We conclude our summary with Table \ref{tab:table1}, which shows the interplay among geometric data (the group generators), phase-space data (charges, fluxes and bracket), and dynamics (the Einstein's tensor). The table summarizes the remarkable property that the renormalized charge bracket \eqref{Lie-R} is equivalent to the asymptotic EEs by showing the detailed relation among the pair of \bmsw generators, its associated renormalized charge bracket, and what components of the Einstein's tensor display the corresponding flux-balance laws.


\begin{table}[htbp]
\setlength{\tabcolsep}{2.pt}
\renewcommand{\arraystretch}{1.8}
\begin{center}
\small
\begin{tabular}{|c|c|c|}
\hline
\textbf{\bmw generators $(\xi, \chi)$} & \textbf{$ \{Q^R_\xi, Q^R_\chi\}+Q^R_{\lbr\xi,\chi\rbr} =0$ \,\,} & \textbf{Einstein's equations}
\\
\hline
\hline
$(\pa_u,\xi_T)$ & $2\textsf E_{ M}-\f14 \bar\Delta \textsf E_{\bF}=0$ & $\xi_T^{\m}G_{\m}^{\;\;r}=0$\\
\hline
$(\xi_T, \pa_u)$ & $2\textsf E_{ M}+ \bD^A \dot {\textsf E}_{{\bU}_A} + \f14 \bar\Delta \textsf E_{\bF} =0$ & $\xi^{u}_TG_{u}^{\;\;r} -\xi^{r}_TG_{u}^{\;\;u} =0 $ \\
\hline
\hline
$( \pa_u,\xi_W)$ & $
\bD^A\textsf E_{\bU_A} +u \left(2\textsf E_{ M}-\f 14 \Delta \textsf E_{\bF}\right)=0 $ & $\xi_W^{\m}G_{\m}^{\;\;r}=0$\\
\hline
$(\xi_W, \pa_u)$ & $
-\bD^A\textsf E_{\bU_A}
+u \left(2\textsf E_{ M}+ \bD^A \dot{\textsf E}_{{\bU}_A} +\f 14 \Delta \textsf E_{\bF}
\right) =0$ & $\xi^{u}_W G_{u}^{\;\;r} -\xi^{r}_W G_{u}^{\;\;u} =0 $ \\
\hline
\hline
$(\pa_u, \xi_Y)$ & $
 \textsf E_{{\bP}_A}+2\bD_A \dot{\textsf E}_{\bar\beta} -2 \bar{F}\textsf E_{\bU_A} -\f12 \bU_A \textsf E_{\bF}=0 $ & $\xi_Y^{\m}G_{\m}^{\;\;r}=0$\\
\hline
$(\xi_Y, \pa_u)$ & $0=0$ & $0 =0 $ \\
\hline
\end{tabular}
\caption{\textsf{\footnotesize{Synoptic overview of the flux-balance laws associated to the pair $(\xi, \chi)$ and their relation to Einstein's equations. We recall that by the property \eqref{brack3} the charge algebra is left unchanged by the renormalization procedure.
}}}
\label{tab:table1}
\end{center}
\end{table}

\section{Geometric interpretation of the expansion coefficients}\label{SecGeo}

Thanks to the coordinate gauge-fixing, all functions appearing in \eqref{eq:FallOff} acquire a precise geometric meaning, which is useful to recall here. This will allow us to further motivate our choice of momentum aspect $\bar P_A$, which unlike the other quantities,  is not universal in the literature.
To that end, let us denote
\begin{align}
{l} := 
\partial_r, \quad l^2=0, \qquad
t :=
\pa_u+U^A\pa_A,\la{NL} \quad t^2 = -2e^{2\b}F, \qquad l_\m t^{\m} = -e^{2\beta}.
\end{align}
$l$ is the null vector tangent to the null geodesics which cut scri along 
the constant $u$ spheres.
$t$ is the unique time-like vector which is transverse to $l$ and tangent 
to the constant $r$ hypersurfaces (see Fig. \ref{Diagram}). Note that, upon compactification,  it becomes a null generator at $\scri$.

\begin{figure}[htbp]
\centering
\includegraphics[height=50mm]{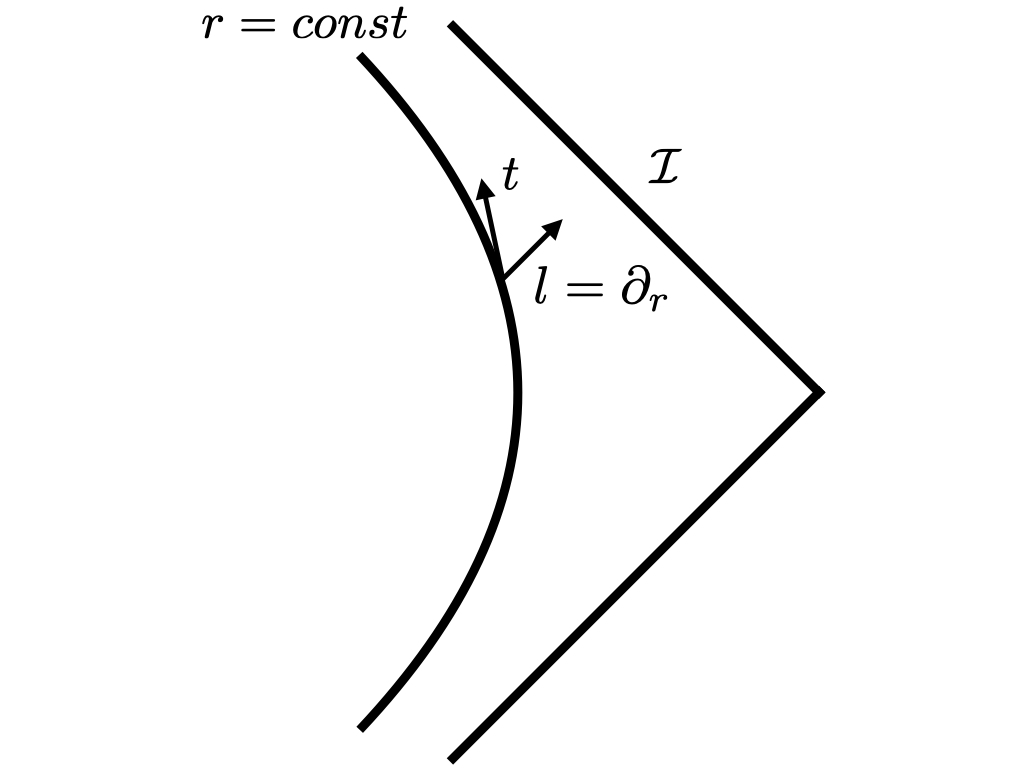}
\caption{\small{\emph{The figure shows an {idealized} $r=const$ hypersurface, the null vector $l$, which 
{can be used to study }
its radial evolution, and the time-like vector $t$   tangent to the hypersurface.}}}\la{Diagram}
\end{figure}

The first relevant geometric quantity is the  extrinsic tensor of the null hypersurfaces normal to $l$. It is given by
\be
S_{AB} := \frac{1}{2r}  \mathcal{L}_{l} g_{AB} = \frac{1}2 \left( 2+ r\pa_r\right)q_{AB}
= \bar{q}_{AB} + \frac{\bC_{AB}}{2 r}+o(r^{-2}).
\ee
From this expression we see that the null geodetic congruences have asymptotic shear $C_{AB}$ and asymptotic expansion $2$, which is just the reflection of $r$ being the area radius.
The second geometric quantity is the extrinsic curvature of
the sphere $S$ within the time-like  hypersurface of constant $r$. It is given by
\begin{align}
K_{AB} &:= \frac{1}{2r} \mathcal{L}_{t} g_{AB} = r \left(\frac12  \pa_u{q}_{AB} +  D_{(A} U_{B)}\right)\nn
\\ &= \frac12 N_{AB} +  \f{1}r \left(\bD_{(A} \bU_{B)} +\frac{\bar{q}_{AB}}8 \pa_u C^2 \right) + o(r^{-1})\,,
\end{align}
where $N_{AB}:=\pa_u C_{AB}$  is the Bondi news tensor and  $C^2:=C_{AB} C^{AB}$. 
From this expression we see that the news tensor is the asymptotic extrinsic curvature. 
As a side remark, we note that the trace $ q^{AB} K_{AB}= {\cal O}(r^{-1})$ is sub-dominant.

The third quantity is the angular momentum aspect $\bP_A$, which we refer to as momentum aspect as explained in the introduction, and which enters the expansion \eqref{UA}. The relevant geometric quantity is the parallel transport of $t$ along $l$, 
\be
\eta_A := r e^{-2\beta} \na_l t_A = r \left(\pa_A \beta + \frac{r^2}{2} e^{-2\beta} q_{AB} \pa_r U^B\right)
= - \bU_A + \frac1r({\bP_A+ 2\pa_A \bar\beta})   + o(r^{-1}).
\ee
This quantity is precisely the spin-1 momentum of the geometric decomposition of the symplectic potential \cite{Parattu:2015gga,Hopfmuller:2016scf,Oliveri:2019gvm}.\footnote{If we introduce a second null vector $n$ transverse to $l$ and such that the space-like spaces orthogonal to the pair $(l,n)$ are tangent to the sphere, we have $\eta_A:= -r n^\m\na_A l_\m$. The quantity $n^\m\na_A l_\m$ contains the two transverse components of the spin-1 momentum, and it is also related to the non-integrability of the $(l,n)$ planes and to the rotational 1-form of the isolated horizon framework; see discussion in \cite{Hopfmuller:2016scf,Oliveri:2019gvm}.}

Finally, from the point of view of the gravity-fluid duality \cite{Rangamani_2009}, it is also useful to consider the velocity 
\be
V^A = r^2 e^{-2\beta} U^A = \bU^A   - \f23 \frac{\bar{q}^{AB}}{r}\left( \bP_B +C_{BC}\bar{U}^C\ + \pa_{B}\bar{\beta}\right)+o(r^{-1}).
\ee
The reader will notice that the four quantities above have been suitably rescaled by appropriate factors of $r$, with respect to their natural geometric definitions. This was done so that their leading orders were finite.

\section{The  BMSW group}\la{sec:WBMS}

In this section we define the \bmw group, its Lie algebra structure,
its boundary and bulk realizations and its relationship with the original BMS group, the generalized BMS group and the extended BMS groupoid.

\subsection{Residual diffeomorphisms}
The residual diffeomorphisms which define the boundary symmetry generators are defined by two sets of requirements, which are to be analyzed separately. These are: 
\begin{itemize}
\item[$(i)$] 
The preservation of the Bondi gauge-fixing \eqref{Bondig}:
\be\label{BMWi}
\pounds_\xi g_{rr}=0, \qquad \pounds_\xi g_{rA}=0, \qquad 
\pa_r \left(g^{AB}\pounds_\xi g_{AB}\right)=0.
\ee
These conditions are solved  
respectively restricting the components of $\xi$ to be of the form
\be
\xi^u=\t\,,\qquad
\xi^A=Y^A -I^{AB} \p_B \t\,, \qquad
 \xi^r= - r \W + \f{r}{2}\left( {D}_A(I^{AB} \p_B \t) + U^A\p_A \t \right),
 \label{BMStransf}
\ee
where $D_A$ is the covariant derivative associated with $q_{AB}$ and 
\be\la{IAB}
I^{AB}:=  \int_r^\infty  \f{\rd r'}{r^{'2}} e^{2\beta}  q^{AB}\,.
\ee
The parameters $\t$, $Y^A$ and $\W$ are, at this stage, arbitrary functions of $(u,\sigma^A)$ but not of $r$,
and the Weyl factor $\W$ is such that
\be\la{dxiq}
\d_\xi \sqrt q= ( D_AY^A-2 \W )\sqrt q\,,
\ee
where  $D_A Y^A $ measures the variation of the sphere area
 element under a diffeomorphism, while  
 $\W$ labels a Weyl rescaling.
 This transformation rule, which distinguishes the rescaling due to tangent diffeomorphisms from the 
 rescaling due to Weyl transformation, justifies with hindsight the parametrization \eqref{BMStransf} of $\xi^r$, which is a slight generalization of the  procedure \cite{BMS}  (see  \cite{Barnich:2009se, Compere:2018ylh}).
 It is important to note that $D_A Y^A =\bD_AY^A$ is independent of $r$ due to the determinant condition in the Bondi gauge. This means that 
 $\pa_r (\d_\xi \sqrt q) =0 $ which is equivalent to the last condition of \eqref{BMWi}. \footnote{This follows from 
\begin{equation}\nn
\d_\xi \ln{\sqrt q} 
=\f1{2}g^{AB}\d_\xi g_{AB}=\f1{2}g^{AB}\cL_\xi g_{AB}
= 2 \frac{\xi^r}{r}+ D_A \xi^A -U^A\pa_A\xi^u\,.
\end{equation}
}

\item[$(ii)$] The preservation of the  boundary conditions \eqref{ExtAC}:
\begin{subequations}\label{BMWii}\begin{align}
& \pounds_\xi g_{ur}=\mathcal{O}(r^{-2}),  &\pounds_\xi g_{uA}&=\mathcal{O}(1) \label{ii1} \\ \label{ii2}
& \pounds_\xi g_{uu}=\mathcal{O}(1),   &\pounds_\xi g_{AB}&=\mathcal{O}(r^2). 
\end{align}\end{subequations}
These impose
\be
\p_u Y^A=0, 
\qquad \p_u \W=0,
\ee
and restrict $\tau$ to be a linear function of $u$ given by
\be
\tau =T+ u \W,\qquad \pa_u T=0,
\qquad  \dot{\tau} = \W,
\qquad \ddot{\tau}=0.
\ee

\end{itemize}

The \bmw vector fields preserving $(i)$-$(ii)$ depend on two arbitrary functions $(T, \W)$  on the 2-sphere, equivalently on a  linear function of time $\tau= T+u\W$,  and on a vector $Y=Y^A\pa_A$ on $S$.
They will be denoted $\xi_{(\t,Y)}$ or $\xi_{(T,\W,Y)}$.
 The gauge parameter $T(\s^A)$ is associated to the asymptotic super-translations,
 $Y^A(\s^A)$ represent  asymptotic diffeomorphisms of the celestial sphere $S$, while $\W(\sigma^A)$ corresponds to Weyl rescaling of the celestial sphere.

All the asymptotic symmetry groups studied so  far in flat space holography  (see  \cite{Ruzziconi:2020cjt} for an overview), such as 
the original BMS group \cite{Bondi62, Sachs:1962wk,Newman:1962cia}, the \emph{extended} BMS group introduced by Barnich--Troessaert \cite{Barnich:2010eb, Barnich:2016lyg,Barnich_2012} and the \emph{generalized} BMS group  put forward by Campiglia  and Laddha \cite{Campiglia:2014yka} 
and studied canonically by Compere et al. \cite{Compere:2018ylh}, 
 appear as a restriction  of the \bmw group.
For instance, the generalized BMS group $\mathrm{Diff}(S)\ltimes \R^S$,
where $\R^S$ denotes the space of functions on $S$,  is obtained from \bmw  by restricting the Weyl factor to be
identified with the divergence of the vector field
\be \label{restrict}
\W^{\bms}= \frac12 \bD_A Y^A\,.
\ee
The original  BMS group $\mathrm{Conf}(S)\ltimes \R^S$ is obtained by 
  fixing also the boundary metric to be the round 2-sphere metric 
\be\label{bondiframe}
\bq_{AB}=\mathring{q}_{AB} 
\ee
and by demanding, in addition, that the vector field belongs to the conformal group of $S$
\be\label{Conf}
\mathring{D}_{\langle A} Y_{B\rangle} =0\,,
\ee 
where $\mathring{D}$ is the covariant derivative of the spherical metric, so that \eqref{bondiframe} is preserved.
The extended BMS group $(\mathrm{Conf}(S)\ltimes \R^S)\times \R^S $, which will be discussed in  more details below,  is obtained by  imposing that sphere diffeomorphisms are conformal \eqref{Conf}, but relaxing the Weyl condition \eqref{restrict} and introducing a field-dependent redefinition of the super-translation parameter $T$ in order to obtain an algebra. We come back to all these relations with more details in Section \ref{sec:LAD} below.

\subsection{Boundary symmetry Lie algebra }\la{sec:LA}

A remarkable consequence of the \bmw extension is that 
the residual diffeomorphisms generate a subalgebra, and not a subalgebroid, of the full diffeomorphism algebra of $\scri$.
This is not obvious from the start, since the \bmw vectors are manifestly field dependent. Therefore the commutator of the field transformations they generate is given by the modified Lie bracket \eqref{Lie-mod}, which in general yields an  algebroid. 
To prove this point, we look at the leading order of the \bmw vector fields on $\scri$. It is given by
\be
\bar \xi_{(\tau,Y)} :=\tau \pa_u   + Y^A\pa_A  -\dot{\tau} r\pa_r\,,
\la{xiB}
\ee
or equivalently 
\be
\bar \xi_{(T,\W,Y)} := T \pa_u   + Y^A\pa_A  + \W( u\pa_u - r\pa_r).
\la{xiA}
\ee
These vector fields are elements of the automorphism group of a line bundle $P\to \scri$ over $\scri $ that we call the {\it scale bundle}.
The translation along the fiber of $P$ is given by the conformal rescaling operation and the asymptotic vector field is an element of its automorphism group, $\bar \xi_{(\tau,Y)}= \bar \xi_{(T,\W,Y)} \in \mathrm{Aut}(P).$
At the infinitesimal level the conformal rescaling is  implemented on $\scri$ by the  operator $r\pa_r$.

The  two expressions \eqref{xiB}, \eqref{xiA} give two different interpretations of the same vector field:
in the first expression we see that $\tau$ labels linear time reparametrization while $\dot{\tau}$ labels a Weyl rescaling.
In the second expression we see that $T$ labels super-translation, while $\W$ labels a super-boost transformation which preserves the normal metric $2\rd u \rd r$.
Demanding that  the boundary symmetry algebra is a Lie algebra, and not an algebroid, means that we have to impose  that $\bar \xi_{(\tau,Y)}$ is field independent, i.e., $\delta \tau= \delta Y =0$.
These conditions mean that  
\be 
\lbr \bar \xi_{(\tau_1,Y_1)}, \bar \xi_{(\tau_2,Y_2)}\rbr = [\bar\xi_{(\t_1,Y_1)},\bar \xi_{(\t_2,Y_2)} ]_{\mathrm{Lie}}\,.
\ee
The vector fields $\bar\xi_{(\tau,Y)}$ are the generators of the boundary symmetry group.
Their Lie commutators give 
\be
\left[\bar\xi_{(\t_1,Y_1)}, \bar\xi_{(\t_2,Y_2)} \right]_{\mathrm{Lie}} = \bar\xi_{(\t_{12},Y_{12})}\,,
\ee
 where 
\bea
& \t_{12}:= \tau_1\dot{\tau}_2- \tau_2\dot{\tau}_1
+ Y_1[\tau_2] -Y_2[\tau_1]\,, \qquad
Y_{12}:=[Y_1,Y_2]_{\mathrm{Lie}}\,. 
\la{comm}
\eea
We thus see that, in the $(\t, Y)$ basis, the infinitesimal generators of the \bmw group at leading order in the Taylor expansion form a Lie algebra given by the semi-direct sum 
\be\la{BMSWalg1}
\textsf{bmsw}
 :=\mathrm{diff}(S) \oright \mathbb{R}^S_u\,,
\ee
with  $ \mathbb{R}^S_u$ denoting functions on the sphere $S$ which are linear in time with a bracket given by the Witt bracket.\footnote{$G^S$ refers to the set of maps from $S$ to the group  $G$.
$\mathbb{R}_u$ can also be identified with the  
subalgebra $\tau \pa_u$  of diffeomorphisms of $\R$ which are linear in $u$.}
Quite remarkably,\footnote{We thank A. Speranza for bringing this result to our attention.} the $\textsf{bmsw}$ symmetry algebra was shown to be the symmetry algebra preserving the so-called intrinsic structure\footnote{ \label{thermalC} An intrinsic structure on a null surface $\cal{N}$ is an equivalence class of  pairs $(n, \kappa)$ where $n=n^a \pa_a$ is a null generator of $\cal{N}$ and $\kappa$ is the null surface gravity. The equivalence relation is under rescaling of the null generator given by 
\be
n \to e^\sigma n, \qquad \kappa \to e^{\sigma}(\kappa +\cL_n \sigma ).
\ee
These transformations insure that the operator $\cL_n+ \kappa $ transforms tensorially. } of any finite null surface in \cite{Chandrasekaran:2018aop}.
This intrinsic structure can be understood as a thermal Carroll structure.
This group also appears as the extended corner symmetry group of a 2-sphere embedded in 
on a 3d hypersurface \cite{Ciambelli:2021vnn}.
The fact that the same symmetry group appears naturally at infinity, although surprising, is not a coincidence. We expect that asymptotic structure and asymptotic symmetry mirror similar structure at finite surfaces.

We can make this {algebra} even more explicit if one uses the explicit parametrization $\tau = T +u \W$.
Then we can rewrite \eqref{BMSWalg1} as a double  semi-direct sum
\be
\textsf{bmsw}
= (\mathrm{diff}(S) \oright \R_\W^S) \oright \R_T^S\,,
\ee
where the first factor is the Weyl group parametrized by $\W$, while the second factor is the 
super-translation group parametrized by $T$ (with $\delta T=\delta\W=0$).
The brackets are 
\be
\W_{12} = 
 Y_1[\W_2] -Y_2[\W_1], \qquad 
T_{12} =   Y_1[T_2] - \W_1 T_2  -(Y_2[T_1]-\W_2 T_1).
\ee
We see that $\W$ transforms as a scalar under $\mathrm{diff}(S)$ while $T$ transforms as a   weight-$1$ section of the scale bundle. A section $\Phi$ of the scale bundle is said to be of scale weight $s$ if it transforms under Weyl rescaling as $ \Phi\to e^{-s\W} \Phi$.
This structure allows one to associate a notion of scale weight to fields on $\scri$, something that has been considered in the literature \cite{Barnich:2010eb, Barnich:2016lyg,Ciambelli:2019bzz}  and that will be explored elsewhere.

We can now formalize the construction \eqref{BMStransf} of the bulk vector field solution of the Bondi gauge condition as the construction of a map $\rho: \mathrm{Aut}(P) \to \Gamma(TM)$ 
given by  \eqref{BMStransf}. The \bmw vector fields $\xi_{(\tau, Y)}$ are  the bulk extension  of $\bar\xi_{(\tau,Y)}$ and they are given at leading orders by
the asymptotic expansion
\be
 \x_{(\tau,Y)}=\rho\left(\bar \xi_{(\tau,Y)}\right)
 = \bar \xi_{(\tau,Y)}+\f1r\xi_{\tau }^1+\f1{r^2} \xi_{\tau }^2+o(r^{-2})\,,\la{rho}
\ee
where
\begin{subequations}
\begin{align}
\xi_\tau^1&= -\p^A\tau \p_A +\f12  \bar \Delta \tau~ r\p_r\,,\\
\xi_\tau^2&=\f1{2} C^{AB}\p_B \tau\p_A -\f12\left( {\bD}_AC^{AB}\p_B \tau +\f12C^{AB}{\bD}_A \p_B \tau \right) r\p_r\,.
\end{align}
\end{subequations}
The map $\rho$ gives an intertwiner  between 
the standard Lie bracket and 
the field-depended modified Lie bracket \eqref{Lie-mod}. 
More precisely, thanks to the map $\rho$, the bulk vectors \eqref{rho} form a faithful representation of the Lie algebra \eqref{BMSWalg1} for the modified Lie bracket \eqref{Lie-mod}. This follows from the property
\be
 \lbr \xi_{(\tau_1,Y_1)}, \xi_{(\tau_2,Y_2)}\rbr=  \lbr \rho(\bar\xi_{(\tau_1,Y_1)}), \rho(\bar\xi_{(\tau_2,Y_2)})\rbr
 =\rho([\bar\xi_{(\t_1,Y_1)},\bar \xi_{(\t_2,Y_2)} ]_{\mathrm{Lie}})\,,
\ee
which yields
\be\la{xi-Lie}
\lbr \xi_{(\tau_1,Y_1)}, \xi_{(\tau_2,Y_2)}\rbr= \xi_{(\t_{12}, Y_{12})}\,,
\ee
with $\t_{12}, Y_{12}$
given by \eqref{comm}. It follows that the bulk vectors in the $(\t, Y)$ basis \eqref{rho} also satisfy the Lie algebra \eqref{BMSWalg1}.

A  remark is in order. 
The construction of the algebra uses explicitly the preferred foliation of $\scri$ introduced by the use of Bondi coordinates. For this reason, the right-hand side of \eqref{BMSWalg1} makes reference to the cross-sections $S$ defined at constant $u$. However, we expect that it is possible to use an intrinsic approach on $\scri$ like in \cite{Ashtekar:1981bq, Chandrasekaran:2018aop}, and define this algebra based on more general universal structure than the one associated to the BMS algebra. In this intrinsic approach, the only notion that is required is the fact that $\scri$ is a fibration, without the need of choosing a specific foliation, and that this fibration is equipped with a thermal connection (see footnote \ref{thermalC}).  This will be investigated elsewhere.

\subsection{Extended BMS and Generalized BMS as Lie sub-algebroids }\la{sec:LAD}
We now want to describe more precisely the relation between the 
\bmw group $(\mathrm{Diff}(S) \ltimes \R_W^S) \ltimes \R_T^S$ and the previous BMS groups.
There are three such groups:
the generalized BMS group $\mathrm{Diff}(S) \ltimes \R_T^S$ \cite{Campiglia:2014yka, Compere:2018ylh},
the extended BMS group $(\mathrm{Conf}(S) \ltimes \R_T^S) \times \R_W^S$  \cite{Barnich:2010eb, Barnich:2016lyg}
and the original BMS group $\mathrm{Conf}(S)  \ltimes \R_T^S$ \cite{Bondi62, Sachs:1962wk}.
There is a key difference between BMSW and all the BMS groups:
The BMS groups depend on a background structure held fixed, while the BMSW group does not 
involve any background structure besides the choice of foliation.

On the one hand, all the BMS groups descend from BMSW, 
which can then be seen as a consistent merging of the previous extensions.
On the other hand, 
there is a subtle point:  to obtain the (original, extended or generalized) BMS algebras, one needs to consider field-dependent vector fields. This creates a potential issue that these sub-algebras are only sub-algebroids and not simply algebras. 
Having a Lie algebroid structure is not admissible if one wants to promote the classical symmetry algebra to a symmetry algebra of quantum gravity. 
One therefore needs to project back this algebroid onto an algebra.
This is what we  investigate and we show how to recover a Lie algebra structure in all three cases.
The consequence of this reparametrization  is that the representation  of the sphere diffeomorphisms is fundamentally different  
for the BMS groups because it involves a mixing of the \bmw diff$(S)$  and Weyl charges.

\begin{table}[htbp]
\centering
\footnotesize
\begin{tabular}{|c|c|c|c|}
\hhline{~---}
\multicolumn{1}{c|}{} & $\xi_T$ & $\xi_W$ & $\xi_Y$  \\
    \hhline{-===}
    \bmsw & \textsf{super-translations} & \textsf{Weyl super-boost} & $\mathrm{Diff(S)}$ \textsf{sphere diffeomorphims} \\
    \hline
\textsf{generalized} \bms & \textsf{super-translations} & $\emptyset$ & \textsf{super-Lorentz: rotations \& boosts } \\
    \hline
    \textsf{extended } \bms & \textsf{super-translations} & $\textsf{super-Weyl} $
     & \textsf{Conformal transformations }  \\
    \hline
\end{tabular}
\label{tab:tableName}
\caption{\textsf{\footnotesize{Comparison of the extended, generalized BMS and BMSW generators. The generalized BMS generators are called super-Lorentz transformations, following  \cite{Compere:2018ylh, Ruzziconi:2020cjt}.
They are obtained from BMSW  by fixing the Weyl rescaling in terms of the sphere diffeomorphism  to preserve the metric scale.
One usually calls super-rotations the area-preserving diffeomorphisms while the 
Lorentz super-boosts refer to  transformations that are divergence-full, $D_A Y^A\neq0$.
In \bmsw we have a finer group structure and the Weyl super-boosts are left free. Note that the Weyl super-boosts are, unlike Lorentz super-boosts, genuine boosts normal to the sphere. The Weyl super-boosts decouple in the extended BMS group.  }}}
\end{table}

\subsubsection{Generalized \bms}
The generalized  \bms algebra,
or \gbms for short,
was first proposed by Campiglia and Laddha \cite{Campiglia:2014yka} as an asymptotic symmetry group to take into account the subleading soft theorem and it was extensively studied by  Compere et al. in  \cite{Compere:2018ylh}.
To obtain  \gbms from \bmsw 
 one has to  redefine the  Weyl  parameter to be field dependent,
\be
 \W^\gbms =\frac12  \bD_AY^A \,.
 \la{bms-para}
\ee
This  parametrization corresponds to a choice where the scale factor is fixed by the symmetry transformations
\be
\delta_{\xi^{\gbms}} \sqrt{\bq} =0. \la{qw}
\ee
With this choice, let us consider  the asymptotic vector fields  
\be\label{bms}
\bar \xi^{\gbms}_{(T,Y)} := \bar \xi_{\left(T ,\frac12  \bD_AY^A ,Y\right)}\,,
\ee
 which become field dependent 
 \be\label{dbms}
\delta \bar \xi^{\gbms}_{(T,Y)} = \f12 Y[\delta \ln{\sqrt{\bar{q}}}] (u\pa_u - r\pa_r).
\ee
As a consequence, while the vectors $\xi_{(\tau,Y)}$ form a Lie algebra, the vectors 
$\bar \xi^{\gbms}_{(T,Y)}$ form a Lie algebroid, unless we take the restriction that $\delta \ln{\sqrt{\bar{q}}}=0$, which is what is implemented for the  generalized \bms algebra \cite{Compere:2018ylh}.   This restriction insures that $\bar\xi^\gbms$ is field independent and therefore that the algebroid is an algebra.

The generalized \bms  Lie algebra bracket is 
\be\la{bms-bra}
\lbr \bar \xi^{\gbms}_{(T_1,Y_1)}, \bar \xi^{\gbms}_{(T_2,Y_2)}\rbr
 = \bar\xi^{\gbms}_{(T_{12}, Y_{12})}\,,
\ee
 with $(T_{12},Y_{12})$ given by 
\begin{subequations}\la{TWY12}
\begin{align}
Y^A_{12}&=  [Y_1,Y_2]_{\mathrm{Lie}}^A
, \\
T_{12} &=Y_1[T_2] -Y_2[T_1] + \f12 T_1\bD_A Y_2^A -\f12T_2 \bD_A Y_1^A \la{Tbra}\,.
\end{align}
\end{subequations}
This algebra is such that  diff$(S)$ does not act on $W$, while its action on $T$ 
is the action on densities of weight  $1/2$.  In other words 
\be
\textsf{gbms} =  (\mathrm{diff}(S) \oright_{1/2} \R_T^S)\,, 
\ee
where $\oright_{\alpha}$ denotes the action of diffeomorphism 
on densities of weight $\alpha$.

The fact that this is an algebra and not an algebroid stems from the fact that the scale factor $\sqrt{\bar{q}}$ is taken to be a background structure 
which is  not part of the phase space and that the algebra action is consistent with the condition $\delta \sqrt{\bq}=0=\delta_{\xi^{\gbms}} \sqrt{\bq}$.

\begin{table}[htbp]
\centering
\footnotesize
\begin{tabular}{|c|c|c|c|c|}
\hhline{~----}
\multicolumn{1}{c|}{} & $\textsf{background structure}$ & $\textsf{restriction}$ & \textsf{parametrisation}& $\textsf{group}$ \\
    \hhline{-====}
    \bmsw & $\emptyset$ & $\emptyset$&$(T,W,Y)$  &$\mathrm{Diff}(S) \ltimes_0 (\R_W^S \ltimes \R_T^S)$ \\
    \hline
\textsf{generalized} \bms & \textsf{scale structure} & $\delta \sqrt{q} =0$ 
& $(T,\tfrac12 D_A Y^A,Y)$  &$\mathrm{Diff}(S) \ltimes_{1/2} \R_T^S$ \\
    \hline
    \textsf{extended } \bms & \textsf{conformal structure} & $\delta [{q}_{AB}] =0 $
     &$(e^{\varphi} t ,\tfrac12 (D_A Y^A-w),Y)$ &$(\mathrm{Conf}(S) \ltimes_{1/2} \R_T^S) \times \R_W^S$ \\
    \hline
\end{tabular}
\label{tab:tableName}
\caption{\textsf{\footnotesize{Comparison of the extended, generalized BMS and BMSW group structure and parametrisation. 
The notation $\ltimes_{\alpha}$ denotes the action of diffeomorphims on densities of weight $\alpha$. $\R_T^S $ refers to super-translations and $\R_W$ to Weyl super-boost.  }}}
\end{table}

\subsubsection{Extended \bms}

The extended \bms algebra, denoted \ebms and  proposed by Barnich  and Troessaert \cite{Barnich:2010eb, Barnich:2016lyg}, does not allow for general diffeomorphism but it allows for a Weyl rescaling of the metric.
It is obtained from the \bmsw algebra  by 
introducing a background conformal  structure\footnote{That is an equivalence class $[q_{AB}]$ of metrics modulo rescaling $q_{AB} \sim e^{2\varphi} q_{AB}$. By the fundamental theorem of Riemann surfaces a conformal structure is equivalent to a complex structure.},  associated with the round sphere metric $\mathring{q}$, and by imposing that\footnote{ Sometimes it is also considered the case where one chooses a north pole $n$ and a south pole $s$  on $S$ and introduces a flat space metric $q^{\mathrm{Flat}}_{AB}$  on $S^*=S\backslash\{n,s\}$. This is possible since $S^*$ has the topology of the cylinder. In this case one can also consider imposing $\bar{q}_{AB} = e^{2\varphi}  q^{\mathrm{Flat}}_{AB}$ on $S^*$. Such metrics are however singular on $S$.}
\be
\bar{q}_{AB} = e^{2\varphi}  \mathring{q}_{AB}.
\ee
Preserving this background-dependent conditions requires modifying the definition of the Weyl factor. One imposes that 
\be
\W^\ebms =\frac12  (\bD_AY^A-w)
\ee
and  demands that it is $w$ which is now field independent $\delta w=0$.
This choice of parametrization corresponds to transformations that preserve the sphere complex structure  
\be
\delta_{\xi^{\ebms}} \sqrt{\bq} = w \sqrt{\bq},\qquad \delta_{\xi^{\ebms}} \varphi = \frac12 w. \la{qw}
\ee
This creates a issue: the boundary vector field is then field dependent and we are at risk of creating an algebroid not an algebra, which would be not suitable for quantization.
There is a way around it, that  is to  redefine also the super-translation generator as 
\be 
{T}^{\ebms} := e^{\varphi} t,
\ee
where $t$ is a field independent parameter, satisfying $\d t=0$.
The  boundary vector field $\bar\xi^{\ebms}_{(t,w,Y)} = \bar\xi_{({T}^{\ebms} ,W^{\ebms},Y)}$ is still field dependent 
 \be\label{ebms}
\delta \bar \xi^{\ebms}_{(t,w,Y)} = (\delta \varphi ) T^{\ebms} \pa_u +  Y[\delta \varphi ] (u\pa_u - r\pa_r)\,,
\ee
 nevertheless one can show that the extended \bms algebroid is in fact a Lie algebra. The extended \bms  Lie algebra bracket is 
\be\la{bms-bra}
\lbr \bar \xi^{\ebms}_{(t_1,w_1,Y_1)}, \bar \xi^{\ebms}_{(t_2,w_2,Y_2)}\rbr
 = \bar\xi^{\ebms}_{(t_{12},w_{12} Y_{12})}\,,
\ee
 with \cite{Barnich:2010eb, Barnich:2019vzx} $(t_{12},w_{12}, Y_{12})$  given by 
 \begin{subequations}\la{TWY12}
\begin{align}
Y^A_{12}&=  [Y_1,Y_2]_{\mathrm{Lie}}^A,\qquad \qquad 
w_{12} =  0, \\
t_{12} &=Y_1[t_2] -\f12 t_2 \mathring{D}_A Y_1^A  
-\left(Y_2[t_1] - \f12 t_1 \mathring{D}_A Y_2^A\right) \la{Tbra}\,.
\end{align}
\end{subequations}
This means that the algebra is 
\be
\textsf{ebms} =  (\mathrm{Conf}_{\mathring{q}}(S) \oright_{1/2} \R_T^S) \oplus \R_\W^S. 
\ee
The main feature of this algebra is the fact that the Weyl rescalings decouple from the conformal transformations and the super-translations.
The Weyl parameter acts as a superselection parameter. Finally, the super-translations  are acted upon by  the conformal generators as density of weight $1/2$, in agreement with the generalized \bms algebra.

Finally, the \bms algebra 
\be
\textsf{bms} =  (\mathrm{Conf}_{\mathring{q}}(S) \oright_{1/2} \R_T^S)
\ee
can be obtained either from the generalized \bms algebra by restricting the diffeomorphisms to be conformal transformations or from the extended \bms algebra by restricting the Weyl transformations to be trivial.

\subsection{Action on the asymptotic phase space}\label{APHaction}

The asymptotic covariant phase space we define is parametrized by functionals of the metric
$\Phi^i(g_{\mu\nu})=(\bar{F}, M, \bar \beta, \bP_A, \bU_A, \bq_{AB}, C_{AB})$.
We are interested in the behaviour  of these functionals under the transformation
$\delta_{(\tau,Y)} \Phi^i = \int \frac{\delta \Phi^i}{\delta g_{\mu\nu}} \cL_{\xi_{(\tau,Y)}}g_{\mu\nu}$ generated by the 
{\bmw group}.

This can be found expanding the action of $\xi_{(\tau,Y)}$ on the metric in inverse powers of $r$, 
\be 
{\pounds}_{\xi_{(\tau,Y)}} g_{\mu\nu} 
= {\pounds}_{\bar \xi_{(\tau,Y)}} g_{\mu\nu}
+
\cL_{\f1r \xi_1+ \f1{r^2} \xi_2+o(r^{-2})}g_{\mu\nu}.
\ee
Accordingly, we can decompose the transformation of the functionals $ \Phi^i$ as
\be
\delta_{(\tau,Y)} \Phi^i= \d_{\bar \xi_{(\tau,Y)}}\Phi^i+\Delta_\xi \Phi^i\,,
\ee
where the homogeneous term  $\d_{\bar \xi_{(\tau,Y)}}$ is determined by $\scri$-component 
$\bar \xi_{(\tau,Y)}$ of the Bondi vector fields \eqref{rho}, while the anomaly term $\Delta_\xi $ is determined by their bulk extension.\footnote{The term `anomaly' refers to the fact that quantities like $C_{AB}$ or $M$  do not  transform as sections of the scale bundle. The additional inhomogeneous terms come from the fact that they are components of a spacetime tensor. These are examples of the type of anomaly transformations defined in \eqref{anodef}.}
An explicit calculation of the homogeneous part and the anomaly contribution yields\footnote{We have included the time derivative term $\p_u$ in the transformation for $\bq_{AB}$ to highlight the overall structure, although our boundary conditions spelled in Section \ref{sec:null-infinity} are such that $\p_u \bq_{AB}=0 $.}
\begin{subequations}\la{del}
\begin{align}
\d_{(\tau,Y)} \bq_{AB}
&=\left[ \tau \pa_u  + \cL_Y - 2 \dot{\tau}\right]  \bq_{AB}\,,
\la{delq}
\\
\d_{(\tau,Y)} C_{AB}
&=\left[ \tau \pa_u  + \cL_Y - \dot{\tau} \right]  C_{AB} - {2\bD_{\langle A} \p_{B\rangle} \tau }\,,
\la{delC}
\\
\d_{(\tau,Y)} N_{AB}
&=\left[ \tau \pa_u  + \cL_Y \right]  N_{AB} - {2\bD_{\langle A} \pa_{B\rangle} \dot\tau }\,,
\label{delN}
\\
\d_{(\tau,Y)} M&=\left[\tau \pa_u +\cL_Y + 3 \dot{\tau} \right]M
+ \left( \f12 \bD_A \dot C^{AB} +\pa^B \bar F \right)\pa_B \tau 
 \cr
&\quad  +  
\f14 N^{AB} \bD_A\pa_B\tau 
+\f14  C^{AB} \bD_A \p_B \dot\tau\,,\label{delM}\\
\d_{(\tau,Y)} \bP_A&=
 [\tau \pa_u +\cL_Y + 2\dot\tau ] \bP_A 
 \cr
 &\quad +  
 3 M  \pa_A \tau -\f18 N_{BC}C^{BC} \pa_A \tau
 + \f12  (C_{A}{}^C N_{BC}) \pa^B \tau 
  \cr
 &\quad + \f34 (\bD_A\bD_C C_B{}^{C}- \bD_B\bD_C C_A{}^{C}) \pa^B \tau
 + \f14 \p_A ( C^{BC} \bD_B\bD_C \tau) \cr
 &\quad + \f12 \bD_{\langle A } \bD_{B\rangle}\tau \bD_C C^{BC} 
 +  C_{AB} \left( \bF\pa^B\tau + \f14 \pa^B \Delta \tau \right)
 \,. \la{delP}
\end{align}
\end{subequations}
These expressions agree with the ones given in \cite{Barnich:2011mi, Compere:2018ylh} if we set $W=\tfrac12 \bD_A Y^A$ and  take into account the relation between the Barnich--Troessaert momentum $N_A$ and the canonical momentum $\bP_A$ given by
$N_A= \bP_A + \pa_A\bar\beta $.
We can also deduce that 
\begin{subequations}
\begin{align}
\d_{(\tau,Y)} \bF&=\left[\tau \pa_u +\cL_Y + 2 \dot{\tau} \right] \bF  + \f12  \bar\Delta \dot{\tau}\label{dR}\,,\\
\d_{(\tau,Y)} \bar\beta&=\left[\tau \pa_u +\cL_Y + 2 \dot{\tau} \right] \bar\beta   
+\f1{8} C^{AB} \bD_A\pa_B \tau\,, \la{delbeta}
\end{align}
\end{subequations}
where the first equality  follows from  the transformation of $g_{uu}$ and the 
 second equality follows from  the transformation of $g_{ur}$.
 Finally the action of the transformations on $g_{uA}$ gives 
 \be
\delta_{(\tau, Y)} \bU_A =  (\tau\pa_u + \cL_{Y} + \dot\tau  ) \bU_A 
   + \f12 (4 \bF  \pa_A\tau +\pa_A \Delta\tau)  +\f12( C_{A}{}^{B} \pa_B\dot\tau - N_{A}{}^{B}\pa_B\tau).\label{delUA}
\ee
Recalling the relation $ \bR(\bq)=4\bF$, we see that the transformation $\d_{(\tau,0)} \bF= 2 \dot{\tau}  \bF  + \f12  \bar\Delta \dot{\tau}$ matches the infinitesimal conformal transformation of the 2d Ricci scalar under the metric rescaling  $\bq_{AB}\rightarrow e^{-2 \dot{\tau} }\bq_{AB}$.
Similarly the variations 
\eqref{delbeta} and \eqref{delUA} can also be obtained through the {on-shell identifications \eqref{asym-EE2}.}

\subsection{Covariant functionals}\label{cov}

The transformation rules reported in the previous section have the general structure
\be
\delta_{(\tau,Y)} {\cal{O}}= [\tau \pa_u +\cL_Y -  s \dot\tau ] {\cal{O}}  + L_{\cal{O}}^A\pa_A\tau +\tilde{L}_{\cal{O}}^A \pa_A\dot\tau +
Q_{\cal{O}}^{AB} \bD_A\pa_B \tau +  
\tilde{Q}_{\cal{O}}^{AB} \bD_A\pa_B \dot\tau.
\ee
The first term is the homogeneous transformation that involves the scale weight $s$ of the functional ${\cal{O}}$.  All scale weights of the different functionals can be found by assigning scale weight $s(\rd s^2)=-2$, while $s(r)=s(\rd r)=-1$ and $s(u)=s(\rd u)=+1$ in the metric expansion, hence the scale weight of $\pa_u $ is $-1$. 
Functionals that transform homogeneously are section of the scale bundle $P$. 
The inhomogeneous terms are of two types, $(L_{\cal{O}}^A , \tilde{L}_{\cal{O}}^A)$ which we call \emph{linear anomalies} and terms $(Q_{\cal{O}}^{AB},  \tilde{Q}_{\cal{O}}^{AB})$ which are the \emph{quadratic anomalies}.
 
 Requiring the absence of quadratic anomalies  turns out to single out important quantities.
A prime example of this is the time derivative of the news tensor, which transforms as
$ \d_{(\tau,Y)} \dot N_{AB}=\left[ \tau \pa_u  + \cL_Y + \dot{\tau} \right]  \dot{N}_{AB}$, i.e., as
a section of weight $-1$ of the scale bundle.
A second example is the \emph{covariant mass aspect} combination 
\be
{\cal M}:= M + \f18 N^{AB} C_{AB}\,,
\la{covM}
\ee
whose transformation is
\be 
\d_{(\tau,Y)}{\cal M} =
\left[\tau \pa_u +\cL_Y + 3 \dot{\tau} \right]{\cal M}
+ M^A \pa_A \tau,
\qquad 
M^A :=  \f12 \bD_B N^{AB} + \pa^A \bF.  
\la{dMcov}
 \ee
The quantity $M^A$ appearing here is itself free of quadratic anomalies. In fact, an explicit calculation gives
\be
 \d_{(\tau,Y)} M^A = \left[\tau \pa_u +\cL_Y + 4\dot{\tau} \right] M^A + 
 \f12 \dot{N}^{AB} \pa_B\tau\,,
 \ee
where we used that $[\pa_A ,\bar\Delta]\dot\tau = -\f12 R \pa_A\dot\tau$.
This shows that if the vacuum structure $M^A=0=\dot{N}_{AB}$ is satisfied,  then the covariant mass aspect $\cal M$ transforms homogeneously.
As a consequence, the manifold of  flat vacua can be defined by the conditions 
\be
 {\cal M}=0, \qquad\dot{N}_{AB}=0, \qquad M^A=0.
\ee
These conditions define an orbit of the \bmw group.

Demanding no quadratic anomaly in the transformation rule for the momentum\footnote{The explicit formula for the transformation will appear in a forthcoming work \cite{FP}.}, we are led to the \emph{covariant momentum aspect} 
\be \label{covariantMom}
{\cal P}_A :=  \bP_A -2  \pa_A\bar\beta-\frac12 \bC_{AB}\bU^B.
\ee
As shown in \cite{Compere:2019gft}, this momentum aspect is the one with the simplest transformation under super-translations. It is also the unique one with no quadratic anomaly, therefore it possesses the simplest transformation under the \bmw group. 

There is a further, independent reason to single out these covariant quantities: They are the leading order contributions to the Weyl scalars. As shown explicitly in Appendix~\ref{AppF}, the covariant momentum aspect is the leading order of the Weyl scalar $\psi_1$, the covariant mass $\cal M$ of (the real part of) $\psi_2$, the vector shift $M_A$ of $\psi_3$, and $\dot N_{AB}$ of $\psi_4$.

\subsection{On the definition of (angular) momentum}

Our analysis has led us to consider two different momentum charge aspects,  $\bP_A$ and ${\cal P}_A $.
They appear in the expansion of  $U^A$ \eqref{UA} as
\begin{align}
U^A&= \f1{r^2} \bU^A -\frac2{3 r^3}\bq^{AB} \left( \bP_B+ \bC_{BC}\bU^C +\pa_B\bar\beta  \right),\cr
 &= \f1{r^2} \bU^A-\f2{ 3r^3} \bq^{AB}\left({\cal P}_B + \f32 \bC_{BC}\bU^C+ 3 \pa_B\bar\b  \right)\,.\label{Uexp}
\end{align}
The momentum aspect $\bP_A$ is the
canonical charge  generating sphere diffeomorphisms for all the   \bms algebra extensions.
It is important to appreciate that these charges are defined at  the finite 
$u=0$ sphere (see \eqref{QYbms} below for the $u=cst$ shift of this  momentum which acquires an extra term linear in $u$).
The fact that $\bP_A$ is the canonical charge aspect of finite cuts is consensual across different approaches in the literature, and its expression coincides in
Ashtekar--Streubel \cite{Ashtekar:1981bq}, Wald--Zoupas \cite{Wald:1999wa}, Barnich--Troessaert \cite{Barnich:2011mi}, Flanagan--Nichols \cite{Flanagan:2015pxa}. 
For this reason we refer to it as the  \emph{canonical momentum}.
On the other hand, the covariant momentum aspect ${\cal P}_A$ appears in Hawking--Perry--Strominger \cite{Hawking:2016sgy} and Compere--Fiorucci--Ruzziconi \cite{Compere:2020lrt}
as charges attached to the limiting spheres at $u=\pm \infty$.
In addition to the  property that this momentum possesses no quadratic anomaly, there is also the fact that it vanishes at timelike infinity, i.e. when $u\rightarrow +\infty$. More generally, it vanishes for vacuum spacetime.

{Even though the covariant and canonical momentum charges are unique in the literature, the parametrization \eqref{Uexp} of the metric coefficient $U^A$ in terms of (angular) momentum, \footnote{We explained in the Introduction why we prefer the name momentum as opposed to angular momentum.}  usually  denoted $N^A$, is far from universal and distinct from the definition of physical charges.
While the latter are clearly more relevant and matter more, 
to help the reader to navigate the different conventions we summarize here    different parametrization used in the literature.
}
To rationalize the discussion,
we consider as in \cite{Compere:2019gft} a two-parameter family  
\be
P_A^{(a,b)} := \bP_A -  \frac{a}{2} C_{AB}\bar{U}^B - 2b   \partial_A \bar{\beta}\,.
\ee
The covariant momentum is $\cP_A=P_A^{(1,1)}$ and the canonical momentum is $\bP_A=P_A^{(0,0)}$.
Most of the literature oscillates between these two parameterization of the metric coefficient: the canonical class, which uses $N_A = \bP_A$, includes
Ashtekar--Streubel \cite{Ashtekar:1981bq}, Wald--Zoupas \cite{Wald:1999wa}, 
while the covariant class, which uses $N_A = \cP_A$, includes
Yau et al. \cite{Chen:2021szm}, Flanagan-Nichols \cite{Flanagan:2015pxa}, Strominger, Hawking--Perry--Strominger \cite{Hawking:2016sgy}.
There are a few authors that consider a different parametrization,
that include Barnich--Troessaert \cite{Barnich:2010eb}, Compere--Fiorucci--Ruzziconi \cite{Compere:2018ylh} and 
M{\"a}dler--Winicour \cite{Winicour16} for $(a,b)=(0,-1/2)$.



Let us note that this discussion was for the BMS momenta, which generate diffeomorphisms combined with a rescaling that preserves the metric determinant. The BMSW canonical momentum, which generates a pure Diff$(S)$, is shifted with respect to the BMS momenta and given by $P_A^{(0,-1)}$, see \eqref{QYR}. 

 \subsection{Vacuum structure}\la{sec:vac}
 As we have seen, the non-radiative sector of the theory is characterized by the conditions 
$ \dot{N}^{\textsf{vac}}_{AB}=0$, and $M^{\textsf{vac}}_A =0$. 
These are solved\footnote{Once one imposes the asymptotic Einstein equation $\bR(\bq) = 4 \bar{F}$.} 
 by  $N_{AB}^{\textsf{vac}} = T_{AB}(\bq)$ where $T_{AB}(\bq)$  is the Liouville  energy-momentum tensor \cite{Compere:2016jwb, Compere:2018ylh} associated with the metric and defined by 
 \be
 \bD^AT_{AB}(\bar{q}) = -\frac12 \pa_B\bR. 
 \ee
It vanishes for the round metric $\mathring{q}$.

We can also understand the vacuum sector as an orbit of the \bmw group (see \cite{Barnich:2016lyg,Barnich:2021dta} for an orbit analysis of the original BMS group).
 To do so we introduce the  \bmw group elements\footnote{ We do not use the $\tau$  parametrization which is more involved as $g_{\tau } = e^{\delta_\tau} = g_{T (e^{\W}-1)/\W, \W}$.}  
 \be
g_{(T,\W,Y)} = e^{\delta_T}e^{\delta_\W} e^{\delta_Y} \,.
\ee
Their action on the gravity phase space is given by exponentiation\footnote{For instance
$$g_{(T,\W,Y)}\cdot  q_{AB}= e^{\cL_Y}\left[ e^{\delta_T}e^{\delta_\W} q_{AB}\right]
= e^{\cL_Y}e^{-2W } \left[ e^{\delta_T} q_{AB}\right]
= e^{\cL_Y}e^{-2W   } e^{T\pa_u} q_{AB}.
$$.} of the infinitesimal action (\ref{delq}, \ref{delC}, \ref{delM}, \ref{dR}) as follows:
 \begin{subequations}
 \begin{align}
 g_{(T,\W,Y)} \cdot \bq_{AB} &=e^{\cL_Y}\left[ e^{- 2 \W} \bq_{AB}\right], \\
g_{(T,\W,Y)} \cdot N_{AB}(u) &=e^{\cL_Y}\left[   N_{AB}(e^\W (u+T) ) -2(\bD_{\langle A}\pa_{B\rangle} \W + \pa_{\langle A}\W\pa_{B\rangle}\W )\right]\,,\\
g_{(T,\W,Y)} \cdot C_{AB}(u) &=e^{\cL_Y}\left[   e^{-\W} C_{AB}(e^\W (u+T) )
- 2  (\bD_{\langle A}\pa_{B\rangle}  T + 2 \pa_{\langle A}\W \pa_{B\rangle}  T)\right.\cr
&\quad \left. -2(u+T) (\bD_{\langle A}\pa_{B\rangle} \W + \pa_{\langle A}\W\pa_{B\rangle}\W ) \right]\,,\\
g_{(T,\W,Y)} \cdot \bR(\bq) &=e^{\cL_Y}\left[  e^{2\W} (\bR(\bq)  + 2 \bar \Delta \W)\right]\,.
\end{align}
\end{subequations}
The vacua $|T,\W,Y\rangle = \hat{g}_{(T,\W,Y)} |0\rangle$ are labelled by a \bmw group element.
In particular, starting from Minkowski spacetime, we see that the orbit of the \bmw group element $g_{(T,W,0)}$ generates a non-zero  value for the news tensor of the vacua given by
\be
N_{AB}^{\textsf{vac}} =T_{AB}(e^{-2\W}\mathring{q})= -2(\bD_{\langle A}\pa_{B\rangle} \W + \pa_{\langle A}\W\pa_{B\rangle}\W )\,.
\la{Nvac}
\ee

Alternatively, it is well known that a generic metric $\bq$ on the sphere can be obtained by the action of 
$g_{(T,\W,Y)}$ on the round metric $\mathring{q}$.
The kernel of this action is $\SU(2)\ltimes \R^S$, so we can also label the vacuum by 
a triple $(\bq_{AB}, T, g )$, where $T$ is a super translation parameter and $g$ is an $\SU(2)$ group element. 

\bigskip

The construction of the \bmw group presented so far applies to any formulation of gravity. We now want to show that the \bmw can be given a canonical formulation in terms of charges, defined using covariant phase space methods, and a suitable renormalization procedure along the lines explained in Section 2. 
The next section will show that one can bypass the problem of integrability of the Hamiltonians, and work uniquely using the Noether charges. This is enough to recover the full \bmw algebra via a centerless bracket at any cross-section of scri, and furthermore to prove that the EEs themselves can be understood as the requirement that this algebra is represented in the phase space. To take these next steps, the choice of a specific Lagrangian and its fundamental variables is a priori necessary.
 In the following, we will specialize to the use of tetrad variables and the Einstein--Cartan formulation, for reasons that we now explain.
 
\section{Tetrad variables}\label{sec:tetrad}

General relativity can be formulated using either the metric or the tetrad as fundamental variables. Both formulations are free of Lagrangian and symplectic anomalies, and the formalism developed in \cite{Freidel:2021cbc} and summarized in Section \ref{sec:summary} can be equally applied to both.
In this paper, we choose to work in tetrad variables. This option is less common in the literature, but there are compelling reasons to do so, and indeed we would like to advocate that the study of boundary symmetries of gravity is better done in tetrad variables.
In addition to the general advantages of working with tetrads, like the technical simplifications of working with forms, and better treatment of matter coupling and first-order version of the action principle, there are specific advantages in terms of covariant phase space methods. These include again simpler expressions to manipulate, but also removing the need of subtraction terms at spatial infinity \cite{Ashtekar:2008jw}, and a convenient framework for isolated and dynamical horizons, e.g. \cite{Ashtekar:2004cn, DiazPolo:2011np}. 
More relevant to the study of future null infinity is the recent result that tetrad variables give access to non-vanishing dual BMS charges \cite{Godazgar:2018qpq, Godazgar:2020kqd,Oliveri:2020xls}.

When using tetrad variables, there is an important aspect to take into account: the additional gauge freedom of internal Lorentz transformations. 
This leads to internal Lorentz charges that are absent in the metric formalism, and a priori different covariant phase spaces. The differences show up for instance in the formulas for the quasi-local charges, 
which been investigated in \cite{DePaoli:2018erh, Oliveri:2019gvm} and \cite{Freidel:2020xyx, Freidel:2020svx, Freidel:2020ayo}; see also \cite{Ashtekar:2008jw, Jacobson:2015uqa,Prabhu:2015vua,Barnich:2016rwk,Frodden:2017qwh} for previous related work.
It is known that equivalence of the charges can be restored for isometries using the Kosmann derivative \cite{Jacobson:2015uqa,Prabhu:2015vua,Barnich:2016rwk} (see discussion in \cite{Oliveri:2019gvm}), but for asymptotic symmetries is it not always the case \cite{Godazgar:2020kqd,Oliveri:2020xls}: at null infinity the standard charges are the same but not the dual ones, thus offering a set-up to recover known BMS results, while at the same time accessing the dual sector.
The exact equivalence can be obtained for all charges including arbitrary diffeomorphisms if one works with a dressed symplectic potential \cite{DePaoli:2018erh, Oliveri:2019gvm}  (see also \cite{Gomes:2018shn,Francois:2020tom}). As explained in \cite{Freidel:2020xyx, Freidel:2020svx}, this choice of potential can be uniquely singled out adding a suitable boundary Lagrangian (see also \cite{Margalef-Bentabol:2020teu,G.:2021qiz}).

Recall from the Introduction that our approach is to focus on Noether charges, and be able to reproduce the algebra of \emph{all}  diffeomorphisms preserving the boundary conditions. In this perspective, the internal Lorentz charges are a key 
component of the corner symmetry group at finite distance, and are also in general non-vanishing at null infinity. While this might seem surprising at first, a moment of reflection shows that this charge contribution is indeed necessary if we are to fix an internal gauge for the frame field which is not preserved by the action of Bondi diffeomorphisms. In fact, we are going to show that the non-vanishing contribution of the internal Lorentz 
symmetry to the Noether charges is crucial in order to  correctly recover the asymptotic Einstein's equations at null infinity.

\subsection{Charges and fluxes}

Let us briefly recall the explicit formulas of the Einstein--Cartan--Holst formulation and its symplectic structure that we will need below.
The Lagrangian is 
\be\la{eq:ECH Lagrangian}
L= \frac12 \Sigma_{IJ}\wedge F^{IJ}(\omega),
\ee
with $ \Sigma_{IJ}= P_{IJKL} e^K\wedge e^L$,
$P_{IJKL}= \frac12 \eps_{IJKL} + \f1\g\eta_{I[K}\eta_{L]J}$ where 
$\g$ is the Barbero-Immirzi or simply Immirzi parameter, and the curvature is $F_{IJ}(\omega)= \rd \omega_{IJ} + \f12 [\omega,\omega]_{IJ}$.
The symplectic potential is given by
\be\la{ECH-pot}
\qquad \theta =\frac12 \Sigma_{IJ}\wedge \delta \omega^{IJ}.
\ee
The action of diffeomorphisms on the phase space variables is given by 
\be
\delta_\xi \Sigma_{IJ}=  {\pounds}_\xi \Sigma_{IJ}= \rd (\iota_\xi \Sigma_{IJ})+\iota_\xi \rd \Sigma_{IJ} , \qquad
\delta_\xi \omega^{IJ}=  {\pounds}_\xi \omega^{IJ}= \rd_\omega (\iota_\xi \omega^{IJ})
+\iota_\xi F^{IJ}. 
\ee
The associated Noether current and charge aspect are  
\be \la{eq:NoetherCharge}
 I_\xi \theta -\iota_\xi L \heq \rd q_\xi, 
\qquad 
q_\xi= \frac12 \Sigma_{IJ}\, \iota_\xi \omega^{IJ} .
\ee
 We  used the vacuum Einstein's equations
$ P_{IJKL} (e^J\wedge F^{KL})\heq 0 $ 
and $\rd_\omega \Sigma^{IJ} \heq 0$.
Since there is no anomaly, the Noetherian flux  is given by 
\bea 
\cF_\xi &=& \iota_\xi \theta+ q_{\delta \xi} \cr
&=& \frac12 \iota_\xi \left(\Sigma_{IJ} \wedge \delta \omega^{IJ}\right) + \frac12\Sigma_{IJ} \wedge  \iota_{\delta \xi} \omega^{IJ}\cr
&=&  \frac12 \iota_\xi \Sigma_{IJ} \wedge \delta \omega^{IJ} + \frac12 \delta \left(\Sigma_{IJ} \iota_\xi \omega^{IJ}\right) 
- \f12 \delta \Sigma_{IJ} \iota_\xi \omega^{IJ}\,.\la{SFluxEC}
\eea
The action of gauge transformations on the phase space is given by 
\be
\delta_\lambda \Sigma_{IJ}=  [\Sigma,\lambda ]_{IJ}, \qquad
\delta_\lambda \omega^{IJ}=  \rd_\omega \lambda^{IJ}.
\ee
The associated Noether current and charge aspect are  given on-shell by 
\be \la{LCharge}
 I_\lambda \theta \heq \rd q_\lambda, 
\qquad 
q_\lambda = \frac12 \Sigma_{IJ}\, \lambda^{IJ},
\ee
while the Noetherian flux  is simply 
\be\la{LFlux}
\cF_\lambda =q_{\delta\lambda}\,. 
\ee

\subsection{Adapted tetrad}

To treat the case of null infinity, it is best to work with a tetrad in doubly-null form, with internal metric $\eta_{01}=-1=-\eta_{23}$ and complex dyad on the sphere.\footnote{This has the advantage of making results from the Newman--Penrose formalism accessible, but all formulas presented in this paper immediately extend to a real dyad and $\eta_{22}=\eta_{33}=1$.}
To avoid the risk of confusion, we distinguish the coframe and frame with a hat,
$e^I=e^I_\mu \rd x^\mu$  and $\hat{e}_I= \hat{e}^\mu_I\pa_\mu$. 
We follow \cite{Godazgar:2020kqd} and choose the following tetrad adapted to the foliation defined by the Bondi coordinates,
\begin{subequations}\label{eq:BondiTetrads}
\begin{align}
e^0 &=e^{2\beta}\rd u\,,\quad &e^1 &= \rd r+  F\rd u\,,\quad &e^i &=r E^i_A \left(\rd \s^A-U^A\rd u\right),
\label{eq:BondiTetrads1}\\
\hat{e}_0&=e^{-2\beta}\left(\p_u-F \p_r+U^A\p_A\right)\,,\quad &\hat{e}_1 &=\p_r\,,\quad &\hat{e}_i &=\f{1}{r} {E}^A_i\p_A. \label{eq:BondiTetrads2}
\end{align}
\end{subequations}
Here $i=2,3$ are the internal indices of the dyad $E^A_i$  on the sphere, with inverse $E_i^A$ and related to the 2d metric by $q_{AB} = E^i_A E^j_B \eta_{ij}$. The dyad is complex with the doubly-null choice, and real if we take $\eta_{ij}=\d_{ij}$.
This choice of tetrad is characterized by the fact that $\hat{e}_i $ are tangent to the sphere and by 
the fact that $\hat{e}_1=\pa_r$.
This eliminates the null rotations around $\hat{e}_0$ and $\hat{e}_1$ and the boost transformations 
$(e^{0},e^1)\to (\l e^{0},\l^{-1} e^1)$.
The only internal gauge freedom left is the tangential frame rotation $ \delta_\theta \hat{e}^{i} =i\theta \epsilon^{i j}\eta_{jk} \hat{e}^k $. 
It can be fixed picking a specific frame or equivalently a 2d spinor on the sphere $z_A$  and demanding that $E^2_A=z_A$ and $E^3_A=\bar{z}_A$.
This tetrad follows the original Newman--Unti adapted tetrad \cite{Newman:1962cia} in taking the first null vector to be the tangent to the null geodesics, but with the difference that the second null vector is not parallel transported along $l$, but chosen so that the dyad is tangent to the sphere, or in other words adapted to the $2+2$ foliation defined by $(u,r)$. See \cite{DePaoli:2017sar} for a comparison of the various choices in the literature and their respective gauge fixings.

In terms of the frames, the vectors  $l$ and $t$ read 
\be
l = \hat{e}_1, \quad t = e^{2\beta}\hat{e}_0 + F \hat{e}_1\,,
\ee
and the torsionless spin connection $\om^{IJ}_\m=e^I_\n\na_\m e^{\n J}$ can be nicely written in terms of the geometric quantities of Section~\ref{SecGeo} 
(see also \cite{Godazgar:2020kqd}),
\begin{subequations}
\begin{align}\label{o10} 
\omega_{10} &= F' \rd u +\f1{r}  \eta_A \left(\rd \s^A -U^A \rd u\right) + 2\beta' \left(\rd r +F \rd u\right), \\
\om_{1i}&= {E^A_i} \left[ \frac{e^{2\beta}}{r^2} \left(   \eta_A  - 2r \pa_A\beta \right)\rd u  -   S_{AB} \left(d\s^B-U^A\rd u\right)\right]\,,\la{o1i}\\
\omega_{0i} &=  -{E^A_i} \left[ \f1r  \p_A F \rd u-  e^{-2\beta }  \left({F} S_{AB}-K_{AB} \right)  \left(\rd\s^B -U^B\rd u\right) + \f{1} {r^2}  \eta_A \left(\rd r +F\rd u\right) \right]\,,
\la{o0i}\\
 \omega_{ij} &=  \left( E^A_{[i} \dot{E}_{j]A}+E^A_{[i} E_{j]B} D_A U^B \right)\rd u + \left( E^A_{[i} E'_{j]A}  \right) \rd r  +  w_{ijA}  \rd\s^A\,.\la{oij}
\end{align}
\end{subequations}
In the last expression, $w_{ijA} = E^{B}_{[i}D_A E_{j]B}$
is the sphere's spin connection associated to the dyad $E^i_A$, and satisfies
$\partial_{[A} E^i_{B]}+  w^i_{\; j [A } E^j{}_{B]}=0$.

For the expansion of the tetrad at future null infinity we use the  expressions \eqref{eq:FallOff}, which we complement with the corresponding one for the dyad, \begin{align}
E^i_A
&=\bar{E}^i_A+  \frac{1}{2r} C_{A}{}^{B}\bar{E}^{i}_B
+  \f1{16 r^2} \bar{E}^{i}_{A}  C_{BC}C^{BC} + o(r^{-2})\,.
\end{align}
Here $\bar{q}_{AB}=\bar{E}^i_A\bar{E}^j_B\eta_{ij}$, and we used the identity \eqref{CC} to simplify the $r^{-2}$ term.

 \subsection{Residual gauge transformations}

The general transformation of the tetrad is a linear combination of diffeomorphisms and internal Lorentz transformations,
 \be\label{dxil}
 \delta_{(\xi,\l)}e^I_\mu = \pounds_\xi e^I_\m 
 - \l^I{}_J e^J_\mu.
 \ee
We have already determined the \bmw vectors fields which preserve the coordinate gauge and metric boundary conditions. 
When working with tetrads we also need to study which internal gauge transformations preserve the adapted form \eqref{eq:BondiTetrads}, in particular the gauge-fixing conditions
 \be
 e_A^0 =0,\qquad e_r^0=0, \qquad e_A^1=0.
\ee
Requiring that \eqref{dxil} preserves these conditions fixes uniquely five gauge parameters in terms of the \bmw vectors $\xi$,
\begin{subequations} \label{Lorentlambda}
\begin{align}
\lambda^{0 i}_{\xi} &= \f{e^{2\beta}}r E^{iA} \pa_A \xi^u,\\
\lambda^{1 i}_{\xi} &= \f{1}r E^{iA} \left(\pa_A \xi^r + F \pa_A \xi^u\right),\\
\lambda^{01}_{\xi} &=  \pa_r\xi^r = \pa_r \left(\f r{2}D_A (I^{AB} \pa_B \tau)+\f r2  U^A \pa_A\tau -r \dot{\tau}\right)\cr
&=- \dot{\tau} +\f1{4r^2} \bD_A (C^{AB} \pa_B \tau)-\f1{2r^2}\bU^A \pa_A\tau+ o(r^{-3}), \la{l01}
\end{align}
\end{subequations}
while $\lambda^{ij}_{\xi}$ is kept free. 
The last term will be the most important to us, it shows that the Weyl boost diffeomorphism needs to be accompanied by a internal boost transformation in the plane normal to the sphere. The strength of this internal boost being given by the boost factor $\pa_r\xi^r$.
The action of the vector fields \eqref{rho} on the frame fields is then
 \be\label{xiext}
 \delta_{\xi}e^I_\m = \pounds_\xi e^I_\m  
 - \lambda_\xi^I{}_J e^J_\m.
 \ee
The explicit asymptotic expansion of the gauge parameters $\lambda^{01}_{\xi}$ corresponding to $\xi_T, \xi_W, \xi_Y$ is given in Appendix \ref{AppExp}.
This expansion is relevant for the construction of Noether charges and fluxes.

\section{Noether charges: energy, Weyl and momentum}\la{sec:charges}
 
As a consequence of \eqref{xiext}, we see that it is not possible to study the asymptotic diffeomorphism charges without simultaneously including the effect of the internal charges.\footnote{This  point was already made in \cite{Godazgar:2020gqd}, however there it had no consequences because they were restricting to the original BMS.}
From the definition of Noether charge \eqref{NQ}, and using the charge aspects associated to the diffeomorphism action \eqref{xiext}, which combines  $\eqref{eq:NoetherCharge}$ with an internal Lorentz symmetry \eqref{LCharge}, one gets
\be\la{ECcharge}
Q_{\xi}= r^2 \int_S \sqrt{q}\, \left(\iota_\xi \omega^{01}+\lambda^{01}_\xi\right). 
\ee
In our present analysis, we are not going to include the dual super-translation charges discovered in \cite{Godazgar:2018qpq, Godazgar:2020kqd}, meaning that we set the Immirzi parameter $1/\g \to 0$.  As revealed in \cite{ Freidel:2020svx}, this limit kills the internal Lorentz rotations, but it still gives us access to the internal Lorentz boosts, namely to the $\lambda^{01}_\xi$ component of the Noether charge \eqref{LCharge}.

We introduce a basis with elements associated to a single parameter and 
 we simply denote $\xi_T :=\xi_{(T,0,0)}$, 
$\xi_\W:=\xi_{(0,\W,0)}$ and $\xi_{Y}:=\xi_{(0,0,Y)}$.
The expansion \eqref{rho} for those vectors gives
\begin{subequations}\la{xi}
\begin{align}
\xi_T&=
T\pa_u -
 \frac1{r}  
\pa^A T \, \pa_A 
+ \f12\left(\bar\Delta T 
-\f1{2r} \bD_A (C^{AB}\pa_BT)
+\f1{r} \bU^A \p_A T 
\right) \pa_r 
+ o(r^{-1})\,, \la{xiT} \\
\xi_\W
&=  u\,  \xi_{T=\W} -\W \, r \pa_r , \la{xiW}\\
  \xi_{Y}
&= 
Y^A \pa_A 
\,. \la{xiY}
\end{align}
\end{subequations}
 Note that in this basis the Weyl transformation  
$\xi_\W$ 
is a radial dilation with parameter $\W$ followed by $u$ times a  super-translation with parameter 
$T=\W$.

We specialize the vector field $\xi$ to be, respectively, the super-translation generator $\xi_T$ \eqref{xiT}, the Weyl super-boost transformation $\xi_\W$ \eqref{xiW}, and the sphere diffeomorphisms $\xi_{Y}$ \eqref{xiY}.
This yields the  following leading order terms\footnote{ All the charge expressions are given modulo $o(1)$.} 
\begin{align} 
Q_T
&= \int_S \sqrt{\bq} \, T\left( M -\f12 \bD_A \bU^A \right)  \,,\la{QT}\\
Q_{\W}&=\int_S \sqrt{\bq}  \,\W \left[-{r^2} +4\bar \beta +  u  \left( M -\f12 \bD_A \bU^A \right) \right]\,,\la{QW}\\
Q_Y&= \int_S \sqrt{\bq}~ Y^A\left(
- r  \bU_A + \bar{P}_A  + 2 \bD_A \bar\beta   
 \right)
 \,.\label{QY}
 \end{align}
One important subtlety is that the \bmsw $\mathrm{Diff}(S)$ or sphere diffeomorphism charge is \emph{not} the same as the \gbms super-Lorentz charge appearing in  \cite{Compere:2018ylh}.
This follows from the relation \eqref{bms} which states that a \gbms vector field is a metric dependent \bmsw vector field.
Explicitly, this means that we can 
write the \gbms vector fields as
\be
 \xi^\gbms_T= \xi_{T}\,,\quad \xi^\gbms_Y=\xi_{Y} +  \xi_{\W=\f12 \bD_A Y^A}\,.\la{xixi}
 \ee
 The main point here is that a \gbms super-Lorentz transformation is the sum of a \bmsw sphere diffeomorphism transformation plus a Weyl super-boost.
 In particular, this implies that the  \gbms super-translation and super-Lorentz charges are given by 
 \be 
 Q^\gbms_{T} = {Q}_{T}, \qquad 
 Q^\gbms_{Y} = Q_Y+ {Q}_{\W=\frac{1}2 \bD_AY^A} .
 \la{QTbms}
 \ee
 This means that  the finer structure of the BMSW group allows us to cleanly disentangle the effect of a sphere diffeomorphism from the effect of a Weyl boost. These two effects are colluded in the original and generalized  BMS groups.
 This observation is essential when discussing the momentum Noetherian flux  in the next section. In particular, the $\mathrm{Diff}(S)$ or sphere diffeomorphism flux is trivial for the BMSW group, as expected for a corner  transformation that do not translate the sphere \cite{Freidel:2020xyx}, while it is not for the generalized BMS group, simply because a \gbms super-Lorentz includes a Weyl boost.
  If one writes the \gbms super-Lorentz charge explicitly one gets the leading asymptotics
\begin{align}
Q^\gbms_{Y} &=   \int_S \sqrt{\bq}~ Y^A\left[
- r  \bU_A + \bar{P}_A 
 - \f{u}{2}  \bD_A  \left(M-\f12\bD_B\bU^B\right)    
 \right]
\,.\la{QYbms}
\end{align}
This shows as promised that $\bP_A$ is the finite and $u$-independent component of the BMS momentum aspect.
This justifies \emph{a posteriori} our parametrization.

\section{Noetherian fluxes: energy, Weyl and momentum}\label{sec:fluxes}

In this section, we provide the asymptotic expressions of the Noetherian fluxes associated to the \bmw vector fields, leaving all details to Appendix~\ref{AppC}.
From the definition of Noetherian flux  \eqref{NF} and recalling that the Einstein--Cartan formulation is anomaly-free, one gets the expression 
\be\la{ECflux}
{\cF}_\xi :=\int_S\left( \iota_\xi \theta + q_{\delta \xi}+ q_{\delta\lambda_\xi} \right)\,,
\ee
where the first contribution is  the symplectic potential flux associated to diffeomorphisms $\eqref{SFluxEC}$, supplemented by the contribution due to the field-dependence of the vector field $\xi$ and the internal Lorentz symmetry \eqref{LFlux}.
It is instructive to split the Noetherian flux  in its three contributions 
and analyze their structure and properties.
The symplectic potential flux is 
\be
\label{Fsympl}
{\cF}^\theta_\xi :=\int_S \iota_\xi \theta\,
=  \int_{S}  \sqrt{q}\, e^{2\beta} r^2\left(\xi^r\theta^u-\xi^u \theta^r \right)\,, 
\ee
where  $\theta=\theta^\m\epsilon_\m$, with $ \epsilon_\m =\iota_{\pa_\m}\epsilon$ and $\epsilon=e^{2\beta} r^2\sqrt{q}\, \rd u\, \rd r\, \rd \s^2$ is the 4-volume form
-- see Appendix~\ref{AppC} for details.
The fact that it depends only on the components $(\xi^u,\xi^r)$ means that it vanishes if $\xi$ is tangential to the sphere, as it is to be expected since  the tangential diffeomorphisms have generators trivially integrable.

Let us analyse separately the expressions of the temporal and radial components.\footnote{Note the possible source of confusion: $\theta^r$ is the ``temporal'' component of the flux because it is multiplied by $\xi^u$. Similarly for the ``radial'' component $\theta^u$.} 
For the temporal flux, we have
\begin{align}\la{tr}
-\sqrt{q}e^{2\beta} r^2\theta^r &= r \sqrt{q}E_i^A \left( e^{2\beta}\delta \omega_{A 0 }{}^i- F \delta \omega_{A 1 }{}^i   \right) 
+ r^2 \sqrt{q}(\delta \omega_{u10}+U^A \delta \omega_{A 10}) \\ & = \sqrt{\bar{q}} \theta^r_{\div} + \sqrt{\bar{q}} \theta^r_{\fin} +o(1)\,,
\end{align}
where the most divergent term is a total derivative plus a total variation, 
 \be
- \sqrt{\bar{q}} \theta^r_{\div}  = 2r\delta (\sqrt{\bar{q}} \bF)  - 
 \f{r}{4} \pa_u \left( \sqrt{\bar{q}}  {C}^{AB} \delta \bar{q}_{AB} \right)\,.
\label{fudiv}
 \ee
This means that it can be renormalized away by a choice of boundary Lagrangian, as shown below in Section \ref{sec:div-pot}. 
The finite contribution to the temporal flux is given by
\begin{align}
\sqrt{\bq}\theta^r_\fin&=
  \delta \left[\sqrt{\bq} ( M+ D_C\bU^C)\right] 
 + \frac14 \sqrt{\bar{q}} N^{AB} \delta C_{AB}
 + \frac12 \sqrt{\bar{q}} \left( \bF  C^{AB}   +D^{\langle A} \bU^{B\rangle}  \right) \delta \bq_{AB} \cr
 &\quad  + \left(M -{ \frac12}  D_C\bU^C - \frac14 N_{CD} C^{CD}\right)  \delta\sqrt{\bq}
 +o(1)\,.
 \la{thetar}
\end{align}
While for the radial flux, we have
 \begin{align}
-e^{2\beta} r^2 \theta^u &= r E_i^A \delta \omega_{A 1 }{}^i -r^2 \delta \omega_{r10} \,. 
\la{tu}
\end{align}
In the case of the radial flux, it is important to appreciate that $r\pa_r$ is an  $\mathcal{O}(1)$-vector,
and that all the vector fields that we are considering contain a term
$\rho r\pa_r$ with $\rho$ finite at $\scri$. 
This means that the divergent term of the radial flux that must be isolated and renormalized is $r \theta^u$.
In light of these considerations, using
the asymptotic expansion of the radial flux (see Eq.~\eqref{thetauB} for its derivation), we can write
\be
\sqrt{\bq}\, e^{2\beta} r^2\theta^u=  \sqrt{\bar{q}} \theta^u_{\div} + \sqrt{\bar{q}} \theta^u_{\fin}+o(r^{-1})\,,
\ee
where the  divergent component is  a total derivative
\be
\sqrt{\bq}\theta^u_\div=\f 12 \p_r \left(r^2 \d \sqrt{\bq}\right)
-\f14 \pa_r \left( r \sqrt{\bq} C^{AB} \d \bar q_{AB} \right)\,,
\label{frdiv}
\ee
and the finite contribution is given by
\bea \label{thetau}
\sqrt{\bq} \theta^u_\fin &=&
\f{1}{4r}  C_{CD}  C^{CD} \delta {\sqrt{\bq}} 
-\f{1}{4r}\sqrt{\bq} C^{AB} \delta  C_{AB}
 -\f 4r \sqrt{\bq} \d \bar \beta\cr 
 &=& 
-\f 4r \sqrt{\bq} \d\textsf{E}_\beta
-\f{1}{4r} \sqrt{\bq}  C^{\langle A}{}_C C^{B\rangle C}\delta  \bq_{AB} \heq 0,
\eea 
where in the last equality we used the identity \eqref{CC}. 
The on-shell vanishing of $\theta^u_\fin$ is necessary to insure that the renormalized symplectic potential
$\int_{\Sigma_u}\theta^u_\fin \epsilon_u$, associated with the {null} slice $\Sigma_u=\{u=\mathrm{cste}\}$, is finite.

The second contribution originating from the field dependence of the vector field, namely
\be
\cF_{\d\xi}:=\int_S q_{\d \xi},
\ee
{though it is nonvanishing, its field space contraction} will not contribute to the flux-balance laws that we will derive in Section \ref{sec:FB2} and it does not contribute any divergent term in the expression for the fluxes.

Finally, the contribution to the Noetherian flux  coming from the field-dependent gauge parameter $\lambda_{\xi}$, denoted as
\be\label{Flambda}
\cF^{\lambda}_{\xi}:=\int_S q_{\d \l_\xi}= r^2 \int_{S}\sqrt{\bq}~\delta \lambda^{01}_{\xi},
\ee
will give nonzero contribution to the flux-balance laws and its expression will be provided for each \bmw generator.

In the following subsections, we provide the asymptotic expressions of the Noetherian fluxes associated to each \bmw generator in \eqref{xiT},  \eqref{xiW},  \eqref{xiY}. 
To lighten the notation, the three contributions to the Noetherian fluxes will be labelled by the symmetry parameter rather than its \bmsw generator. For example, in the case of super-translation, we denote $\cF^{\theta}_{T}:= \cF^{\theta}_{\xi_T}$ and $\cF_{\d \xi_T} := \cF_{\d T}$.

\subsection{Energy Noetherian flux }
Let us consider the super-translation generator $\xi_T$. The energy Noetherian flux reads as 
\be
{\cF}_T ={\cF}^\theta_T+ \cF^{\lambda}_{T}+\cF_{\d T}.
\ee
The first contribution, the simplectic flux \eqref{Fsympl} for $\xi=\xi_T$, displays $r$-divergent terms coming from \eqref{fudiv} and \eqref{frdiv}, and $\mathcal{O}(1)$ terms from \eqref{thetar} and \eqref{thetau}. Collecting these contributions, one obtains that
\begin{align}
\cF^\theta_T &= r\int_{S}   \left[\,2 T  \delta \left( \sqrt{\bq} \bF\right) +\f12 \bar\Delta T \delta \sqrt{\bq}\,  - \frac{1}4  \sqrt{\bq} \,T \left( N^{AB} \delta \bq_{AB}\right)\right] \cr
&\quad -   \int_{S}T \delta \left[ \sqrt{\bq}\left( M+ \bD_C\bU^C\right) \right]
-\frac14 \int_{S} \sqrt{\bar{q}} T \left( N^{AB} \delta C_{AB}\right) \cr
&\quad - \int_{S}  \left[  T\left(M  -{ \frac12}  \bD_A\bU^A - \frac14 N_{CD} C^{CD}\right) -\f12  
\bU^A \p_A T  +\f1{4} \bD_A \left(C^{AB}\pa_BT
 \right) \right]\delta\sqrt{\bq}\cr
&\quad -  \frac12 \int_{S} \sqrt{\bar{q}}  \left(\f14  \bar\Delta T  ~C^{AB} + T \bF  C^{AB}   +T D^{\langle A} \bU^{B\rangle}   \right) \delta \bq_{AB}+o(1)\la{Ftt}
\,.
\end{align}

The second contribution to the energy Noetherian flux  is 
\be
\cF_{\d T}=\int_S \sqrt{\bq}\, \d \bq^{AB} \p_B T \bU_A+ o(1)\,.\la{FdT}
\ee

The third and final contribution, namely that one coming from the field-dependent gauge parameter $\lambda$ in \eqref{Flambda}, for $\lambda^{01}_\xi = \lambda^{01}_{\xi_T}$ (see \eqref{l01} and \eqref{lambdaT}), reads as
\be
 \cF_{T}^{\l} = \f12 \int_S \sqrt{\bq}\left\{\f1{2} \d \left[\bD_A \left(C^{AB} \pa_B T\right)\right]-\d \left(\bU^A \pa_AT\right) \right\} + o(1)\,.
 \la{qlt}
\ee

We emphasize that the divergent part of the energy flux comes uniquely from the symplectic term in the first line of \eqref{Ftt}. For later convenience, let us highlight it and denote it as
\be
{\cF}_T^\div= r\int_{S}   \left[\,2 T  \delta \left( \sqrt{\bq} \bF\right) +\f12 \bar\Delta T \delta \sqrt{\bq}\,  - \frac{1}4  \sqrt{\bq} \,T \left( N^{AB} \delta \bq_{AB}\right)\right]\,.\la{FTdiv}
\ee

\subsection{Weyl Noetherian flux}

Let us consider the Weyl rescaling generator $\xi_\W$. The Weyl Noetherian flux is given by
\be
{\cF}_\W ={\cF}^\theta_\W+ \cF_{\W}^{\l}+\cF_{\d \W}. \la{Fw}
\ee
We follow the same computational steps as in the previous case of the energy Noetherian flux. For the Weyl Noetherian flux , considering $\xi=\xi_W$, its symplectic contribution can be written as
\begin{align}
\cF^\theta_\W &=u\, \cF^\theta_{T=\W }-r  \int_S \sqrt{\bq}\, \W \theta^u\cr
&=  u\, \cF^\theta_{T=\W }-r^2 \int_{S} \,\W\,  \delta {\sqrt{\bq}} +\f r4\int_{S} \sqrt{\bq} \, \W\, C^{AB} \d \bar q_{AB}\cr
&\quad -\f{1}{4} \int_{S}  \, \W \left( C_{AB}  C^{AB}\delta {\sqrt{\bq}}
- \sqrt{\bq} \,  C^{AB} \delta  C_{AB}
 -16  \sqrt{\bq}\, \d \bar \beta\right)
 +o(1)\,.
 \la{Ftw}
\end{align}

The second contribution is $\cF_{\d W}=u \cF_{\d T}$, with $T=W$  in \eqref{FdT}.
Finally, the contribution originating from the field-dependent gauge parameter $\lambda$ in \eqref{Flambda}, for $\lambda^{01}_\xi = \lambda^{01}_{\xi_W}$ (see \eqref{l01} and \eqref{lambdaW}) takes the form
\be
 \cF^\l_\W = u \cF^\l_{T=\W} =
 \f u2\int_S \sqrt{\bq}\left\{\f1{2} \d \left[\bD_A \left(C^{AB} \pa_B \W\right)\right]-\d \left(\bU^A \pa_A \W \right)\right\} + o(1)\,.
 \la{qlw}
\ee

Also in the Weyl case, the radial divergences come from the Weyl symplectic flux, that in turn come from the divergences in the energy symplectic flux. They are given by the following expression
\begin{align}
{\cF}_\W^\div&= u {\cF}_{T=\W}^\div - r  \int_S \sqrt{\bq}\, \W\, \theta_{\div}^u\cr
&= 2ur\int_{S}   \left[\,\W\,  \delta \left( \sqrt{\bq} \bF\right) +\f14 \bar\Delta \W\, \delta \sqrt{\bq}\,  - \frac{1}8  \sqrt{\bq} \,\W \left( N^{AB} \delta \bq_{AB}\right)\right]\cr
&\quad -{r^2}\int_{S} \,\W\,  \delta {\sqrt{\bq}} +\f r4\int_{S} \sqrt{\bq} \, \W\, C^{AB} \d \bar q_{AB}\,.
\la{FWdiv}
\end{align}

\subsection{Momentum Noetherian flux}

Let us consider  the ${\rm diff}(S)$ generator $\xi_Y$. The momentum Noetherian flux is
\be
{\cF}_Y ={\cF}^\theta_Y+ {\cF}^\l_Y+\cF_{\d Y}\,, \la{Fy}
\ee
and it is immediate to see that 
\be
{\cF}_Y =0\,,\la{Fy0}
\ee
as the sphere diffeomorphism generator $\xi_Y$ has only the tangential component $Y^A \pa_A $.

It is illustrative to complete   the comparison with the generalized \bms algebra. Let us write the corresponding Noether fluxes  in terms of the \bmw ones derived above. By means of the relation \eqref{xixi} between generalized BMS and BMSW vectors fields, it is immediate to see that
\be
\cF^\gbms_{T} = {\cF}_{T}\,,\qquad
\cF^\gbms_{Y} = {\cF}_{W=\frac{1}2 \bD_AY^A}\,.
\ee
This highlights how  the momentum Noether charge \eqref{QY}
is  Hamiltonian  in the \bmw group, while it is not for the  generalized BMS group of \cite{Compere:2018ylh}. This striking feature is a reflection of the fact that, in the Bondi frame, Weyl super-boost transformations are equivalent to conformal rescalings and the 
Lie algebra basis $(\t, Y)$ allows us to filter this component of the $\mathrm{diff}(S) $ action out of the momentum charge. On the other hand, the condition $W^\gbms=\frac{1}2 \bD_AY^A$ imposed in \cite{Compere:2018ylh} to keep the scale factor constant (see Section \ref{sec:LAD}) mixes the Weyl super-boost and the sphere diffeomorphism fields, so that the resulting momentum charge is now a combination of the two and it has non-vanishing flux. In the BMS case, one recovers an Hamiltonian charge only when restricting to  area-preserving diffeomorphisms $\bD_A Y^A=0$, namely to the rotational part of the charge.

 \section{Einstein's equations from flux-balance laws}\la{sec:FB2}

We are now ready to prove the main result of the paper, namely the remarkable fact that the  asymptotic Einstein's equations can be obtained from the  flux-balance law \eqref{Flux1}, where the bracket on the RHS is defined by \eqref{BTb} in terms of the Noether charges \eqref{ECcharge} and fluxes \eqref{ECflux}.
Before doing so, we recall the explicit expressions of the asymptotic Einstein's equations to make their connection with the flux-balance laws  manifest.

\subsection{Einstein's equations in Bondi gauge}\la{AppEE}

It follows from considering the vector field $l = \pa_r$ and the one-form $\underline{n} = \dd r$ (see also Section \ref{SecGeo} and Fig.\ref{Diagram}),
that the Einstein's equations can be grouped in four sets of equations 
\begin{subequations}
\begin{align}
G_{\m \underline{n}} = G_{\m}^{\;\;r} &=0 \qquad \quad &\text{(constraints equations)}\\
G_{\m {l}} = G_{\m r} = -e^{2\beta}G_{\m}^{\;\;u} 
&=0 \qquad \quad &\text{(radial evolution equations)} \\
G_{\langle AB \rangle}&
=0 \qquad \quad &\text{(propagating equations)}\\
q^{CD}G_{CD}&=0 \qquad \quad &\text{(``trivial'' equation)}
\end{align}
\end{subequations}
Among the constraints equations, we notably have in $G_{u}^{\;\;r} =0$ and $G_{A}^{\;\;r} =0$, respectively, the (retarded time) evolution for the Bondi mass and the Bondi angular momentum aspects.

We list the Einstein tensor with mixed indices $G^{\;\;\n}_{\m}$ obtained in the Bondi gauge. In the following, $\textsf{E}_{\Phi}$ stands for the asymptotic Einstein's equations for the fields $\{\bF, \bar \beta, \bU, M, \bP_A\}$ defined in \eqref{asym-EE}, \eqref{EEM} and \eqref{EEP}.

The four radial evolution equations $G^{\;\;u}_{\mu}=0$ give
\begin{subequations}
\begin{align}
G^{\;\;u}_{u} &= -\frac{1}{2r^2}\textsf E_{\bF} -\frac{1}{r^3}~\bD^A \textsf E_{\bU_A}+  o(r^{-3}) ,\\
G^{\;\;u}_{r} &= \frac{8}{r^4}\textsf E_{\bar\beta}+o(r^{-4}) ,\\
G^{\;\;u}_{A} &=  \frac{1}{r^2}\textsf E_{\bU_A}- \frac{4}{r^3}\bD_A \textsf E_{\bar\beta}+ o(r^{-3}).
\end{align}
\end{subequations}
In addition, we have the  constraints equations $G^{\;\;r}_{\mu}=0$, with
\begin{subequations}
\begin{align}
G^{\;\;r}_{u} &= \frac{2}{r^2}\left(\textsf E_{ M}+ \f12 \bD^A \dot{\textsf E}_{{\bU}_A} \right) +  o(r^{-2}) ,\\
G^{\;\;r}_{r} &= -\frac{1}{2r^2}\textsf E_{\bF}-\frac{1}{r^3}~\bD^A \textsf E_{\bU_A}+  o(r^{-3}) ,\\
G^{\;\;r}_{A} &= -\frac{1}{r}\dot{\textsf E}_{\bU_A}+\frac{1}{r^2}\left(\textsf E_{{\bP}_A}+2\bD_A \dot{\textsf E}_{\bar\beta}  -2 \bar{F}\textsf E_{\bU_A}  -\f12 \bU_A \textsf E_{\bF}\right) +o(r^{-2}).
\end{align}
\end{subequations}
The remaining Einstein's equations are 
\begin{subequations}
\begin{align}
G_{\langle AB \rangle } &= 
-\frac{1}{2r}\textsf E_{\bF} C_{AB} +\f1{r^2} \left( \dot{E}_{AB} + \cdots \right)+ o(r^{-2}) ,\\
g^{CD}G_{CD}&=\frac{8}{r^3} \dot{\textsf E}_{\bar \beta}+ o(r^{-3}).
\end{align}
\end{subequations}
{We also recall that $G_{\mean{AB}}={\cal R}_{\mean{AB}}$, where the sphere components of the 4d Ricci tensor and the 4d Ricci scalar are given respectively by}
\begin{align}
{\cal R}_{AB} &= \f12 \textsf E_{\bF}\bq_{AB}+\frac{1}{r}\bD^{C}\textsf E_{\bU_C}\bq_{AB} + o(r^{-1}),\\
{\cal R} &= 2 g^{ur} {\cal R} _{ur} + g^{rr}{\cal R} _{rr}+2g^{rA}{\cal R} _{rA} + g^{AB}{\cal R} _{AB}\cr
& = \frac{1}{r^2}\textsf E_{\bF} +\frac{1}{r^3}\left(2\bD^A \textsf E_{\bU_A}-8\dot{\textsf E}_{\bar \beta}\right) + o(r^{-3})\,,\la{R}
\end{align}

  Two comments are in order here. First, the propagating equations, $G_{\langle AB \rangle } =0$, contain the evolution equation for the symmetric and traceless field $E_{AB}$, that appears in the asymptotic expansion of $q_{AB}$ (see Section \ref{sec:null-infinity} above) and enters the Weyl {scalar $\psi_0$}
  (see Appendix \ref{AppF}). The dots in the $\mathcal{O}(r^{-2})$ coefficient stand for additional terms that are not important for the present discussion (see more on this in the Conclusions Section \ref{sec:Conc}).
Second,  in the expressions of the Einstein tensor above, we have set the boundary condition $\pa_u \bar{q}_{AB}=0$ in Eq.~\eqref{asym-EE} to streamline the presentation of the Einstein tensor components. However, the reader can find it useful to know that, without that condition, the leading orders of $G^{\;\;\nu}_{\mu}$ change as follows: $G^{\;\;u}_{u} = \mathcal{O}(r^{-1})$, $G^{\;\;u}_{r} = \mathcal{O}(r^{-1})$, $G^{\;\;r}_{u} = \mathcal{O}(1)$, $G^{\;\;r}_{A} = \mathcal{O}(1)$ and $G_{AB} = \mathcal{O}(r)$.

\subsection{Flux-balance laws}
We are now ready to evaluate the flux balance laws  \eqref{TheBra}, which can be written as 
 \be\la{TheBra2}
 \delta_\xi Q_\chi - I_\chi \cF_\xi
  + \int_S \iota_\xi \iota_\chi L+Q_{\lbr\xi,\chi\rbr}=
 r^2 \int_S  \sqrt{q} \,e^{2 \beta} (\xi^u\chi^\m G_{\m}{}^r -\xi^r\chi^\m G_{\m}{}^u ) \,,
 \ee
{To make the flux-balance laws manifest, we use the fact that $\chi=\hat u:=\p_u\in$ \bmw, and its corresponding variation $\d_{\hat u}$ is a time derivative, thus providing a term $\dot Q_\xi$ in \eqref{TheBra2}.}
The expression of the Noether charge is given by
\be\la{Qchi}
Q_{\hat u}= \int_S \sqrt{\bar{q}} M\,,
\ee
and the expression of the Noetherian flux  reads as
\begin{align}
\cF_{{\hat u}}&= r \int_{S} \left( 2 \delta (  \sqrt{\bq} \bar{F})
- \frac{1}4   \sqrt{\bq} N^{AB} \delta \bq_{AB} 
\right)
 \cr
&\quad -
\int_{S} \delta \left(\sqrt{\bq}( M    + \bar  D_A\bU^A)\right) - \int_S \left(M -{ \frac12}  D_C\bU^C - \frac14 N_{CD} C^{CD}\right)   \delta \sqrt{\bq}\cr
&\quad - \frac12 \int_S \sqrt{\bar{q}} \left( \bF  C^{AB}   +D^{\langle A} \bU^{B\rangle}\right) \delta \bq_{AB}-\f 14\int_{S}\sqrt{q} N^{AB} \d  C_{AB} +o(1)\,.\la{Fchi}
\end{align}

In our proof, we concentrate on the $(T, \W, Y)$ basis \eqref{xiT}, \eqref{xiW}, \eqref{xiY} which satisfy the Lie algebra \eqref{xi-Lie}.
 This allows us to readily identify which residual diffeomorphism transformation relates the associated
 charge bracket  to the (holographically) equivalent asymptotic Einstein's equations.
Specifically, we will specialize the pair $(\xi, \chi)$ to $(\xi_{T,W,Y}, \pa_u)$ and the {flipped order} $( \pa_u, \xi_{T,W,Y})$.
The computation uses the Noether charges derived in Section \ref{sec:charges} and the Noetherian fluxes obtained in Section \ref{sec:fluxes}.

For later convenience, let us write the Lagrangian as 
\be
L = \frac{1}{2}\epsilon {\cal R}\,,
\ee 
and use the 4d Ricci scalar in \eqref{R} to compute the inner products of the Lagrangian form with time translations $\pa_u$ and the three \bmsw generators
\begin{subequations}
\begin{align}
\int_S \iota_{\xi_T}\iota_{\hat u} L &= \f14 \int_S \sqrt{\bq}~ \bar{\Delta}T~\textsf{E}_{\bF}  +o (1),
\la{LT}\\
\int_S \iota_{\xi_\W}\iota_{\hat u} L &=-\f12 \int_S \sqrt{\bq}\left[ \W\, \left(r\textsf E_{\bF} + 2\bD^A \textsf E_{\bU_A} - 8\dot{\textsf E}_{\beta}\right)-\frac{u}{2}\bar{\Delta}\W\, \textsf E_{\bF} \right] + o(1),
\la{LW}
\\
\int_S \iota_{\xi_Y}\iota_{\hat u} L &=0\,.
\la{LY}
\end{align}
\end{subequations}
These expressions will be shortly needed to evaluate the brackets \eqref{BTb} in the flux-balance relations \eqref{TheBra}.

\subsubsection{Energy flux-balance}

We apply the general expression \eqref{TheBra2} to write the energy flux-balance law, respectively, with $(\xi, \chi) = (\pa_u, \xi_T)$ and $(\xi, \chi) = (\xi_T,\pa_u)$. They are
\begin{subequations}\la{FBT-off}
\begin{align} \label{FBT-off1}
 \d_{\hat u} Q_T-I_{T}\cF_{\hat u}+
 \int_S \iota_{\hat u} \iota_{\xi_T}L 
+ Q_{\lbr{\hat u},\xi_T\rbr}&= \int_S \sqrt{\bq}\, T\left(2\textsf E_{ M}
-\f14 \bar\Delta \textsf E_{\bF}
\right) +o(1)\,,\\
  \d_T Q_{\hat u}-I_{\hat u}\cF_T
  +
 \int_S\iota_{\xi_T} \iota_{\hat u} L 
  +Q_{\lbr\xi_T, {\hat u}\rbr}&=\int_S \sqrt{\bq}\,  T\left(2\textsf E_{ M}+  \bD^A \dot {\textsf E}_{{\bU}_A}
 + \f14 \bar\Delta \textsf E_{\bF}
  \right)  +o(1)\,. \label{FBT-off2}
\end{align}
\end{subequations}

We first focus on the first expression \eqref{FBT-off1}, for which the energy charge bracket
\be\la{FBT}
\{Q_{\hat u}, Q_{\xi_T}\}:= \d_{\hat u} Q_T-I_{T}\cF_{\hat u} + \int_S \iota_{\hat u} \iota_{\xi_T}L= -Q_{\lbr{\hat u},\xi_T\rbr}
\ee
is equivalent to combination of  asymptotic  Einstein's equations
\be
B:= 2\textsf E_{ M}
-\f14 \bar\Delta \textsf E_{\bF}=0\,.\la{FBE1}
\ee
We provide the proof by separately computing all the terms in Eq.~\eqref{FBT}.
By means of Eq.~\eqref{QT}, the variation with respect to time of the energy Noether charge is given by
\be\label{QuT}
 \d_{\hat u} Q_T= \int_S  \sqrt{\bq} T\left( \dot M -\f12 \bD_A \dot \bU^A \right) \,.
\ee
We then use the expression of the flux $\cF_{\hat u}$ in ~\eqref{Fchi} and the field variations under super-translations in \eqref{dT}
to compute the field space contraction on $\cF_{\hat u}$
\begin{align} \label{ITF}
I_{T}\cF_{\hat u}&=-
\int_{S}\sqrt{q} \delta_T M    
-\f 14\int_{S}\sqrt{q} N^{AB} \d_T  C_{AB}\cr
&=-
\int_{S}\sqrt{q}\, T \left(
\dot M - \bar\Delta \bF -\f34  \bD_A\bD_B N^{AB}
+\f14  N^{AB}N_{AB}
{+\f12 \bD^A\dot{\textsf E}_{\bU_A}}
\right)\,.
\end{align}
Finally, by means of the commutators \eqref{comm}, we have $\lbr{\hat u},\xi_T\rbr=0$, and hence $Q_{\lbr{\hat u},\xi_T\rbr }=0$. Therefore, it is straightforward to see that the energy flux-balance law \eqref{FBT}, after adding up Eqs.~\eqref{QuT}, \eqref{ITF}, and \eqref{LT}, yields  the condition \eqref{FBE1}.

From \eqref{FBT-off2}, we see that   flipping the order of the vector fields in the energy charge bracket \eqref{FBT},  which can be read off from the right-hand side of \eqref{FBT-off2}, 
\be\la{FBE-rev}
\d_T Q_{\hat u}-I_{\hat u} \cF_T+\int_S \iota_{\xi_T}\iota_{\hat u} L= Q_{\lbr{\hat u},\xi_T\rbr}\,,
\ee
is expected to be equivalent to the following linear combination of asymptotic Einstein's equations
\be
B':= 2\textsf E_{ M}+  \bD^A \dot {\textsf E}_{{\bU}_A}
 + \f14 \bar\Delta \textsf E_{\bF}=0\,.\la{FBE2}
\ee

We can  prove this statement again by  evaluating the different terms in \eqref{FBE-rev} separately. First, we compute the variation of $Q_{\hat u}$ under super-translations  
\bea
\d_T Q_{\hat u}=
\int_S\sqrt{\bq}\,T\left(\dot M-\bar\Delta  \bF -\f14 \bD_A\bD_B N^{AB} +{ \f12\bD_A\dot{\textsf E}_{\bU_A}}\right)\,,
\eea
then the field contraction of the energy Noetherian flux 
\bea
I_{\hat u} \cF_T=
 -   \int_{S}\sqrt{\bq}\, T   \left( \dot M+ \bD_A\dot \bU^A+\f14 N^{AB}N_{AB}
 -\f12\bD_A \dot\bU^A\right)\,.
\eea
By means of \eqref{LT} and $Q_{\lbr{\hat u},\xi_T\rbr }=0$,  we obtain  exactly \eqref{FBE2}.

We will see in a the next subsection how both \eqref{FBE1} and \eqref{FBE2} are automatically satisfied by the asymptotic Einstein's equations implied by the Weyl flux-balance law.

\subsubsection{Weyl flux-balance}

In the case of conformal transformations, the flux-balance law \eqref{TheBra2} with $(\xi, \chi) = (\pa_u, \xi_W)$ and $(\xi, \chi) = (\xi_W,\pa_u)$ is given, respectively, by
\begin{subequations}\la{FBW-off}
\begin{align}
 \d_{\hat u} Q_\W&-I_{\W}\cF_{\hat u}+ \int_S  \iota_{\hat{u}} \iota_{\xi_{\W}} L +Q_{\lbr{\hat u},\xi_\W\rbr}\cr
 &= \int_S \sqrt{\bq}\,\W\, \left[
  \frac{r}{2}\textsf E_{\bF}
+\bD^A\textsf E_{\bU_A} 
  +2u 
  \left(\textsf E_{ M}
-\f 18 \Delta   \textsf E_{\bF}
 \right)  
  \right]+o(1)\,,
 \la{FBW-off1}
  \\
  \d_\W Q_{\hat u}&-I_{\hat u}\cF_\W+ \int_S \iota_{\xi_{\W}} \iota_{\hat{u}} L+Q_{\lbr\xi_\W, {\hat u}\rbr}\cr
  &=  \int_S \sqrt{\bq}\, \W\,\left[
 -  \frac{r}{2}\textsf E_{\bF}
-  \bD^A\textsf E_{\bU_A} 
  +2u 
  \left(\textsf E_{ M}
 + \f12 \bD^A \dot{\textsf E}_{{\bU}_A} 
 +\f 18 \Delta   \textsf E_{\bF}
 \right)   \right]+o(1).
 \la{FBW-off2}
\end{align}
\end{subequations}

We focus on the first Weyl charge bracket
\be\la{FBW}
 \d_{\hat u} Q_\W-I_{\W}\cF_{\hat u}+ \int_S  \iota_{\hat{u}} \iota_{\xi_{\W}} L= -Q_{\lbr{\hat u},\xi_\W\rbr},
\ee
and show that it is equivalent to the following combination of the asymptotic Einstein's equations
\be
A+u B=0\,,
\la{FBW-1}
\ee
 with
\begin{align}
A&:=
  \frac{r}{2}\textsf E_{\bF}
+\bD_A\textsf E_{\bU_A}\,,
\end{align}
and $B$ given in \eqref{FBE1}, as expected from the RHS of \eqref{FBW-off1}.

By means of Eq.~\eqref{QW}, we compute the time variation of the Weyl Noether charge
\be \label{QuW}
 \d_{\hat u} Q_\W=
 4 \int_S \sqrt{\bq} \,\W\, \dot{\bar \beta}
+ u  \int_S \sqrt{\bq}\,\W\,\left( \dot M
-\f12  \bD_A\dot  \bU^A\right)\,.
\ee
We then use the field variations under the Weyl rescalings in \eqref{dW} 
to compute the field contraction on $\cF_{\hat u}$
\begin{align} \label{IWF}
I_{\W}\cF_{\hat u}
 &=\int_S\sqrt{\bq}\, \W\, \left(    M
  -\f14  \bD_A \bD_B C^{AB}
   -  \bD_A \bU^A
 -\f14  N^{AB} C_{AB}
  {+\f12 \bD_A \textsf E_{\bU_A}}
 \right)\cr
 &\quad- u \int_S\sqrt{\bq}\, \W\, \left(\dot M- \bar \Delta \bF
 -\f 34 \bD_A \bD_B N^{AB} 
 +\f14  N^{AB}  N_{AB}
 {+\f12 \bD_A \dot{\textsf E}_{\bU_A}}
 \right)\,.
\end{align}
Finally, by means of the commutators \eqref{comm}, we have $\lbr{\hat u},\xi_\W\rbr=\xi_{T=\W }$,
so that
\be
Q_{\lbr{\hat u},\xi_\W\rbr}=  \int_S \sqrt{\bq} \, \W\,\left( M -\f12 \bD_A \bU^A \right)\,.
\ee
Therefore, upon substituting Eqs.~\eqref{QuW}, \eqref{IWF}, and \eqref{LW}, the Weyl flux-balance law \eqref{FBW} yields exactly the condition \eqref{FBW1}.

On the other hand, to expand the flipped version of \eqref{FBW}, namely
\be
 \d_\W Q_{\hat u} -I_{\hat u}\cF_\W+ \int_S \iota_{\xi_{\W}} \iota_{\hat{u}} L=-Q_{\lbr\xi_\W, {\hat u}\rbr}\,,
 \la{FBW-2}
\ee
we first compute the variation under Weyl transformations of the $Q_{\hat u}$ and obtain
\begin{align}
\d_\W Q_{\hat u}&=\int_S\sqrt{\bq}\,\W\,\left(  M+\f14\bD_A \bD_B C^{AB}
 {-\f12 \bD_A \textsf E_{\bU_A}}
\right)\cr
&\quad+u \int_S\sqrt{\bq}\,\W\,\left(  \dot M- \bar \Delta \bF
  -\f14  \bD_A  \bD_B N^{AB} 
   {+\f12 \bD_A \dot{\textsf E}_{\bU_A}}
   \right)
\,,
\end{align}
and the field contraction of the Weyl Noetherian flux 
\begin{align}
I_{\hat u}\cF_\W &=\f{1}{4}\int_{S} \sqrt{\bq} \, \W\, C^{AB} N_{AB}
 +4 \int_{S}\sqrt{\bq} \,\W\, \dot{ \bar \beta}\,\cr
 &\quad-u  \int_{S} \sqrt{\bq}  \, \W\, \left ( \dot M    +\f12 \bD_A \dot \bU^A +\f14 N^{AB} N_{AB}\right)\,.
\end{align}
Therefore, the flipped Weyl flux-balance law \eqref{FBW2} yields a condition of the form $A'+uB'=0$, where

\be
A' :=
 - \frac{r}{2}\textsf E_{\bF}
- \bD_A\textsf E_{\bU_A}\,,
\ee
and $B'$ given in \eqref{FBE2}, consistently with the RHS of \eqref{FBW-off2}.

We thus conclude that the  flux-balance laws \eqref{FBW-off} for Weyl rescaling transformations yield two asymptotic Einstein's equations in Eq.~\eqref{asym-EE}, namely $A=-A'=0$ implies 
\be
\la{FBW1}
\boxed{
\bar{D}_A {\bU}^A= -\f 12\bD_A \bD_B C^{AB}\,.}
\ee
and
\be
\la{FBW2}
\boxed{
\bF=\f {\bR(\bq)}4\,.}
\ee

By plugging \eqref{FBW1}, \eqref{FBW2} into the condition $B=0$ (or equivalently $B'=0$), we recover
the evolution equation for the Bondi mass  \cite{Barnich:2011mi, Compere:2018ylh}
\be
 \boxed{
 \dot M=
 \f14\bD_A\bD_B  \dot C^{AB}
   -\f 18 N^{AB} N_{AB} 
+\f18\bar \Delta  \bar R
\,.
}\la{EEE}
\ee
These equations immediately satisfy the conditions  obtained in Eqs.~\eqref{FBE1}, \eqref{FBE2} from the energy balance law. Therefore,  the asymptotic Einstein's equations ${\textsf E}_{\bF}=0, \textsf E_{ M}=0, \textsf E_{{\bU}_A}=0 $ are indeed recovered from the energy and the Weyl flux-balance laws.

\subsubsection{Momentum flux-balance}

In the case of sphere diffeomorphisms, 
the momentum flux-balance law with $(\xi, \chi) = (\pa_u, \xi_Y)$ reads as\footnote{The flipped version is trivial; in fact, $\delta_Y Q_{\hat u}=0$, $\cF_Y=0$ and the other terms in the flux-balance give vanishing contribution.}
\begin{align}
 \d_{\hat u} Q_Y&-I_{Y}\cF_{\hat u}+ \int_S\iota_{\hat{u}}  \iota_{\xi_{Y}}  L+Q_{\lbr{\hat u},\xi_Y\rbr}=\cr
 &= \int_S \sqrt{\bq} \,Y^A \left[-r \dot{\textsf E}_{\bU_A}
+  \textsf E_{{\bP}_A}+2\p_A \dot{\textsf E}_{\bar\beta}  -2 \bar{F}\textsf E_{\bU_A}  -\f12 \bU_A \textsf E_{\bF}
  \right]. \la{FBY-on}
\end{align}
Be aware that the divergent contribution in $G_{A}{}^r$, namely $-r \dot{\textsf E}_{\bU_A}$ in the right-hand side of the above equation, is exactly cancelled out by the divergent contributions appearing in $\delta_{\hat u} Q_Y$ and $I_Y \cF_{\hat u}$ given below. Consistently, in the renormalization procedure carried out in Section \ref{sec:div-pot}, it is shown that the Einstein's equation ${\textsf E}_{\bU_A}=0$ needs to be imposed in order to obtain a finite expression for the momentum charge.

With these considerations in mind, 
we see from \eqref{FBY-on} that on-shell of the  asymptotic  Einstein's for $\bU^A$, $\bF$ and $M$ just derived, the momentum charge bracket 
\be
 \d_{\hat u} Q_Y-I_{Y}\cF_{\hat u}+ \int_S\iota_{\hat{u}}  \iota_{\xi_{Y}}  L=-Q_{\lbr{\hat u},\xi_Y\rbr}
\la{FBY}
\ee
is expected to be equivalent to the extra Einstein's equations
\begin{align}\la{EEY-on}
 \textsf E_{{\bP}_A}+2\p_A \dot{\textsf E}_{\bar\beta} =0\,,
\end{align}
as suggested by the RHS of \eqref{FBY-on}.

We verify this statement proceeding as above.
By means of Eq.~\eqref{QY}, we compute the time variation of the momentum charges
\be
 \d_{\hat u} Q_Y=
 - r \int_S \sqrt{\bq}\, Y^A\dot \bU_A + 
 \int_S\sqrt{\bq} \, Y^A\left( \dot \bP_A 
 + 2 \p_A \dot{ \bar\beta} \right) \,.
\ee
We then make use of the field variations under the vector $Y^A$ in \eqref{dY}
which in particular imply that 
\be
\int_S \delta_Y (  \sqrt{\bq} \bar{F})=0\,,\quad 
\int_S \delta_Y (  \sqrt{\bq} M)=0\,,\quad 
\int_S \delta_Y (  \sqrt{\bq} \bD_A \bU^A)=0\,, 
\ee
to compute the field contraction on $\cF_{\hat u}$
\begin{align}
I_{Y}\cF_{\hat u}
& = \f{r}{2} \int_S \sqrt{\bar{q}}\,  Y_A \bD_B N^{AB} 
 \cr
 &\quad
 + \int_S \sqrt{\bq} \, Y^A \p_A M
+ \int_S \sqrt{\bar{q}}\, Y^A C_{AB} \bD^B \bF \cr
&\quad + \frac14 \int_S\sqrt{\bar{q}}\, Y^C\left[ \bD_{B} ( N^{AB}  C_{CA} -C^{AB}  N_{CA})
- N^{AB}\bD_C C_{AB}\right]\cr
&\quad+\f12 \int_S \sqrt{\bar{q}}\, Y^A\left(  \bD_B \bD^{ B} \bU_{A}- \bD_B \bD_{ A} \bU^{B}
 \right)
 +o(1)\,.
\end{align}
 In the derivation of the expression above, we have used
 \be
 \bD_B\bD_A\bU^B= \bD_A\bD_B \bU^B  +2\bF \bU_A\,,
\ee
which follows from the commutator $[\bD_A, \bD_C]$ applied to the vector $\bU^A$,
and the relation \eqref{NC}.
Finally, by means of the commutators \eqref{comm}, we have $\lbr{\hat u},\xi_Y\rbr=0\,$, implying that $Q_{\lbr{\hat u},\xi_Y\rbr} =0$.

Therefore, the momentum charge bracket \eqref{FBY}, on-shell of the asymptotic equations \eqref{FBW1}, which removes the divergent contributions, and \eqref{FBW2}, yields 
\begin{empheq}[box=\fbox]{align}
\dot \bP_A 
 + 2 \p_A \dot{ \bar\beta} 
 &=
 \p_A M
+  C_{AB} \bD^B \bF \cr
&\quad + \frac14 \bD_{B} ( N^{AB}  C_{AC} -C^{AB}  N_{AC})
- \frac14 N^{BC}\bD_A C_{BC}\cr
&\quad+\f14 \left(\bD_B D_{ A} \bD_C C^{BC} -\bD_B D^{ B}\bD^C C_{AC}
 \right)
\,,\la{PEE}
\end{empheq}
  which can be rewritten in the form \eqref{EEY-on}
as expected. We thus recover the momentum evolution equation \eqref{EEP} (see also \cite{Barnich:2011mi, Compere:2018ylh}) in linear combination with the asymptotic Einstein's equation for $\bar \beta$ in \eqref{asym-EE2}. It is straightforward to verify that the reversed version of the momentum balance law \eqref{FBY}  trivially gives $0=0$. {The reader might also appreciate the fact that the combination $\dot \bP_A 
 + 2 \p_A \dot{ \bar\beta}$ is indeed the (time derivative of the) renormalized \bmw momentum aspect \eqref{QYR}. Its derivation is provided in the next section.}

This completes the derivation of   the asymptotic Einstein's equations at null infinity from the flux-balance laws defined by the bracket \eqref{Flux1} for an open Hamiltonian system.

\section{Charge and flux renormalization}\la{sec:div-pot}

We now want to investigate the renormalization procedure for the charges and for the 
symplectic flux outlined  in Section \ref{sec:Ren}. The limit to infinity is taken by first considering the spacetime boundary ${\cal B}=\Sigma_-\cup \Delta\cup \Sigma_+$, with $\Sigma_\pm, \Delta $ codimension-1 hypersurfaces respectively given by $u=u_\pm, r= r_\Delta $, with $u_+>u_-$ and then taking the limit\footnote{ The limit $u_\pm\rightarrow \pm\infty$ will be considered elsewhere.} $ r_\Delta\rightarrow +\infty$. 
The different components of the symplectic symplectic potential  
are given by
\be
\Theta_{\pm}
=\int_{\Sigma_{\pm}}\!\!\!\!\!\theta=
\int_{\va \Sigma_\pm}\!\!\!\!\! r^2\sqrt{q}\,e^{2\beta} \theta^u \rd r\,\rd^2\sigma,\qquad
\Theta_{\Delta}= -\int_{\Delta} \!\!\!\theta=r^2
\int_{\va \Delta}\! \sqrt{q}\,e^{2\beta} \theta^r \rd u\, \rd^2\sigma\,,
\ee
$\Theta_\pm$ represents the symplectic potential on different time slices, while $\Theta_\Delta$ represents the symplectic potential 
flux leaking through the boundary.
The fact that the Lagrangian variation vanishes on-shell implies the conservation equation
\be
\Theta_+ \heq \Theta_- + \Theta_{\Delta}.
\ee
We can use the results \eqref{fudiv}, \eqref{frdiv} to write the divergent part of $\Theta_{\cal B}$ as
\be
\theta_\div=\rd \vartheta_{\div}
-\f r 2 \d\left( \sqrt{\bq}\, {\textsf E}_{\bF}\right )\rd u \rd^2\s\,,
\ee 
where the corner symplectic potential is 
\be
\vartheta_{\div} =  \left(  \frac{r^2}{2}\delta(\sqrt{\bq})  -\f{r}4  \sqrt{\bq} C^{AB} \d \bar q_{AB}    \right)  \rd^2\sigma
+ r\vartheta^A \epsilon_{AB} \rd \sigma^B \wedge \rd u\,,
\ee
and we have introduced the vector valued variational form
\be\label{vartt}
 \vartheta_A := \frac12 \sqrt{q} \bD^B \left(\delta \bq_{AB}  - q_{AB} q^{CD}\delta q_{CD}\right)\,.
\ee
This form appears in the variation of the scalar curvature
\be\label{varR}
\frac12 \delta \left(\sqrt{\bq}\bR(\bq)\right) 
= \p_A \vartheta^A.
\ee

Therefore, by means of the Einstein's equation ${\textsf E}_{\bF}=0$,\footnote{Imposition of this equation is not necessary for the renormalization of the potential, but for the energy and Weyl charges and fluxes, as otherwise, according to \eqref{ren1}, an extra divergent term would appear in the renormalized expressions \eqref{Qren}, \eqref{Fren}.} we can  define the  renormalized symplectic potential in the form  \eqref{ren1} with $\elld=0$ on-shell of the boundary condition $\p_u\bq_{AB}=0$, namely
\be
\theta^R =\theta - \rd \vartheta_{\div}\,.
\ee
This yields the renormalized symplectic  2-form \eqref{Omega}.
According to \eqref{Qren} and \eqref{Fren}, 
the renormalized charge and  flux are given by
\begin{align}
Q^R_\xi &= Q_\xi - \int_S I_\xi \vartheta_{\div}\,,\la{Qren2} \\
\cF_\xi^{R}& =\cF_\xi - \int_S \d_\xi \varthetad\,.\la{Fren2}
\end{align}

\subsection{Renormalized charges}

We can  compute the  renormalized charge aspects given by \eqref{Qren2}.
In the case of the energy aspect, we use \eqref{dT} and \eqref{QT} to compute
\be
Q_T^R=
Q_T
 - \int_S  I_{\xi_T}  \vartheta_{\mathrm{div}} 
=
\int_{S}  \sqrt{\bar{q}}  T  \left (M - \f12 \bD_A \bU^A \right)\,.\la{QRT}
\ee

In the case of the Weyl aspect, we use \eqref{dW} and \eqref{QW} to compute
\be
Q^R_\W=Q_\W - \int_S  I_{\xi_\W}  \vartheta_{\mathrm{div}} 
=
\int_S \sqrt{\bq} \,\W\, \left[4 \bar \beta
+ u   \left(M
-\frac 1{2} \bD_A \bar{U}^A\right)
\right]
 \,.\la{QRW}
\ee
Notice that the internal Lorentz contribution to the charge \eqref{LCharge}, namely $\int_\Sigma I_\lambda \theta $, does not contribute to the charge renormalization  term as $\int_SI_\lambda \varthetad=0$.
It is also interesting to appreciate that the divergence of the Weyl aspect \eqref{QW} is entirely due to the Lorentz charge contribution \eqref{lambdaW}. 
The renormalization term $ \int_S  I_{\xi_W}  \vartheta_{\mathrm{div}}$  for the Weyl aspect cancels exactly this internal Lorentz  divergence.  This highlights again the importance of taking into account also the  internal $\SL(2,\C)$ contribution to the charges to get  finite expressions.

In the case of the momentum aspect, we use \eqref{dY}, \eqref{QY}  and the Einstein's equation ${\textsf E}_{\bU_A}=0$ to compute
\be
Q_Y^R =
Q_Y
 - \int_S  I_{\xi_Y}  \vartheta_{\mathrm{div}} 
=  \int_S \sqrt{\bq}~ Y^A\left(
 \bar{P}_A  + 2 \pa_A \bar\beta   
 \right)\,.\la{QRY}
\ee
We thus see how the renormalization procedure removes exactly the radially divergent terms in the Noether charge, but leaves the finite terms unmodified.
The renormalized BMS charges  
$Q^{R-\gbms}_{(T, Y)}= Q^{R}_{(T, W=\f12 \bD_A Y^A, Y)}$ are given by
\be\la{Q-0}
Q^{R-\gbms}_{(T, Y)}=
\int_S\sqrt{\bq} \left[
\tau \left(M
-\frac 1{2} \bD_A \bar{U}^A\right)
+Y^A  \bar{P}_A     
 \right]\,,
\ee
where $\tau=T+\f u 2 \bD_A Y^A$.

\subsection{Relation to Barnich--Troessaert charges}\la{sec:shift}

One puzzle we face is  that the Noetherian  charges that we constructed from the covariant Lagrangian are not the same as the one considered by  Barnich--Troessaert \cite{Barnich:2011mi} or Flanagan--Nichols \cite{Flanagan:2015pxa} and studied further by Compere et al. \cite{Compere:2018ylh, Compere:2020lrt}.
Given the emphasis we have put on the Noetherian split's relevance, one has to wonder whether these previous charges are obtainable from a choice of boundary Lagrangian. We now show that this is indeed the case: the BT  charges are recovered from our covariant Noether charges after adding a non-covariant boundary Lagrangian.

To that end, let us consider a family of boundary Lagrangians, parametrized by one parameter $\alpha\in \R$,
\be
\ell_\alpha := \sqrt{\bq}
\left(M+\f \alpha 8 C^{AB} N_{AB}\right)
\rd u\rd^2\sigma\,.
\la{ella}
\ee
This gives
the corner symplectic potential 
\be
\vartheta_\alpha= \f \alpha 8 
\sqrt{\bq}\,  \left(C^{AB}\d C_{AB} \right) \rd^2\sigma \,.
\ee

Applying the general formula \eqref{trans1}
and thanks to the identity \eqref{CC},
the shift in the charge $Q_{(\xi;\alpha)}^{R}-Q^R_\xi$ due to the presence of the boundary Lagrangian is thus  given by 
\be
 \int_S
(i_\xi \ell_\alpha-I_\xi\vartheta_\alpha) = 
\int_S\sqrt{\bq}\left[\t  M + \f \alpha 8\left( \dot{\tau}- \frac12 \bD_CY^C\right) 
 C^{AB} C_{AB} +\f{\a}4 C^{AB}\bD_A\p_B\t 
 \right]\,.
\ee
Applying this to the super translation charge and integrating by part  gives
\be
Q_{(T;\alpha)}^{R} =\int_S\sqrt{\bq}\, T \left(2 M-\f12(1+\alpha) \bD_A \bU^A\right)\,.
\ee
For the Weyl charge one obtains
\be
Q_{(W;\alpha)}^{R}=\int_S \sqrt{\bq} \,\W\, \left[4(1-\alpha) \bar \beta
+ u  \left(2 M-\f12(1+\alpha) \bD_A \bU^A\right)
\right]\,,
\ee
where we have used the asymptotic EEs for $\bU^A$ and $\bar \beta$.
A similar calculation shows that the shifted expression of the renormalized sphere diffeomorphism charge is
\begin{align}
Q_{(Y;\alpha)}^{R}
=  \int_S \sqrt{\bq}~ Y^A\left(
 \bar{P}_A  + 2(1- \alpha) \pa_A \bar\beta   
 \right)\,.
\end{align}
The full Noether charge can then be written as
\begin{align}
Q^{R}_{(T, W, Y;\alpha)}=\int_S\sqrt{\bq} \Bigg[&
(T+uW) \left(2 M-\f12(1+\alpha) \bD_A \bU^A\right)+
\, 2(1-\alpha)(2\W -\bD_AY^A) \bar \beta +Y^A
 \bar{P}_A   
\Bigg]\,.
\end{align}
We can thus see that the charge considered by Barnich--Troessart \cite{Barnich:2011mi} or Flanagan--Nichols \cite{Flanagan:2015pxa}  
is recovered for $\alpha=-1$ once we set $W=\f12 \bD_A Y^A$, namely
\be\la{Q-1}
{Q}^{\textsf{BT}}_{(T, Y)}= Q^{ R}_{(T, W=\f12 \bD_A Y^A, Y;\, -1)}=
\int_S\sqrt{\bq} \left[
2\tau M
+Y^A  \bar{P}_A     
 \right]\,,
\ee
where $\tau=T+\f u 2 \bD_A Y^A$.

As pointed out in Section \ref{sec:FR-rel} (and derived in \cite{Freidel:2021cbc}), the non-covariance of the boundary Lagrangian \eqref{ella} introduces an extra contribution, $K_{(\xi,\chi)}$, in the bracket \eqref{bran} in terms of the boundary Lagrangian anomaly. This extra contribution is given in \eqref{K} and it can be computed
using the anomalous transformation terms (that can be read off from \eqref{del})
\begin{align}
\Delta_{\tau} C_{AB}
&=- {2\bD_{\langle A} \p_{B\rangle} \tau }\,,
\la{delC1}
\\
\Delta_{\tau} N_{AB}
&=- {2\bD_{\langle A} \pa_{B\rangle} \dot\tau }\,,
\label{delN1}
\\
\Delta_{\tau} M&=
 \left( \f12 \bD_A N^{AB} +\pa^B \bF \right)\pa_B \tau 
  +  
\f14 N^{AB} \bD_A\pa_B\tau 
+\f14  C^{AB} \bD_A \p_B \dot\tau\,.\label{delM1}
\end{align}

This yields
\bea
\Delta_{\tau} \ell_{-1}&=&  \sqrt{\bq}
\left(\Delta_{\tau}  M-\f 1 8( \Delta_{\tau}  C^{AB} N_{AB}+C^{AB} \Delta_{\tau} N_{AB} )\right)
\rd u\wedge \rd^2\sigma
\cr
&=&\sqrt{\bq} \left[
\pa^A \bF \pa_A \tau 
   +  
\f12 \bD_A (N^{AB} \pa_B\tau )
+\f12  C^{AB} \bD_A \p_B \dot\tau
\right]
\rd u\wedge \rd^2\sigma\,,
\eea
from which (we use the notation $\iota_\tau:= \iota_{\xi_\tau}$)
\bea
 K_{(\tau_1,\tau_2)}&=& \int_S\left(   \iota_{{\tau_1}}  \Delta_{{\tau_2}}\ell_{-1} -\iota_{{\tau_2}}  \Delta_{{\tau_1}}\ell_{-1}\right)\cr
&=&\f14 \int_S \sqrt{\bq} \left[
\tau_1 \pa^A \bR \pa_A \tau_2
+  \tau_1C^{AB} \bD_A \p_B \bD_C Y^C_2-1\leftrightarrow 2
\right]\,.
 \eea
This expression for the cocycle, first proposed in \cite{Chandrasekaran:2020wwn}, matches exactly the one for the  2-cocycle in the   Barnich--Troessaert  bracket for the charges \eqref{Q-1} derived in 
\cite{Compere:2020lrt}.

This establishes cleanly that the BT charge is obtained from the addition of the Lagrangian $\ell_{-1}$.
It is important to note however that $\ell_{-1}$ is not integrable on $\scri$ and therefore this polarization does not seem to be accessible if we want  agreement between Hamiltonian and Lagrangian formulations.
The non-integrability of $\ell_{-1}$ comes from the fact that when $u\to \infty$ we have that $\cM\to 0$, which means that 
\be 
\ell_{\alpha} \to\f18(\alpha-1) N^{AB}_{\textsf{vac}} C_{AB}^{\textsf{vac}}.
\ee
The only value for which this converges to $0$ is $\alpha=1$.
This means that the BT split of charge and flux cannot accommodate the presence of a non-trivial asymptotic Weyl frame.

\subsection{Renormalized Noetherian fluxes}

We can finally compute the  renormalized Noetherian fluxes given by \eqref{Fren2}.
In the case of the energy Noetherian flux, by means of the \eqref{xiT}, we have
\begin{align}
 \cF_T ^R&=  \cF_T- \int_S \d_T \varthetad \cr
&=  \cF_T
+ \f{r}4 \int_S \sqrt{\bq}  TN^{AB} \d \bar q_{AB} 
+ \f{r}2 \int_S \sqrt{\bq} T  \bD_A\bD_B \d \bar q^{AB} 
+  \f{r}2 \int_S   \bar \Delta T   \d\sqrt{\bq}\cr
&=  \cF_T
+ \f{r}4 \int_S \sqrt{\bq}  TN^{AB} \d \bar q_{AB} 
- 2r \int_S  T  \d(\sqrt{\bq} \bF)
-  \f{r}2 \int_S   \bar \Delta  T  \d\sqrt{\bq}
 \,,\la{FT-ren}
\end{align}
where we have used \eqref{FBW2}, \eqref{dq}, \eqref{CCq}, and the relation
\be
\f12 \sqrt{\bq} \bD_A\bD_B\d \bq^{AB}=-\f12 \d(\sqrt{\bq}\bR(\bq))-\bar\Delta \d\sqrt{\bq}\,,
\ee
which follows from \eqref{varR}.

In the case of the Weyl symplectic flux, by means of the \eqref{xiW}, similar manipulations for the energy symplectic flux show that we have 
\begin{align}
\cF_\W ^R&=  \cF_\W- \int_S \d_\W \varthetad \cr
    &=\cF_\W
     -  \f{r}4 \int_S  C_{AB} \d \bar q^{AB}\d_\W \sqrt{\bq}
    -  \f{r}4 \int_S \sqrt{\bq}  \d_\W C_{AB} \d \bar q^{AB}
    -  \f{r}4 \int_S \sqrt{\bq}  C_{AB} \d \d_\W \bq^{AB} -  \f{r^2}2 \int_S \d( \d_\W \sqrt{\bq})\cr
     &=\cF_\W
     -\f{r}4 \int_S \sqrt{\bq}  \W\, C^{AB} \d \bar q_{AB}
       +r^2 \int_S     \W\, \d \sqrt{\bq} \cr
   &  \quad+ \f{ru}4 \int_S \sqrt{\bq}    \W\, N^{AB} \d \bar q_{AB} 
-2ru \int_S    \W\,  \d(\sqrt{\bq} \bF)
-  \f{ru}2 \int_S   \bar \Delta    \W\,  \d\sqrt{\bq} \,.\la{FW-ren}
\end{align}

In the case of the momentum symplectic flux, by means of the \eqref{xiY}, it is straightforward to see that 
\be
\cF_Y ^R=  \cF_Y- \int_S \d_Y \varthetad \,,
 = 0\,
\ee
where we use the fact that $ \cF_Y=0$ and that $\d_Y $ has no anomaly and therefore it can be replaced by $\cL_Y$.
    
We thus see that the renormalization procedure for the charges $Q_\xi$ and  fluxes $\cF_\xi$ computed above 
removes the divergent terms in the Noetherian expressions obtained from \eqref{ECcharge} and \eqref{ECflux} and does not pick up any new finite term (see the expressions \eqref{FTdiv},  \eqref{FWdiv} for the divergent parts of the fluxes).  
Similarly for  $Q^R_{\hat u}=Q_{\hat u}$ and
\be
 \cF_{\hat u}^R=  \cF_{\hat u}-\int_{S}\d_{\hat u} \vartheta_{\mathrm{div}}
  = \cF_{\hat u}
  + \f r4 \int_{S}  \sqrt{\bar{q}}\, N^{AB}\delta \bar{q}_{AB}\,.
\la{Fchi-ren}
 \ee

It follows immediately that the derivation of the asymptotic Einstein's equations  presented in  Section \ref{sec:FB2} goes through in exactly the same way if working at null infinity to begin with and using the renormalized Noetherian quantities for the bracket \eqref{BTb}. This is no surprise of course, as it represents  an explicit check of the invariance property of the bracket \eqref{Flux1} under the shift \eqref{trans1}.
Finally we can also evaluate the shifted renormalized flux associated with the boundary Lagrangian \eqref{ella}.This can be easily done following the procedure outlined so far.

\section{Conclusions}\la{sec:Conc}

In this work, we have achieved five  interconnected results:
\begin{enumerate}
\item We have shown that, under the boundary conditions proposed by \cite{Campiglia:2014yka, Compere:2018ylh}, the asymptotic symmetry group of flat space null infinity is given by the BMSW group, which contains super-translations, $\mathrm{Diff(S)}$ transformations, and Weyl super-boosts.
We also have shown that the BMS group is obtained as field-dependent reduction of BMSW.
\item Following \cite{Freidel:2021cbc}, we have given a generalization of the   Barnich--Troessaert bracket and shown that demanding that this bracket provides a representation of the symmetry algebra implies 5 asymptotic Einstein's equations: ${\textsf E}_{\bF}$ and $D_A {\textsf E}_{\bU^A}$  as well as the 
energy and momenta conservation equation ${\textsf E}_M$ and  ${\textsf E}_{\bP_A}+2\dot{{\textsf E}}_{\bar\beta}$. 
These results are summarized in
table \ref{tab:table1}.
\item We have provided the holographic renormalization of the symplectic potential and the BMSW charges.
We have shown that the renormalized symplectic potential is finite provided the asymptotic equation ${\textsf E}_{\bar \beta}$  is satisfied. The renormalized charges are also finite provided ${\textsf E}_{\bar{F}}, {\textsf E}_{\bU^A}$ are satisfied. This result extends the analysis performed by Compere et al. in \cite{Compere:2018ylh}
by including Weyl transformations and the 2d metric determinant as a phase space variable.
\item We have shown that we can obtain the Barnich--Troessaert \cite{Barnich:2011mi} BMS charges from a Noetherian split associated with a non-covariant Lagrangian defined in \eqref{ella}.
\item Finally, we have shown that the vacuum structure of asymptotically flat gravity is labeled by a BMSW group element.
This contains a super-translation and a Weyl label similar to the one revealed by Comp\`ere and Long \cite{Compere:2016jwb}. In addition it contains a Diff$(S)$ label.
\end{enumerate}
  
These results emphasize the importance of choosing a Noetherian split associated with a  choice of Lagrangian. They also generalize to asymptotic infinity the results of \cite{Freidel:2021cbc} obtained at a finite distance for the extended corner symmetry group.
In particular, the remarkable connection, through the flux laws \eqref{TheBra}, between Einstein's equations and the canonical representation of the symmetry algebra (be it finite or asymptotic)   could be understood as the core element of (local) holography. This suggests a deep connection between finite and asymptotic symmetry algebra (see also \cite{Wieland:2020gno} for such a connection). 

One could extend our results in several ways:
first, one could try to relax the condition $\pa_u \bq_{AB}=0$ and authorize a time-dependent background metric. Essential elements of this generalizations have been worked out in \cite{Barnich:2011mi}, but in a restricted context that does not allow the full $\mathrm{Diff(S)}$ symmetry.

Also, our analysis reveals that the bigger the symmetry group, the more Einstein's equations can be recast as identities of canonical brackets or flux-balance laws.
It thus suggests that the maximal symmetry group should be defined as the one that recovers all Einstein's equations, not a subset of them. 
Our study is still missing 
the two purely space-like Einstein's equations\footnote{Recall that we obtained five EEs from the charge bracket and two more from the renormalization procedure, while $G^A{}_A=0$ follows trivially from these.} $G_{\langle AB\rangle}=0$. 
This means that we expect the full symmetry group to be bigger than \bmw. We conjecture that the full extension includes dual super-translations associated to the dual energy charges \cite{Godazgar:2018qpq}. 
One way to reveal these symmetry transformations  is to introduce the Immirzi parameter by considering the  Einstein--Cartan--Holst action  and repeating the analysis performed here for the dual component of its symplectic potential \eqref{ECH-pot}. This will be addressed in a forthcoming work.

One issue that needs to be understood better is the nature of the limit $u\to \pm \infty$ of the Noether charges. In this limit one pushes the sphere $S$ to spacelike and timelike infinities, which are fixed by the symmetry transformations and therefore we do not expect the presence of non-trivial fluxes. The charges that survive this limit are the covariant mass and the covariant momentum which are related to the Weyl tensor. This corresponds to the proposal of \cite{Compere:2020lrt}.
It is not clear however if these covariant charges can be obtained as limits of the Noetherian ones.

At the same time, it would be interesting to understand the connection between the BMSW symmetry algebra and the soft theorems more deeply. Especially, what is the role of the newly revealed Weyl charge in this context, and whether the new Diff$(S)$ vacua label could be related to  a new memory effect associated with the  spin memory effect as a vacuum transition \cite{Himwich:2019qmj, Compere:2019odm}.

And finally, now that we have a \emph{bona fide} $\mathrm{Diff(S)}$ symmetry group and not just a super-Lorentz symmetry acting on the gravitational phase space, one can wonder whether this means that there exists a fluid description of the asymptotic conservation equations that extends what is found at the corner \cite{Donnelly:2020xgu}.

\section*{Acknowledgement}

We would like to thank Glenn Barnich, Geoffrey Comp\`ere, Adrien Fiorucci, Romain Ruzziconi, Anthony Speranza for helpful discussions and insights. 
R.~O.~thanks Geoffrey Comp\`ere, Ali Seraj, Bernard Whiting for collaborations on related topics over the past years, and is grateful to Constantinos Skordis for his support and advice.
We thank A. M. for lifting our spirits.
R.~O.~is  funded  by  the  European  Structural  and  Investment  Funds (ESIF) and the Czech Ministry of Education, Youth and Sports (MSMT), Project CoGraDS-CZ.02.1.01/0.0/ 0.0/15003/0000437.
Research at Perimeter Institute is supported in part by the Government of Canada through the Department of Innovation, Science and Economic Development Canada and by the Province of Ontario through the Ministry of Colleges and Universities. This project has received funding from the European Union's Horizon 2020 research and innovation programme under the Marie Sklodowska-Curie grant agreement No. 841923. 
%

\appendix

\section{Asymptotic expansions}\la{AppExp}

In this Appendix we collect all asymptotic expansions used to derive the results in the main text.

\vspace{0.2cm}
{\bf Metric coefficients}
\begin{align}
q^{AB} & = \bar{q}^{AB} -\frac{1}{r}C^{AB} + \frac{1}{4r^2} \bar{q}^{AB} C_{CD}C^{CD} + \mathcal{O}(r^{-2})\,,\\
D_A V^A&=\frac1{\sqrt{q}}\pa_A ( \sqrt{q} V^A) = 
\frac1{\sqrt{\bar{q}}}\pa_A ( \sqrt{\bar{q}} V^A)= \bar{D}_A V^A,\\
I^{AB}&= \int_r^\infty  \f{\rd r'}{r^{'2}} e^{2\beta}  q^{AB}
=\f1r \bq^{AB} - \f1{2r^2} C^{AB} +
\f1{16 r^2} {\bq^{AB}} C^{CD}C_{CD}
+ o(r^{-3}).
\end{align}

\vspace{0.2cm}
{\bf Tetrad coefficients}
\begin{align}
E^i_A
&=\bar{E}^i_A+  \frac{1}{2r} C_{A}{}^{B}\bar{E}^{i}_B
+  \f1{16 r^2} \bar{E}^{i}_{A}  C_{BC}C^{BC} + o(r^{-2})\,,\\
\hat E^A_i&=\bar{E}^A_i - \frac{1}{2r} C^{A}{}_{B}\bar{E}^{B}_i 
+  \f1{16 r^2} \bar{E}_i^{A}  C_{BC}C^{BC}+o(r^{-2})\,. 
\end{align}

\vspace{0.2cm}
{\bf Spin connection}
\begin{align}
\omega_{10} 
&= \left[\frac{M}{r^2} +\mathcal{O}(r^{-2})\right]\rd u +\left[- \frac{4\bar{\beta}}{r^3} + o(r^{-3})\right]\rd r + \left[-\frac{\bU_A}{r} + \frac{1}{r^2}\left(\bP_A +2\bD_A \bar{\beta} \right) \right] \rd x^A\,.\label{omega10exp} 
\end{align}

\vspace{0.2cm}
{\bf Gauge parameters } 
\begin{align} 
\lambda_T^{01}&=\frac{1}{2r^2} \left[\f12 \bD_A \left(C^{AB} \pa_B T\right)-\bU^A \pa_AT \right]+o(r^{-2})\,, \la{lambdaT}\\
 \lambda^{01}_\W&= -\W+ \f u {2r^2}\left[\f12 \bD_A \left(C^{AB} \pa_B \W \right)-\bU^A \pa_A \W\right] +o(r^{-2})\,,\la{lambdaW}\\
   \lambda^{01}_Y&= 0 \,.\la{lambdaY}
 \end{align}

\section{Derivation the symplectic  flux}\la{AppC}

In this appendix, we compute the symplectic flux 
${\cF}_\xi^\theta :=\int_S \iota_\xi \theta$.
We first expand the contraction of the EC symplectic potential $\eqref{ECH-pot}$ with a generic vector field $\xi$.
We use the convention that $\epsilon_{01ij}=\epsilon_{ij}$ and the notation $\xi^a e_a^I  = \xi^I$. We have
\bea
{\cF}^\theta_\xi&=&
\f{1}{4}\epsilon_{IJKL}
\int_{S} 
 \left(
 2i_\xi e^K e^L\wedge \delta \omega^{IJ}
 +e^K\wedge e^L  \iota_\xi\delta \omega^{IJ}
 \right)\cr
 &=&\frac12 
 \int_{S} 
 \left[
   r \epsilon_{IJK \ell} \xi^K E^\ell_A (\delta \omega_{B}{}^{IJ}) 
 + r^2 \epsilon_{ij} E^i_A E^j_B ( \iota_\xi\delta \omega^{01} )
 \right]\epsilon^{AB} \rd^2\s
\cr
&=&
 r  \int_{S} 
   \epsilon_{ij} E^j_A \epsilon^{AB} \left[
       \xi^i  (\delta \omega_{B}{}^{01}) 
  -    \xi^1  (\delta \omega_{B}{}^{0i })
  +  \xi^0 (\delta \omega_{B}{}^{1i})
   \right] \rd^2\s
 + r^2 \int_{S_\infty}  \sqrt{q}  ( \iota_\xi\delta \omega^{01} )
\rd^2\s \cr 
&=&  r \int_{S}  \sqrt{q}
 \left[  E^{iA} \Big( \xi^0 \delta \omega_{A 0 i} - \xi^1   \delta \omega_{A 1 i }
  - \xi_i \delta \omega_{A 10} \Big)
 + r   \iota_\xi\delta \omega_{10} 
 \right] \rd^2\s\,.\la{Fluxmain}  
 \eea 
In the last equality we have used  that $\epsilon_{CA} \epsilon^{AB} = -\sqrt{q} \delta_C^B$ and $\epsilon_{ij} \epsilon^{AB} E^i_A = \sqrt{q} E_j^B$.

Next, we use the identities
 \be
 \xi^i E_i^A = r(\xi^A-  U^A \xi^u),\qquad
 \xi^0 = e^{2\beta} \xi^u,
 \qquad
 \xi^1= \xi^r+F\xi^u,
 \ee
to evaluate the symplectic flux as
 \begin{align}
 \cF^\theta_\xi &=
 r \int_{S}  \sqrt{q}
    \xi^u  
 \left[ E_i^A \left( e^{2\beta}\delta \omega_{A 0 }{}^i- F \delta \omega_{A 1  }{}^i   \right)
 + r U^A \delta \omega_{A 10} \right]\cr
 &\quad - r \int_{S}  \sqrt{q}  \xi^r  \left(E_i^A \delta \omega_{A 1 }{}^i \right)
 + r^2\int_S  \sqrt{q} \left( \iota_\xi \delta \omega_{10} -  \xi^A \delta \omega_{A 10} \right)\,. \la{Flux2a} 
 \end{align}
This   can  be rewritten as in Eq.~\eqref{Fsympl}, namely
 \begin{align}\la{Flux2}
 \cF^\theta_\xi &=
 \int_{S}  \sqrt{q}(\xi^r\theta^u-\xi^u \theta^r )\cr
&= r \int_{S}  \sqrt{q}
    \xi^u  
 \left[ E_i^A \left( e^{2\beta}\delta \omega_{A 0 }{}^i- F \delta \omega_{A 1  }{}^i   \right) + r \left( \delta\omega_{u10}
 + U^A \delta \omega_{A 10}\right) \right]\cr
 &\quad -  r \int_{S}  \sqrt{q}  \xi^r  
  \left(E_i^A \delta \omega_{A 1 }{}^i -r  \delta\omega_{r10}\right)\,.
 \end{align}

\subsection{Asymptotic expansion of the temporal flux}

We want to expand the temporal flux
\be \label{Fluxu}
-\theta^r =
 r  
 \Big[ E_i^A \left( e^{2\beta}\delta \omega_{A 0 i}- F \delta \omega_{A 1 i }   \right)
 + r (\delta \omega_{u10}+U^A \delta \omega_{A 10}) \Big]\,. 
\ee

 First, we focus on the last two terms
 \bea
   r^2(\delta \omega_{u 10} + U^A \delta \omega_{A 10} )
 &=&  r^2\delta (\omega_{u 10} + U^A  \omega_{A 10}) - r^2\delta U^A \omega_{A 10}\cr
&=&  \delta \left(r^2F' + 2r^2\beta' F \right) 
- r\delta U^A P_A\cr
&=&  \delta M 
 +  o(1)\,,\label{Fluxu1}
 \eea
 where we have used Eq.~\eqref{omega10exp} and the asymptotic expansions \eqref{eq:FallOff}.
 
Now, we expand two terms of \eqref{Fluxu}. They give
\bea  
 e^{2\beta}  \delta \omega_{B0i}-F  \delta \omega_{B1i}
 &=& e^{2\beta}  \delta[ e^{-2\beta } E_i^A \left({F} S_{AB} -K_{AB} \right)]
+  F \delta (E_i^A  S_{AB}) \cr
 &=& \delta F E_i^A S_{AB}  
 - \delta (E_i^A  K_{AB})
 + 2 F \delta (E_i^A  S_{AB})
 - 2\delta \beta E_i^A \left({F} S_{AB} -K_{AB} \right)\,. \nn
 \eea
 We can contract this with $E^{Bi}$,  using that $
  \delta E_i^{(A} E^{B)i} = \frac12 \delta q^{AB}  $ 
 and denoting $S= q^{AB} S_{AB}$ and $K= q^{AB} K_{AB}$.
 In particular, we use the following relation
 \be 
 E^{Bi}\delta [E_i^A  K_{AB}] 
 = \frac12 K_{AB} \delta q^{AB} 
 + q^{AB} \delta K_{AB}
 = \delta K - \frac12 K_{AB} \delta q^{AB}
 \ee
 to get
 \begin{align}
   r(e^{2\beta}  \delta \omega_{B0i}-F  \delta \omega_{B1i})E^{Bi}
  &=
 r \left[S \delta F    
 + \frac12 K_{AB} \delta q^{AB} 
 -  \delta K  
 + 2 F \delta S
 -  F  S_{AB} \delta q^{AB} \right]\cr 
 &\quad - 2r \delta \beta  \left({F} S  -K \right)\,.
 \label{Fluxu2}
  \end{align}
 
Therefore, we have
 \bea
-\theta^r  &=& r\Big[ E_i^A \left( e^{2\beta}\delta \omega_{A 0 }{}^i- F \delta \omega_{A 1 }{}^i   \right)
 + r (\delta \omega_{u10}+U^A \delta \omega_{A 10}) \Big] \cr
 &=&  \left[ \delta M +o(1)\right]
 + r \left[ \delta (SF-K)    
 + \frac12 K_{AB} \delta q^{AB} 
 +  F \delta S
 -  F  S_{AB} \delta q^{AB} \right]\,.
 \la{exp}
 \eea
 
To perform the asymptotic expansion, we need the following expressions 
 \begin{subequations}
 \begin{align}
 q^{AB}&= \bar{q}^{AB}-\f1r \bC^{AB}+o(r^{-1})\,,\\
 rK_{AB} &=  \left( \frac12 D_C\bU^C + \frac14 N_{CD} C^{CD}\right) \bar{q}_{AB} + \frac{r}2 N_{\langle AB \rangle} + D_{\langle A} \bU_{B\rangle}+o(1)\,, \\
 r K&=   D_C\bU^C +o(1)\,, \\
S_{AB} &=  \bar{q}_{AB} + \frac{\bC_{AB}}{2 r}  +o(r^{-1})\,, \\
S&= 2 + o(r^{-1})\,,
 \end{align}
 \end{subequations}
 where $\langle AB \rangle$ denote the symmetric traceless component.
By adding \eqref{Fluxu1} and \eqref{Fluxu2}, we get the asymptotic expression of the temporal flux   
 \bea
-\theta^r 
  &=&  
r\left( 2 \delta \bar{F} -\bar{F}\bar{q}_{AB} \delta \bq^{AB}
+ \frac{1}4 N_{AB} \delta \bq^{AB} 
  \right)-  \frac14 N_{AB} \delta C^{AB}
 + \frac12 \left( \bar{D}_{\langle A} \bU_{B\rangle} + \bF  C_{AB} \right) \delta \bq^{AB} \cr
& & - \left( \frac12  D_C\bU^C + \frac14 N_{CD} C^{CD}+2M \right)\delta \ln\sqrt{q} -  \delta ( M    +   D_C\bU^C)
  +o(1)\,, 
\label{Fluxu3}
 \eea
 where we used $\bar F\bar q_{AB}  \delta \bar C^{AB}= 
  \bar F \delta \bar q^{AB}  \bar C_{AB}$ and 
 $ \delta \ln\sqrt{q} = -\frac12 \bar{q}_{AB}\delta \bq^{AB} $.
  
To further simplify the above expression, we use the following variations
\begin{subequations}\la{dprop}
\begin{align}
\d \bq^{AB}&=-\bq^{AC} \bq^{BD} \d \bq_{CD}\,,\la{dq}\\
\d C^{AB}&=\d( \bq^{AC} \bq^{BD} C_{CD})=
 C_{C}{}^B\d \bq^{AC}+ C^A{}_{C} \d \bq^{BC} +\bq^{AC} \bq^{BD} \d C_{CD}\,,\la{CCq}
\end{align}
\end{subequations}
and the properties that a symmetric and traceless $2\times 2$ matrix $C_{AB}$ obey
  \begin{align}
C_{AC} C^C{}_B&=\f12 \bq_{AB} C_{CD} C^{CD}\,,\la{CC}\\
N_{AC} N^C{}_B&=\f12 \bq_{AB} N_{CD} N^{CD}\,,\\
N_{AC} C^C{}_B&=-C_{AC} N^C{}_B+\bq_{AB} N_{CD} C^{CD}\,,\la{NC}\\
\dot N_{AC} C^C{}_B&=-C_{AC} \dot N^C{}_B+\bq_{AB} \dot N_{CD} C^{CD}\,.
\end{align}
Note, in particular, that Eq.~\eqref{NC} implies
 \be
  N^{\langle A}{}_C C^{B \rangle C} \delta \bq_{AB}=0\,.
 \ee

This gives us the final expression for the time flux
\begin{align}
-\sqrt{\bq}\theta^r &= 2 \delta \left( r \sqrt{\bar{q}} {\bF}\right)  - 
 \f{1}{4}\p_u \left( r\sqrt{\bar{q}} {C}^{AB} \delta \bar{q}_{AB}\right) \cr
&\quad - \delta \left[\sqrt{\bq} ( M+ D_C\bU^C)\right] 
 -
   \frac14 \sqrt{\bar{q}} N^{AB} \delta C_{AB}
 - \frac12 \sqrt{\bar{q}} \left( \bF  C^{AB}   +D^{\langle A} \bU^{B\rangle}  \right) \delta \bq_{AB} \cr
 &\quad  - \left(M -{ \frac12}  D_C\bU^C - \frac14 N_{CD} C^{CD}\right)  \delta\sqrt{\bq}
  +o(1)\,,
 \la{thetarB}
\end{align}
 where we have used the asymptotic condition that $ \pa_u{\bar{q}}_{ AB }=0$ 
 and that $N_{AB}=\pa_u C_{AB}$ to rewrite the divergent contributions in the first line as a total time derivative.

\subsection{Asymptotic expansion of the radial flux}

Likewise it was done for the temporal flux, we can show that the radial flux \eqref{tu} is given by 
\begin{align}
\sqrt{\bq}\theta^u&=  -  r \sqrt{\bq} \left(E^{Ai} \delta (\omega_{1 i })_A - r \delta (\omega_{10})_r\right)\cr
&=  r  \sqrt{\bq}E^{Ai}\d \left( E^B_i S_{BA}\right) +2r^2 \sqrt{\bq} \d \beta'  \cr
&=r\sqrt{\bq}\left(\delta S - \frac12 S_{AB} \delta q^{AB}\right) -\f 4r\sqrt{\bq}  \d \bar \beta \cr
&= \frac{r}{2}\sqrt{\bq} \bar{q}^{AB} \delta q_{AB} 
+\f 12\sqrt{\bq}\bq_{AB} \delta  C^{AB}
 -\f14\sqrt{\bq} C_{AB} \d \bar q^{AB}\cr
&\quad-\f{1}{2r}\sqrt{\bq} \bq_{AB} \delta (  C^A{}_C  C^{CB}-\f14\bar{q}^{AB}  C_{CD}  C^{CD})\cr
&\quad+\f{1}{4r}\sqrt{\bq} C_{AB} \delta  C^{AB}
 -\f 4r \sqrt{\bq} \d \bar \beta +o(r^{-1})\cr
  &=r \delta {\sqrt{\bq}}
-\f14\sqrt{\bq} C^{AB} \d \bar q_{AB}\cr
&\quad+\f{1}{4r}  C_{CD}  C^{CD} \delta {\sqrt{\bq}} 
-\f{1}{4r}\sqrt{\bq} C^{AB} \delta  C_{AB}
 -\f 4r \sqrt{\bq} \d \bar \beta
 +o(r^{-1})\,,
 \la{thetauB}
 \end{align}
where  we used again the properties \eqref{dprop}, \eqref{CC},
and
 the fact that $\sqrt{\bar q}\bar{q}_{AB}\delta \bq^{AB}=-2 \d \sqrt{\bar q}$.

\section{Variations of fields} \label{Appvariation}
In this appendix, we list the field variations under the \bmsw generators of those fields that enter the Noether charges and fluxes. This list of variations is derived from the field variations in Section \ref{APHaction} and it is useful to perform the computation outlined in Section \ref{sec:FB2} when performing the field variations of the Noether charges and the field contractions on the Noether charges.

Field variations under super-translation, written off-shell of the ${\textsf E}_{\bU_A}$ equation of motion:
\begin{subequations}\la{dT}
\begin{align}
\d_T \bq_{AB}&=0\,,\\
\d_T \bF&=0\,,\\
\d_T M&= T\dot M + \pa^A T \p_A \bF +\f14 N^{AB}\bD_A\bD_B T+\f12 \p_AT\left({\bD}_B N^{AB}{ -\dot{\textsf E}_{\bU_A}}\right)\,,\\
\d_T C_{AB}&= T N_{AB}-2\bD_A\bD_B T+\bq_{AB} \bar \Delta T\,,\\
\d_T (\bD_A \bU^A)&=\bD_A \d_T \bU^A\,.
\end{align}
\end{subequations}

Field variations under the Weyl rescaling, also off-shell of  ${\textsf E}_{\bU_A}=0$:
\begin{subequations}\la{dW}
\begin{align}
\d_{\W} \bq_{AB}&=-2\W\, \bq_{AB}\,,\la{dWq}\\
\d_{\W} \sqrt{\bq}&=-2\W\, \sqrt{\bq}\,,\la{dWq2}\\
\d_{\W} \bF&=2 \W\,\bF  +\f12  \bar\Delta \W \,,\la{dWF}\\
\d_{\W} M&= \W \,\dot M+3\W\,  M +u \pa^A \W\, \p_A \bF \cr
&\quad +\f14 C^{AB}  \bD_A\p_B \W
+ \f u2\left( \bD_AN^{AB}  { -\dot{\textsf E}_{\bU_B}}\right)\p_B \W\cr
&\quad+\f u4N^{AB} \bD_A\bD_B \W
{+\f1{2} \textsf E_{\bU_A} \p_A \W}
\,,\\
\d_{\W} C_{AB}&=-\W\, C_{AB} +u \W \,N_{AB}
+u  \bar\Delta \W\,   \bq_{AB}
-2u \bD_A \bD_B \W\,,\\
\d_{\W} N_{AB}&=
u \W\, \dot N_{AB}
+ \bar\Delta \W   \bq_{AB}
-2 \bD_A \bD_B \W\,,\\
\d_\W (\bD_A\bU^A)&=\bD_A(\d_\W \bU^A)-2\bU^A \p_A \W\,.
\end{align}
\end{subequations}

Field variations under the sphere diffeomorphism $Y^A$ 
\begin{subequations}\la{dY}
\begin{align}
\d_Y \bq_{AB}&= 2\bD_{(A} Y_{B)} \,,\\
\d_Y \sqrt{\bq}&=\bD_A Y^A \sqrt{\bq} \,,\\
\d_Y \bF &= Y^A \bD_A \bF \,,\\
\d_Y M&= Y^A\bD_A M \,,\\
\d_Y C_{AB}&= 
Y^C\bD_C C_{AB} +2C_{C(A} \bD_{B)} Y^C
\,,\\
\d_Y (\bD_A\bU^A)
&=-\bD_A Y^A\bD_C\bU^C 
+\f1{\sqrt{\bq}}\p_C \left(\d_Y(\sqrt{\bq}\bU^C) \right)\,.
\end{align}
\end{subequations}

\section{Weyl scalars in Bondi gauge}\la{AppF}

To compute the Weyl scalars, we recall the frame field in Eq.~\eqref{eq:BondiTetrads2}
\begin{equation}
\hat{e}_0 = e^{-2\beta}\left(\partial_u -F \partial_r +U^A \partial_A\right), \quad \hat{e}_1 =\partial_r, \quad \hat{e}_2=\frac{1}{r}\hat{E}^A_2\partial_A, \quad \hat{e}_3=\frac{1}{r}\hat{E}^A_{3}\partial_A.
\end{equation}
We also recall that $q^{AB} =  \hat{E}^A_i \hat{E}^B_j \eta^{ij} = 2\hat{E}^{(A}_2 \hat{E}^{B)}_3$ and $\epsilon^{AB} = \hat{E}^A_i \hat{E}^B_j \epsilon^{ij} =2\hat{E}^{[A}_2 \hat{E}^{B]}_3 $.

The five Weyl scalars are defined by contracting the Weyl tensor $ W_{\mu \nu \rho \sigma}$ with the frame field above. It is more useful, in order to highlight the covariant feature of these quantities, to introduce the following definitions
\begin{subequations}
\begin{align}
(\Psi_0)_{AB} &=-3\left(E_{AB}-\frac{1}{16}C_{AB}C_{CD}C^{CD}\right) ,  & (\Psi_4)_{AB} &= -\frac{1}{2}\dot{N}_{AB}
,\\
(\Psi_1)_{A} &= - \cP_A ,  &(\Psi_3)_{A} &= -\frac{1}{2} \left( \f12 \bD_A \bar{R}  + \bD_B N_{\;A}^B\right),\\
\text{Re}\Psi_2 &= -\left( M  +\f18 C^{AB}N_{AB}\right),  &\text{Im}\Psi_2 &=-\epsilon^{AB} \left(\frac{1}{4}\bD_{A}\bD_{C} C^C_{\;B} +\frac{1}{8}C_{A}{}^{C}N_{CB} \right).
\end{align}
\end{subequations}
The first column contains the charge aspects 
$(\Psi_{0AB},\Psi_{1A},\text{Re}\Psi_2)$ that satisfy an evolution equation on $\scri$.
The second column contains the components of the Weyl tensor that provide information about the radiative nature of asymptotic infinity. 
Note that $(\Psi_4)_{AB}$ and $(\Psi_0)_{AB}$ are symmetric and tracefree tensors, $(\psi_3)_{A}$ and $(\psi_1)_{A}$ are vectors, while $\text{Re}\psi_2$ and $\text{Im}\psi_2$ are scalar quantities.
Moreover, it is worth emphasizing that $(\Psi_3)_A $ is $M_A$, $\text{Re}\Psi_2$ is the covariant mass $\mathcal{M}$, $\text{Im}\Psi_2$ is the covariant dual mass $\tilde{\mathcal{M}}$, and $(\Psi_1)_A$ is the covariant momentum $\cP_A$ introduced in section~\ref{cov} to be those quantities that transform under the \bmw without any quadratic anomaly.

An explicit computation of the leading asymptotic values of the Weyl scalars shows that
\begin{subequations}
\begin{align}
\psi_4 & :=W_{\hat{0} \hat{3} \hat{0} \hat{3}}  = \frac{1}{r} \hat{\bE}^A_{3}\hat{\bE}^B_{3} (\Psi_4)_{AB} + o(r^{-1}),\\
\psi_3 & :=W_{\hat{0} \hat{3} \hat{0} \hat{1}} =  \frac{1}{r^2}  \hat{\bE}^A_{3} (\Psi_3)_{A}+ o(r^{-2}),  \\
\text{Re}\psi_2 & :=W_{\hat{1} \hat{0} \hat{1} \hat{0}} =\frac{1}{r^3}\text{Re}\Psi_2+ o(r^{-3}),\\
i \text{Im}\psi_2 & :=W_{\hat{1} \hat{0} \hat{2} \hat{3}} = \frac{1}{r^3}i\text{Im}\Psi_2+ o(r^{-3}),\\
\psi_1 & :=W_{\hat{1} \hat{0} \hat{1} \hat{2}} =\frac{1}{r^4}\hat{\bE}^A_{2}(\bar\Psi_1)_{A}  + o(r^{-4}),\\
\psi_0 & :=W_{\hat{1} \hat{2} \hat{1} \hat{2}} =  \frac{1}{r^5}\hat{\bE}^A_{2} \hat{\bE}^B_{2} (\bar\Psi_0)_{AB}+ o(r^{-5}).
\end{align}
\end{subequations}

\newpage
\bibliographystyle{bib-style2}
\bibliography{biblio-fluxes}

\providecommand{\href}[2]{#2}\begingroup\raggedright\begin{thebibliography}{100}

\bibitem{Noether:1918zz}
E.~Noether, \emph{{Invariant Variation Problems}},
  \href{http://dx.doi.org/10.1080/00411457108231446}{\emph{Gott. Nachr.}
  {\bfseries 1918} (1918) 235--257},
  [\href{https://arxiv.org/abs/physics/0503066}{{\ttfamily physics/0503066}}].

\bibitem{Bondi:1960jsa}
H.~Bondi, \emph{{Gravitational Waves in General Relativity}},
  \href{http://dx.doi.org/10.1038/186535a0}{\emph{Nature} {\bfseries 186}
  (1960) 535--535}.

\bibitem{BMS}
H.~Bondi, M.~G.~J. van~der Burg and A.~W.~K. Metzner, \emph{{Gravitational
  waves in general relativity. 7. Waves from axisymmetric isolated systems}},
  \href{http://dx.doi.org/10.1098/rspa.1962.0161}{\emph{Proc. Roy. Soc. Lond.}
  {\bfseries A269} (1962) 21--52}.

\bibitem{Sachs62}
R.~Sachs, \emph{{On the characteristic initial value problem in gravitational
  theory}}, {\emph{J.Math.Phys.} {\bfseries 3} (1962) 908--914}.

\bibitem{Newman:1962cia}
E.~T. Newman and T.~W.~J. Unti, \emph{{Behavior of Asymptotically Flat Empty
  Spaces}}, \href{http://dx.doi.org/10.1063/1.1724303}{\emph{J. Math. Phys.}
  {\bfseries 3} (1962) 891}.

\bibitem{Winicour16}
T.~M{\"a}dler and J.~Winicour, \emph{{Bondi-Sachs Formalism}},
  \href{http://dx.doi.org/10.4249/scholarpedia.33528}{\emph{Scholarpedia}
  {\bfseries 11} (2016) 33528},
  [\href{https://arxiv.org/abs/1609.01731}{{\ttfamily 1609.01731}}].

\bibitem{Weinberg:1965nx}
S.~Weinberg, \emph{{Infrared photons and gravitons}},
  \href{http://dx.doi.org/10.1103/PhysRev.140.B516}{\emph{Phys. Rev.}
  {\bfseries 140} (1965) B516--B524}.

\bibitem{Christodoulou:1991cr}
D.~Christodoulou, \emph{{Nonlinear nature of gravitation and gravitational wave
  experiments}},
  \href{http://dx.doi.org/10.1103/PhysRevLett.67.1486}{\emph{Phys. Rev. Lett.}
  {\bfseries 67} (1991) 1486--1489}.

\bibitem{ThorneB}
K.~S.~T. Braginsky, \emph{Gravitational-wave bursts with memory and
  experimental prospects}, {\emph{Nature} {\bfseries 327} (1987) 123--125}.

\bibitem{Strominger:2013jfa}
A.~Strominger, \emph{{On BMS Invariance of Gravitational Scattering}},
  \href{http://dx.doi.org/10.1007/JHEP07(2014)152}{\emph{JHEP} {\bfseries 07}
  (2014) 152}, [\href{https://arxiv.org/abs/1312.2229}{{\ttfamily 1312.2229}}].

\bibitem{Strominger:2014pwa}
A.~Strominger and A.~Zhiboedov, \emph{{Gravitational Memory, BMS
  Supertranslations and Soft Theorems}},
  \href{http://dx.doi.org/10.1007/JHEP01(2016)086}{\emph{JHEP} {\bfseries 01}
  (2016) 086}, [\href{https://arxiv.org/abs/1411.5745}{{\ttfamily 1411.5745}}].

\bibitem{Strominger:2017zoo}
A.~Strominger, \emph{{Lectures on the Infrared Structure of Gravity and Gauge
  Theory}},  [\href{https://arxiv.org/abs/1703.05448}{{\ttfamily 1703.05448}}].

\bibitem{Compere:2018aar}
G.~Comp\`ere and A.~Fiorucci, \emph{{Advanced Lectures on General Relativity}},
   [\href{https://arxiv.org/abs/1801.07064}{{\ttfamily 1801.07064}}].

\bibitem{Cachazo:2014fwa}
F.~Cachazo and A.~Strominger, \emph{{Evidence for a New Soft Graviton
  Theorem}},  [\href{https://arxiv.org/abs/1404.4091}{{\ttfamily 1404.4091}}].

\bibitem{Pasterski:2015tva}
S.~Pasterski, A.~Strominger and A.~Zhiboedov, \emph{{New Gravitational
  Memories}}, \href{http://dx.doi.org/10.1007/JHEP12(2016)053}{\emph{JHEP}
  {\bfseries 12} (2016) 053},
  [\href{https://arxiv.org/abs/1502.06120}{{\ttfamily 1502.06120}}].

\bibitem{Campiglia:2014yka}
M.~Campiglia and A.~Laddha, \emph{{Asymptotic symmetries and subleading soft
  graviton theorem}},
  \href{http://dx.doi.org/10.1103/PhysRevD.90.124028}{\emph{Phys. Rev. D}
  {\bfseries 90} (2014) 124028},
  [\href{https://arxiv.org/abs/1408.2228}{{\ttfamily 1408.2228}}].

\bibitem{Flanagan:2015pxa}
E.~E. Flanagan and D.~A. Nichols, \emph{{Conserved charges of the extended
  Bondi-Metzner-Sachs algebra}},
  \href{http://dx.doi.org/10.1103/PhysRevD.95.044002}{\emph{Phys. Rev. D}
  {\bfseries 95} (2017) 044002},
  [\href{https://arxiv.org/abs/1510.03386}{{\ttfamily 1510.03386}}].

\bibitem{Compere:2018ylh}
G.~Comp\`{e}re, A.~Fiorucci and R.~Ruzziconi, \emph{{Superboost transitions,
  refraction memory and super-Lorentz charge algebra}},
  \href{http://dx.doi.org/10.1007/JHEP11(2018)200}{\emph{JHEP} {\bfseries 11}
  (2018) 200}, [\href{https://arxiv.org/abs/1810.00377}{{\ttfamily
  1810.00377}}].

\bibitem{Himwich:2019qmj}
E.~Himwich, Z.~Mirzaiyan and S.~Pasterski, \emph{{A Note on the Subleading Soft
  Graviton}}, \href{http://dx.doi.org/10.1007/JHEP04(2021)172}{\emph{JHEP}
  {\bfseries 04} (2021) 172},
  [\href{https://arxiv.org/abs/1902.01840}{{\ttfamily 1902.01840}}].

\bibitem{Barnich:2009se}
G.~Barnich and C.~Troessaert, \emph{{Symmetries of asymptotically flat 4
  dimensional spacetimes at null infinity revisited}},
  \href{http://dx.doi.org/10.1103/PhysRevLett.105.111103}{\emph{Phys. Rev.
  Lett.} {\bfseries 105} (2010) 111103},
  [\href{https://arxiv.org/abs/0909.2617}{{\ttfamily 0909.2617}}].

\bibitem{Barnich:2011mi}
G.~Barnich and C.~Troessaert, \emph{{BMS charge algebra}},
  \href{http://dx.doi.org/10.1007/JHEP12(2011)105}{\emph{JHEP} {\bfseries 12}
  (2011) 105}, [\href{https://arxiv.org/abs/1106.0213}{{\ttfamily 1106.0213}}].

\bibitem{Donnay:2020guq}
L.~Donnay, S.~Pasterski and A.~Puhm, \emph{{Asymptotic Symmetries and Celestial
  CFT}}, \href{http://dx.doi.org/10.1007/JHEP09(2020)176}{\emph{JHEP}
  {\bfseries 09} (2020) 176},
  [\href{https://arxiv.org/abs/2005.08990}{{\ttfamily 2005.08990}}].

\bibitem{Kapec:2014opa}
D.~Kapec, V.~Lysov, S.~Pasterski and A.~Strominger, \emph{{Semiclassical
  Virasoro symmetry of the quantum gravity $ \mathcal{S}$-matrix}},
  \href{http://dx.doi.org/10.1007/JHEP08(2014)058}{\emph{JHEP} {\bfseries 08}
  (2014) 058}, [\href{https://arxiv.org/abs/1406.3312}{{\ttfamily 1406.3312}}].

\bibitem{Kapec:2016jld}
D.~Kapec, P.~Mitra, A.-M. Raclariu and A.~Strominger, \emph{{2D Stress Tensor
  for 4D Gravity}},
  \href{http://dx.doi.org/10.1103/PhysRevLett.119.121601}{\emph{Phys. Rev.
  Lett.} {\bfseries 119} (2017) 121601},
  [\href{https://arxiv.org/abs/1609.00282}{{\ttfamily 1609.00282}}].

\bibitem{Pasterski:2016qvg}
S.~Pasterski, S.-H. Shao and A.~Strominger, \emph{{Flat Space Amplitudes and
  Conformal Symmetry of the Celestial Sphere}},
  \href{http://dx.doi.org/10.1103/PhysRevD.96.065026}{\emph{Phys. Rev. D}
  {\bfseries 96} (2017) 065026},
  [\href{https://arxiv.org/abs/1701.00049}{{\ttfamily 1701.00049}}].

\bibitem{Donnay:2015abr}
L.~Donnay, G.~Giribet, H.~A. Gonzalez and M.~Pino, \emph{{Supertranslations and
  Superrotations at the Black Hole Horizon}},
  \href{http://dx.doi.org/10.1103/PhysRevLett.116.091101}{\emph{Phys. Rev.
  Lett.} {\bfseries 116} (2016) 091101},
  [\href{https://arxiv.org/abs/1511.08687}{{\ttfamily 1511.08687}}].

\bibitem{Donnay:2016ejv}
L.~Donnay, G.~Giribet, H.~A. Gonz\'alez and M.~Pino, \emph{{Extended Symmetries
  at the Black Hole Horizon}},
  \href{http://dx.doi.org/10.1007/JHEP09(2016)100}{\emph{JHEP} {\bfseries 09}
  (2016) 100}, [\href{https://arxiv.org/abs/1607.05703}{{\ttfamily
  1607.05703}}].

\bibitem{Donnay:2019jiz}
L.~Donnay and C.~Marteau, \emph{{Carrollian Physics at the Black Hole
  Horizon}}, \href{http://dx.doi.org/10.1088/1361-6382/ab2fd5}{\emph{Class.
  Quant. Grav.} {\bfseries 36} (2019) 165002},
  [\href{https://arxiv.org/abs/1903.09654}{{\ttfamily 1903.09654}}].

\bibitem{Hopfmuller:2016scf}
F.~Hopfmuller and L.~Freidel, \emph{{Gravity Degrees of Freedom on a Null
  Surface}}, \href{http://dx.doi.org/10.1103/PhysRevD.95.104006}{\emph{Phys.
  Rev.} {\bfseries D95} (2017) 104006},
  [\href{https://arxiv.org/abs/1611.03096}{{\ttfamily 1611.03096}}].

\bibitem{Wieland:2017zkf}
W.~Wieland, \emph{{New boundary variables for classical and quantum gravity on
  a null surface}},
  \href{http://dx.doi.org/10.1088/1361-6382/aa8d06}{\emph{Class. Quant. Grav.}
  {\bfseries 34} (2017) 215008},
  [\href{https://arxiv.org/abs/1704.07391}{{\ttfamily 1704.07391}}].

\bibitem{Chandrasekaran:2018aop}
V.~Chandrasekaran, E.~E. Flanagan and K.~Prabhu, \emph{{Symmetries and charges
  of general relativity at null boundaries}},
  \href{http://dx.doi.org/10.1007/JHEP11(2018)125}{\emph{JHEP} {\bfseries 11}
  (2018) 125}, [\href{https://arxiv.org/abs/1807.11499}{{\ttfamily
  1807.11499}}].

\bibitem{Wieland:2020gno}
W.~Wieland, \emph{{Null infinity as an open Hamiltonian system}},
  [\href{https://arxiv.org/abs/2012.01889}{{\ttfamily 2012.01889}}].

\bibitem{DonnellyFreidel}
W.~Donnelly and L.~Freidel, \emph{{Local subsystems in gauge theory and
  gravity}}, \href{http://dx.doi.org/10.1007/JHEP09(2016)102}{\emph{JHEP}
  {\bfseries 09} (2016) 102},
  [\href{https://arxiv.org/abs/1601.04744}{{\ttfamily 1601.04744}}].

\bibitem{Freidel:2015gpa}
L.~Freidel and A.~Perez, \emph{{Quantum gravity at the corner}},
  \href{http://dx.doi.org/10.3390/universe4100107}{\emph{Universe} {\bfseries
  4} (2018) 107}, [\href{https://arxiv.org/abs/1507.02573}{{\ttfamily
  1507.02573}}].

\bibitem{Freidel:2016bxd}
L.~Freidel, A.~Perez and D.~Pranzetti, \emph{{Loop gravity string}},
  \href{http://dx.doi.org/10.1103/PhysRevD.95.106002}{\emph{Phys. Rev.}
  {\bfseries D95} (2017) 106002},
  [\href{https://arxiv.org/abs/1611.03668}{{\ttfamily 1611.03668}}].

\bibitem{Freidel:2018pvm}
L.~Freidel and E.~R. Livine, \emph{{Bubble networks: framed discrete geometry
  for quantum gravity}},
  \href{http://dx.doi.org/10.1007/s10714-018-2493-y}{\emph{Gen. Rel. Grav.}
  {\bfseries 51} (2019) 9}, [\href{https://arxiv.org/abs/1810.09364}{{\ttfamily
  1810.09364}}].

\bibitem{Freidel:2019ees}
L.~Freidel, E.~R. Livine and D.~Pranzetti, \emph{{Gravitational edge modes:
  from Kac-Moody charges to Poincar{\'e} networks}},
  \href{http://dx.doi.org/10.1088/1361-6382/ab40fe}{\emph{Class. Quant. Grav.}
  {\bfseries 36} (2019) 195014},
  [\href{https://arxiv.org/abs/1906.07876}{{\ttfamily 1906.07876}}].

\bibitem{Freidel:2020xyx}
L.~Freidel, M.~Geiller and D.~Pranzetti, \emph{{Edge modes of gravity - I:
  Corner potentials and charges}},
  [\href{https://arxiv.org/abs/2006.12527}{{\ttfamily 2006.12527}}].

\bibitem{Freidel:2020svx}
L.~Freidel, M.~Geiller and D.~Pranzetti, \emph{{Edge modes of gravity - II:
  Corner metric and Lorentz charges}},
  [\href{https://arxiv.org/abs/2007.03563}{{\ttfamily 2007.03563}}].

\bibitem{Freidel:2020ayo}
L.~Freidel, M.~Geiller and D.~Pranzetti, \emph{{Edge modes of gravity. Part
  III. Corner simplicity constraints}},
  \href{http://dx.doi.org/10.1007/JHEP01(2021)100}{\emph{JHEP} {\bfseries 01}
  (2021) 100}, [\href{https://arxiv.org/abs/2007.12635}{{\ttfamily
  2007.12635}}].

\bibitem{Donnelly:2020xgu}
W.~Donnelly, L.~Freidel, S.~F. Moosavian and A.~J. Speranza,
  \emph{{Gravitational Edge Modes, Coadjoint Orbits, and Hydrodynamics}},
  [\href{https://arxiv.org/abs/2012.10367}{{\ttfamily 2012.10367}}].

\bibitem{Livine:2021qwx}
E.~R. Livine, \emph{{Loop Quantum Gravity Boundary Dynamics and SL(2,C) Gauge
  Theory}},  [\href{https://arxiv.org/abs/2101.07565}{{\ttfamily 2101.07565}}].

\bibitem{Chen:2021vrc}
Q.~Chen and E.~R. Livine, \emph{{Loop Quantum Gravity's Boundary Maps}},
  [\href{https://arxiv.org/abs/2103.08409}{{\ttfamily 2103.08409}}].

\bibitem{Freidel:2021cbc}
L.~Freidel, R.~Oliveri, D.~Pranzetti and S.~Speziale, \emph{{Extended corner
  symmetry, charge bracket and Einstein's equations}},
  [\href{https://arxiv.org/abs/2104.12881}{{\ttfamily 2104.12881}}].

\bibitem{Troessaert:2015nia}
C.~Troessaert, \emph{{Hamiltonian surface charges using external sources}},
  \href{http://dx.doi.org/10.1063/1.4947177}{\emph{J. Math. Phys.} {\bfseries
  57} (2016) 053507}, [\href{https://arxiv.org/abs/1509.09094}{{\ttfamily
  1509.09094}}].

\bibitem{Wald:1999wa}
R.~M. Wald and A.~Zoupas, \emph{{A General definition of 'conserved quantities'
  in general relativity and other theories of gravity}},
  \href{http://dx.doi.org/10.1103/PhysRevD.61.084027}{\emph{Phys. Rev. D}
  {\bfseries 61} (2000) 084027},
  [\href{https://arxiv.org/abs/gr-qc/9911095}{{\ttfamily gr-qc/9911095}}].

\bibitem{Chandrasekaran:2020wwn}
V.~Chandrasekaran and A.~J. Speranza, \emph{{Anomalies in gravitational charge
  algebras of null boundaries and black hole entropy}},
  \href{http://dx.doi.org/10.1007/JHEP01(2021)137}{\emph{JHEP} {\bfseries 01}
  (2021) 137}, [\href{https://arxiv.org/abs/2009.10739}{{\ttfamily
  2009.10739}}].

\bibitem{Barnich:2010eb}
G.~Barnich and C.~Troessaert, \emph{{Aspects of the BMS/CFT correspondence}},
  \href{http://dx.doi.org/10.1007/JHEP05(2010)062}{\emph{JHEP} {\bfseries 05}
  (2010) 062}, [\href{https://arxiv.org/abs/1001.1541}{{\ttfamily 1001.1541}}].

\bibitem{Barnich:2007bf}
G.~Barnich and G.~Compere, \emph{{Surface charge algebra in gauge theories and
  thermodynamic integrability}},
  \href{http://dx.doi.org/10.1063/1.2889721}{\emph{J. Math. Phys.} {\bfseries
  49} (2008) 042901}, [\href{https://arxiv.org/abs/0708.2378}{{\ttfamily
  0708.2378}}].

\bibitem{Barnich:2013axa}
G.~Barnich and C.~Troessaert, \emph{{Comments on holographic current algebras
  and asymptotically flat four dimensional spacetimes at null infinity}},
  \href{http://dx.doi.org/10.1007/JHEP11(2013)003}{\emph{JHEP} {\bfseries 11}
  (2013) 003}, [\href{https://arxiv.org/abs/1309.0794}{{\ttfamily 1309.0794}}].

\bibitem{Compere:2020lrt}
G.~Comp\`ere, A.~Fiorucci and R.~Ruzziconi, \emph{{The $\Lambda$-BMS$_4$ charge
  algebra}}, \href{http://dx.doi.org/10.1007/JHEP10(2020)205}{\emph{JHEP}
  {\bfseries 10} (2020) 205},
  [\href{https://arxiv.org/abs/2004.10769}{{\ttfamily 2004.10769}}].

\bibitem{Adami:2020ugu}
H.~Adami, M.~M. Sheikh-Jabbari, V.~Taghiloo, H.~Yavartanoo and C.~Zwikel,
  \emph{{Symmetries at null boundaries: two and three dimensional gravity
  cases}}, \href{http://dx.doi.org/10.1007/JHEP10(2020)107}{\emph{JHEP}
  {\bfseries 10} (2020) 107},
  [\href{https://arxiv.org/abs/2007.12759}{{\ttfamily 2007.12759}}].

\bibitem{Ruzziconi:2020wrb}
R.~Ruzziconi and C.~Zwikel, \emph{{Conservation and Integrability in
  Lower-Dimensional Gravity}},
  [\href{https://arxiv.org/abs/2012.03961}{{\ttfamily 2012.03961}}].

\bibitem{Alessio:2020ioh}
F.~Alessio, G.~Barnich, L.~Ciambelli, P.~Mao and R.~Ruzziconi, \emph{{Weyl
  charges in asymptotically locally AdS$_3$ spacetimes}},
  \href{http://dx.doi.org/10.1103/PhysRevD.103.046003}{\emph{Phys. Rev. D}
  {\bfseries 103} (2021) 046003},
  [\href{https://arxiv.org/abs/2010.15452}{{\ttfamily 2010.15452}}].

\bibitem{Fiorucci:2020xto}
A.~Fiorucci and R.~Ruzziconi, \emph{{Charge Algebra in Al(A)dS$_n$
  Spacetimes}},  [\href{https://arxiv.org/abs/2011.02002}{{\ttfamily
  2011.02002}}].

\bibitem{Compere:2019gft}
G.~Comp\`ere, R.~Oliveri and A.~Seraj, \emph{{The Poincar\'e and BMS
  flux-balance laws with application to binary systems}},
  \href{http://dx.doi.org/10.1007/JHEP10(2020)116}{\emph{JHEP} {\bfseries 10}
  (2020) 116}, [\href{https://arxiv.org/abs/1912.03164}{{\ttfamily
  1912.03164}}].

\bibitem{Kijowski1976ACS}
J.~D. Kijowski and W.~Szczyrba, \emph{A canonical structure for classical field
  theories}, {\emph{Communications in Mathematical Physics} {\bfseries 46}
  (1976) 183--206}.

\bibitem{Gawdzki1991ClassicalOO}
K.~Gaw{\c e}dzki, \emph{{Classical origin of quantum group symmetries in
  Wess-Zumino-Witten conformal field theory}}, {\emph{Communications in
  Mathematical Physics} {\bfseries 139} (1991) 201--213}.

\bibitem{Crnkovic:1986ex}
C.~Crnkovic and E.~Witten, \emph{{Covariant description of canonical formalism
  in geometrical theories}}, pp.~676--684.
\newblock Three Hundred Years of Gravitation, Cambridge: Cambridge University
  Press, 1987, pp. 676--684, 1986.

\bibitem{Ashtekar:1990gc}
A.~Ashtekar, L.~Bombelli and O.~Reula, \emph{The covariant phase space of
  asymptotically flat gravitational fields},  in \emph{Mechanics, Analysis and
  Geometry: 200 Years After Lagrange} (M.~Francaviglia, ed.), North-Holland
  Delta Series, pp.~417 -- 450.
\newblock Elsevier, Amsterdam, 1991.
\newblock \href{http://dx.doi.org/10.1016/B978-0-444-88958-4.50021-5}{DOI}.

\bibitem{Lee:1990nz}
J.~Lee and R.~M. Wald, \emph{{Local symmetries and constraints}},
  \href{http://dx.doi.org/10.1063/1.528801}{\emph{J. Math. Phys.} {\bfseries
  31} (1990) 725--743}.

\bibitem{Campiglia:2020qvc}
M.~Campiglia and J.~Peraza, \emph{{Generalized BMS charge algebra}},
  \href{http://dx.doi.org/10.1103/PhysRevD.101.104039}{\emph{Phys. Rev. D}
  {\bfseries 101} (2020) 104039},
  [\href{https://arxiv.org/abs/2002.06691}{{\ttfamily 2002.06691}}].

\bibitem{Ciambelli:2021vnn}
L.~Ciambelli and R.~G. Leigh, \emph{{Isolated Surfaces and Symmetries of
  Gravity}},  [\href{https://arxiv.org/abs/2104.07643}{{\ttfamily
  2104.07643}}].

\bibitem{Bonga:2018gzr}
B.~Bonga and E.~Poisson, \emph{{Coulombic contribution to angular momentum flux
  in general relativity}},
  \href{http://dx.doi.org/10.1103/PhysRevD.99.064024}{\emph{Phys. Rev. D}
  {\bfseries 99} (2019) 064024},
  [\href{https://arxiv.org/abs/1808.01288}{{\ttfamily 1808.01288}}].

\bibitem{Ashtekar:2019rpv}
A.~Ashtekar, T.~De~Lorenzo and N.~Khera, \emph{{Compact binary coalescences:
  The subtle issue of angular momentum}},
  \href{http://dx.doi.org/10.1103/PhysRevD.101.044005}{\emph{Phys. Rev. D}
  {\bfseries 101} (2020) 044005},
  [\href{https://arxiv.org/abs/1910.02907}{{\ttfamily 1910.02907}}].

\bibitem{Elhashash:2021iev}
A.~Elhashash and D.~A. Nichols, \emph{{Definitions of (super) angular momentum
  in asymptotically flat spacetimes: Properties and applications to
  compact-binary mergers}},
  [\href{https://arxiv.org/abs/2101.12228}{{\ttfamily 2101.12228}}].

\bibitem{Chen:2021szm}
P.-N. Chen, M.-T. Wang, Y.-K. Wang and S.-T. Yau, \emph{{Supertranslation
  invariance of angular momentum}},
  [\href{https://arxiv.org/abs/2102.03235}{{\ttfamily 2102.03235}}].

\bibitem{Compere:2021inq}
G.~Comp\`ere and D.~A. Nichols, \emph{{Classical and Quantized
  General-Relativistic Angular Momentum}},
  [\href{https://arxiv.org/abs/2103.17103}{{\ttfamily 2103.17103}}].

\bibitem{Compere:2016jwb}
G.~Comp\`ere and J.~Long, \emph{{Vacua of the gravitational field}},
  \href{http://dx.doi.org/10.1007/JHEP07(2016)137}{\emph{JHEP} {\bfseries 07}
  (2016) 137}, [\href{https://arxiv.org/abs/1601.04958}{{\ttfamily
  1601.04958}}].

\bibitem{Ashtekar:1981sf}
A.~Ashtekar, \emph{{Asymptotic Quantization of the Gravitational Field}},
  \href{http://dx.doi.org/10.1103/PhysRevLett.46.573}{\emph{Phys. Rev. Lett.}
  {\bfseries 46} (1981) 573--576}.

\bibitem{Ashtekar:1981bq}
A.~Ashtekar and M.~Streubel, \emph{{Symplectic Geometry of Radiative Modes and
  Conserved Quantities at Null Infinity}},
  \href{http://dx.doi.org/10.1098/rspa.1981.0109}{\emph{Proc. Roy. Soc. Lond.
  A} {\bfseries 376} (1981) 585--607}.

\bibitem{Ciambelli:2018wre}
L.~Ciambelli, C.~Marteau, A.~C. Petkou, P.~M. Petropoulos and K.~Siampos,
  \emph{{Flat holography and Carrollian fluids}},
  \href{http://dx.doi.org/10.1007/JHEP07(2018)165}{\emph{JHEP} {\bfseries 07}
  (2018) 165}, [\href{https://arxiv.org/abs/1802.06809}{{\ttfamily
  1802.06809}}].

\bibitem{Ciambelli:2019lap}
L.~Ciambelli, R.~G. Leigh, C.~Marteau and P.~M. Petropoulos, \emph{{Carroll
  Structures, Null Geometry and Conformal Isometries}},
  \href{http://dx.doi.org/10.1103/PhysRevD.100.046010}{\emph{Phys. Rev. D}
  {\bfseries 100} (2019) 046010},
  [\href{https://arxiv.org/abs/1905.02221}{{\ttfamily 1905.02221}}].

\bibitem{Herfray:2020rvq}
Y.~Herfray, \emph{{Asymptotic shear and the intrinsic conformal geometry of
  null-infinity}}, \href{http://dx.doi.org/10.1063/5.0003616}{\emph{J. Math.
  Phys.} {\bfseries 61} (2020) 072502},
  [\href{https://arxiv.org/abs/2001.01281}{{\ttfamily 2001.01281}}].

\bibitem{Herfray:2021xyp}
Y.~Herfray, \emph{{Tractor geometry of asymptotically flat space-times}},
  [\href{https://arxiv.org/abs/2103.10405}{{\ttfamily 2103.10405}}].

\bibitem{Reisenberger:2007ku}
M.~P. Reisenberger, \emph{{The Poisson bracket on free null initial data for
  gravity}},
  \href{http://dx.doi.org/10.1103/PhysRevLett.101.211101}{\emph{Phys. Rev.
  Lett.} {\bfseries 101} (2008) 211101},
  [\href{https://arxiv.org/abs/0712.2541}{{\ttfamily 0712.2541}}].

\bibitem{Reisenberger:2012zq}
M.~P. Reisenberger, \emph{{The symplectic 2-form for gravity in terms of free
  null initial data}},
  \href{http://dx.doi.org/10.1088/0264-9381/30/15/155022}{\emph{Class. Quant.
  Grav.} {\bfseries 30} (2013) 155022},
  [\href{https://arxiv.org/abs/1211.3880}{{\ttfamily 1211.3880}}].

\bibitem{Parattu:2015gga}
K.~Parattu, S.~Chakraborty, B.~R. Majhi and T.~Padmanabhan, \emph{{A Boundary
  Term for the Gravitational Action with Null Boundaries}},
  \href{http://dx.doi.org/10.1007/s10714-016-2093-7}{\emph{Gen. Rel. Grav.}
  {\bfseries 48} (2016) 94},
  [\href{https://arxiv.org/abs/1501.01053}{{\ttfamily 1501.01053}}].

\bibitem{Reisenberger:2018xkn}
M.~P. Reisenberger, \emph{{The Poisson brackets of free null initial data for
  vacuum general relativity}},
  \href{http://dx.doi.org/10.1088/1361-6382/aad569}{\emph{Class. Quant. Grav.}
  {\bfseries 35} (2018) 185012},
  [\href{https://arxiv.org/abs/1804.10284}{{\ttfamily 1804.10284}}].

\bibitem{Hopfmuller:2018fni}
F.~Hopfm{\"u}ller and L.~Freidel, \emph{{Null Conservation Laws for Gravity}},
  [\href{https://arxiv.org/abs/1802.06135}{{\ttfamily 1802.06135}}].

\bibitem{Oliveri:2019gvm}
R.~Oliveri and S.~Speziale, \emph{{Boundary effects in General Relativity with
  tetrad variables}},
  \href{http://dx.doi.org/10.1007/s10714-020-02733-8}{\emph{Gen. Rel. Grav.}
  {\bfseries 52} (2020) 83},
  [\href{https://arxiv.org/abs/1912.01016}{{\ttfamily 1912.01016}}].

\bibitem{Campiglia:2015yka}
M.~Campiglia and A.~Laddha, \emph{{New symmetries for the Gravitational
  S-matrix}}, \href{http://dx.doi.org/10.1007/JHEP04(2015)076}{\emph{JHEP}
  {\bfseries 04} (2015) 076},
  [\href{https://arxiv.org/abs/1502.02318}{{\ttfamily 1502.02318}}].

\bibitem{Freidel:2019ofr}
L.~Freidel, E.~R. Livine and D.~Pranzetti, \emph{{Kinematical Gravitational
  Charge Algebra}},
  \href{http://dx.doi.org/10.1103/PhysRevD.101.024012}{\emph{Phys. Rev. D}
  {\bfseries 101} (2020) 024012},
  [\href{https://arxiv.org/abs/1910.05642}{{\ttfamily 1910.05642}}].

\bibitem{Penna:2015gza}
R.~F. Penna, \emph{{BMS invariance and the membrane paradigm}},
  \href{http://dx.doi.org/10.1007/JHEP03(2016)023}{\emph{JHEP} {\bfseries 03}
  (2016) 023}, [\href{https://arxiv.org/abs/1508.06577}{{\ttfamily
  1508.06577}}].

\bibitem{Penna:2017bdn}
R.~F. Penna, \emph{{Near-horizon BMS symmetries as fluid symmetries}},
  \href{http://dx.doi.org/10.1007/JHEP10(2017)049}{\emph{JHEP} {\bfseries 10}
  (2017) 049}, [\href{https://arxiv.org/abs/1703.07382}{{\ttfamily
  1703.07382}}].

\bibitem{anderson1992introduction}
I.~M. Anderson, \emph{Introduction to the variational bicomplex},
  {\emph{Contemporary Mathematics} {\bfseries 132} (1992) }.

\bibitem{Francois:2020tom}
J.~Fran\c{c}ois, \emph{{Bundle geometry of the connection space, covariant
  Hamiltonian formalism, the problem of boundaries in gauge theories, and the
  dressing field method}},
  \href{http://dx.doi.org/10.1007/JHEP03(2021)225}{\emph{JHEP} {\bfseries 03}
  (2021) 225}, [\href{https://arxiv.org/abs/2010.01597}{{\ttfamily
  2010.01597}}].

\bibitem{Sachs:1962wk}
R.~K. Sachs, \emph{{Gravitational waves in general relativity. 8. Waves in
  asymptotically flat space-times}},
  \href{http://dx.doi.org/10.1098/rspa.1962.0206}{\emph{Proc. Roy. Soc. Lond.}
  {\bfseries A270} (1962) 103--126}.

\bibitem{Bondi62}
H.~Bondi, M.~van~der Burg and A.~Metzner, \emph{Gravitational waves in general
  relativity. 7. waves from axisymmetric isolated systems},
  {\emph{Proc.Roy.Soc.Lond.} {\bfseries A269} (1962) 21--52}.

\bibitem{Barnich:2016lyg}
G.~Barnich and C.~Troessaert, \emph{{Finite BMS transformations}},
  \href{http://dx.doi.org/10.1007/JHEP03(2016)167}{\emph{JHEP} {\bfseries 03}
  (2016) 167}, [\href{https://arxiv.org/abs/1601.04090}{{\ttfamily
  1601.04090}}].

\bibitem{Blanchet87}
L.~Blanchet, \emph{Radiative gravitational fields in general relativity ii.
  asymptotic behaviour at future null infinity},
  \href{http://dx.doi.org/10.1098/rspa.1987.0022}{\emph{Proceedings of the
  Royal Society of London. A. Mathematical and Physical Sciences} {\bfseries
  409} (1987) 383--399}.

\bibitem{Blanchet:2020ngx}
L.~Blanchet, G.~Comp\`ere, G.~Faye, R.~Oliveri and A.~Seraj, \emph{{Multipole
  expansion of gravitational waves: from harmonic to Bondi coordinates}},
  \href{http://dx.doi.org/10.1007/JHEP02(2021)029}{\emph{JHEP} {\bfseries 02}
  (2021) 029}, [\href{https://arxiv.org/abs/2011.10000}{{\ttfamily
  2011.10000}}].

\bibitem{deHaro:2000vlm}
S.~de~Haro, S.~N. Solodukhin and K.~Skenderis, \emph{{Holographic
  reconstruction of space-time and renormalization in the AdS / CFT
  correspondence}},
  \href{http://dx.doi.org/10.1007/s002200100381}{\emph{Commun. Math. Phys.}
  {\bfseries 217} (2001) 595--622},
  [\href{https://arxiv.org/abs/hep-th/0002230}{{\ttfamily hep-th/0002230}}].

\bibitem{Papadimitriou:2005ii}
I.~Papadimitriou and K.~Skenderis, \emph{{Thermodynamics of asymptotically
  locally AdS spacetimes}},
  \href{http://dx.doi.org/10.1088/1126-6708/2005/08/004}{\emph{JHEP} {\bfseries
  08} (2005) 004}, [\href{https://arxiv.org/abs/hep-th/0505190}{{\ttfamily
  hep-th/0505190}}].

\bibitem{Compere:2008us}
G.~Compere and D.~Marolf, \emph{{Setting the boundary free in AdS/CFT}},
  \href{http://dx.doi.org/10.1088/0264-9381/25/19/195014}{\emph{Class. Quant.
  Grav.} {\bfseries 25} (2008) 195014},
  [\href{https://arxiv.org/abs/0805.1902}{{\ttfamily 0805.1902}}].

\bibitem{Freidel:2019ohg}
L.~Freidel, F.~Hopfm\"uller and A.~Riello, \emph{{Asymptotic Renormalization in
  Flat Space: Symplectic Potential and Charges of Electromagnetism}},
  \href{http://dx.doi.org/10.1007/JHEP10(2019)126}{\emph{JHEP} {\bfseries 10}
  (2019) 126}, [\href{https://arxiv.org/abs/1904.04384}{{\ttfamily
  1904.04384}}].

\bibitem{AshtekarReula}
A.~Ashtekar, L.~Bombelli and O.~Reula, \emph{The covariant phase space of
  asymptotically flat gravitational fields},  in \emph{Analysis, geometry and
  mechanics: 200 years after Lagrange} (M.~Francaviglia and D.~Holm, eds.),
  North-Holland, 1991.

\bibitem{Barnich:2001jy}
G.~Barnich and F.~Brandt, \emph{{Covariant theory of asymptotic symmetries,
  conservation laws and central charges}},
  \href{http://dx.doi.org/10.1016/S0550-3213(02)00251-1}{\emph{Nucl. Phys.}
  {\bfseries B633} (2002) 3--82},
  [\href{https://arxiv.org/abs/hep-th/0111246}{{\ttfamily hep-th/0111246}}].

\bibitem{Rangamani_2009}
M.~Rangamani, \emph{Gravity and hydrodynamics: lectures on the fluid-gravity
  correspondence},
  \href{http://dx.doi.org/10.1088/0264-9381/26/22/224003}{\emph{Classical and
  Quantum Gravity} {\bfseries 26} (Oct, 2009) 224003}.

\bibitem{Ruzziconi:2020cjt}
R.~Ruzziconi, \emph{{On the Various Extensions of the BMS Group}}.
\newblock PhD thesis, Universit{\'e} libre de Bruxelles, 9, 2020.
\newblock \href{https://arxiv.org/abs/2009.01926}{{\ttfamily 2009.01926}}.

\bibitem{Barnich_2012}
G.~Barnich and P.-H. Lambert, \emph{A note on the newman-unti group and the bms
  charge algebra in terms of newman-penrose coefficients},
  \href{http://dx.doi.org/10.1155/2012/197385}{\emph{Advances in Mathematical
  Physics} {\bfseries 2012} (2012) 1--16}.

\bibitem{Ciambelli:2019bzz}
L.~Ciambelli and R.~G. Leigh, \emph{{Weyl Connections and their Role in
  Holography}},
  \href{http://dx.doi.org/10.1103/PhysRevD.101.086020}{\emph{Phys. Rev. D}
  {\bfseries 101} (2020) 086020},
  [\href{https://arxiv.org/abs/1905.04339}{{\ttfamily 1905.04339}}].

\bibitem{Barnich:2019vzx}
G.~Barnich, P.~Mao and R.~Ruzziconi, \emph{{BMS current algebra in the context
  of the Newman\textendash{}Penrose formalism}},
  \href{http://dx.doi.org/10.1088/1361-6382/ab7c01}{\emph{Class. Quant. Grav.}
  {\bfseries 37} (2020) 095010},
  [\href{https://arxiv.org/abs/1910.14588}{{\ttfamily 1910.14588}}].

\bibitem{FP}
L.~Freidel and D.~Pranzetti, ``{\it Gravity from symmetry}.'' 2021, to appear.

\bibitem{Hawking:2016sgy}
S.~W. Hawking, M.~J. Perry and A.~Strominger, \emph{{Superrotation Charge and
  Supertranslation Hair on Black Holes}},
  \href{http://dx.doi.org/10.1007/JHEP05(2017)161}{\emph{JHEP} {\bfseries 05}
  (2017) 161}, [\href{https://arxiv.org/abs/1611.09175}{{\ttfamily
  1611.09175}}].

\bibitem{Barnich:2021dta}
G.~Barnich and R.~Ruzziconi, \emph{{Coadjoint representation of the BMS group
  on celestial Riemann surfaces}},
  [\href{https://arxiv.org/abs/2103.11253}{{\ttfamily 2103.11253}}].

\bibitem{Ashtekar:2008jw}
A.~Ashtekar, J.~Engle and D.~Sloan, \emph{{Asymptotics and Hamiltonians in a
  First order formalism}},
  \href{http://dx.doi.org/10.1088/0264-9381/25/9/095020}{\emph{Class. Quant.
  Grav.} {\bfseries 25} (2008) 095020},
  [\href{https://arxiv.org/abs/0802.2527}{{\ttfamily 0802.2527}}].

\bibitem{Ashtekar:2004cn}
A.~Ashtekar and B.~Krishnan, \emph{{Isolated and dynamical horizons and their
  applications}}, \href{http://dx.doi.org/10.12942/lrr-2004-10}{\emph{Living
  Rev. Rel.} {\bfseries 7} (2004) 10},
  [\href{https://arxiv.org/abs/gr-qc/0407042}{{\ttfamily gr-qc/0407042}}].

\bibitem{DiazPolo:2011np}
J.~Diaz-Polo and D.~Pranzetti, \emph{{Isolated Horizons and Black Hole Entropy
  In Loop Quantum Gravity}},
  \href{http://dx.doi.org/10.3842/SIGMA.2012.048}{\emph{SIGMA} {\bfseries 8}
  (2012) 048}, [\href{https://arxiv.org/abs/1112.0291}{{\ttfamily 1112.0291}}].

\bibitem{Godazgar:2018qpq}
H.~Godazgar, M.~Godazgar and C.~Pope, \emph{{New dual gravitational charges}},
  \href{http://dx.doi.org/10.1103/PhysRevD.99.024013}{\emph{Phys. Rev. D}
  {\bfseries 99} (2019) 024013},
  [\href{https://arxiv.org/abs/1812.01641}{{\ttfamily 1812.01641}}].

\bibitem{Godazgar:2020kqd}
H.~Godazgar, M.~Godazgar and M.~J. Perry, \emph{{Hamiltonian derivation of dual
  gravitational charges}},
  \href{http://dx.doi.org/10.1007/JHEP09(2020)084}{\emph{JHEP} {\bfseries 20}
  (2020) 084}, [\href{https://arxiv.org/abs/2007.07144}{{\ttfamily
  2007.07144}}].

\bibitem{Oliveri:2020xls}
R.~Oliveri and S.~Speziale, \emph{{A note on dual gravitational charges}},
  \href{http://dx.doi.org/10.1007/JHEP12(2020)079}{\emph{JHEP} {\bfseries 12}
  (2020) 079}, [\href{https://arxiv.org/abs/2010.01111}{{\ttfamily
  2010.01111}}].

\bibitem{DePaoli:2018erh}
E.~De~Paoli and S.~Speziale, \emph{{A gauge-invariant symplectic potential for
  tetrad general relativity}},
  \href{http://dx.doi.org/10.1007/JHEP07(2018)040}{\emph{JHEP} {\bfseries 07}
  (2018) 040}, [\href{https://arxiv.org/abs/1804.09685}{{\ttfamily
  1804.09685}}].

\bibitem{Jacobson:2015uqa}
T.~Jacobson and A.~Mohd, \emph{{Black hole entropy and Lorentz-diffeomorphism
  Noether charge}},
  \href{http://dx.doi.org/10.1103/PhysRevD.92.124010}{\emph{Phys. Rev. D}
  {\bfseries 92} (2015) 124010},
  [\href{https://arxiv.org/abs/1507.01054}{{\ttfamily 1507.01054}}].

\bibitem{Prabhu:2015vua}
K.~Prabhu, \emph{{The First Law of Black Hole Mechanics for Fields with
  Internal Gauge Freedom}},
  \href{http://dx.doi.org/10.1088/1361-6382/aa536b}{\emph{Class. Quant. Grav.}
  {\bfseries 34} (2017) 035011},
  [\href{https://arxiv.org/abs/1511.00388}{{\ttfamily 1511.00388}}].

\bibitem{Barnich:2016rwk}
G.~Barnich, P.~Mao and R.~Ruzziconi, \emph{{Conserved currents in the Cartan
  formulation of general relativity}},  in \emph{{About Various Kinds of
  Interactions: Workshop in honour of Professor Philippe Spindel Mons, Belgium,
  June 4-5, 2015}}, 2016.
\newblock \href{https://arxiv.org/abs/1611.01777}{{\ttfamily 1611.01777}}.

\bibitem{Frodden:2017qwh}
E.~Frodden and D.~Hidalgo, \emph{{Surface Charges for Gravity and
  Electromagnetism in the First Order Formalism}},
  \href{http://dx.doi.org/10.1088/1361-6382/aa9ba5}{\emph{Class. Quant. Grav.}
  {\bfseries 35} (2018) 035002},
  [\href{https://arxiv.org/abs/1703.10120}{{\ttfamily 1703.10120}}].

\bibitem{Gomes:2018shn}
H.~Gomes and A.~Riello, \emph{{Unified geometric framework for boundary charges
  and particle dressings}},
  \href{http://dx.doi.org/10.1103/PhysRevD.98.025013}{\emph{Phys. Rev. D}
  {\bfseries 98} (2018) 025013},
  [\href{https://arxiv.org/abs/1804.01919}{{\ttfamily 1804.01919}}].

\bibitem{Margalef-Bentabol:2020teu}
J.~Margalef-Bentabol and E.~J.~S. Villase\~nor, \emph{{Geometric formulation of
  the Covariant Phase Space methods with boundaries}},
  \href{http://dx.doi.org/10.1103/PhysRevD.103.025011}{\emph{Phys. Rev. D}
  {\bfseries 103} (2021) 025011},
  [\href{https://arxiv.org/abs/2008.01842}{{\ttfamily 2008.01842}}].

\bibitem{G.:2021qiz}
F.~Barbero, J.~Margalef-Bentabol, V.~Varo and E.~J.~S. Villase\~nor,
  \emph{{Covariant phase space for gravity with boundaries: metric vs tetrad
  formulations}},  [\href{https://arxiv.org/abs/2103.06362}{{\ttfamily
  2103.06362}}].

\bibitem{DePaoli:2017sar}
E.~De~Paoli and S.~Speziale, \emph{{Sachs\textquoteright{} free data in real
  connection variables}},
  \href{http://dx.doi.org/10.1007/JHEP11(2017)205}{\emph{JHEP} {\bfseries 11}
  (2017) 205}, [\href{https://arxiv.org/abs/1707.00667}{{\ttfamily
  1707.00667}}].

\bibitem{Godazgar:2020gqd}
H.~Godazgar, M.~Godazgar and M.~J. Perry, \emph{{Asymptotic gravitational
  charges}},
  \href{http://dx.doi.org/10.1103/PhysRevLett.125.101301}{\emph{Phys. Rev.
  Lett.} {\bfseries 125} (2020) 101301},
  [\href{https://arxiv.org/abs/2007.01257}{{\ttfamily 2007.01257}}].

\bibitem{Compere:2019odm}
G.~Comp\`ere, \emph{{Infinite towers of supertranslation and superrotation
  memories}},
  \href{http://dx.doi.org/10.1103/PhysRevLett.123.021101}{\emph{Phys. Rev.
  Lett.} {\bfseries 123} (2019) 021101},
  [\href{https://arxiv.org/abs/1904.00280}{{\ttfamily 1904.00280}}].

\end{thebibliography}\endgroup

\end{document}